%% file: 4th_arxiv.tex
\documentclass[authoryear,preprint,review,12pt]{elsarticle}

\long\def\comment#1{}

\newif\ifdvipdfmgraphicx
        \dvipdfmgraphicxfalse   

\newif\ifUseOverleaf
        \UseOverleaffalse

\input{pream4}

 \def\myfigwidth{0.98\textwidth}    

\journal{Planetary and Space Science}

\begin{document}

\selectlanguage{english}

\begin{frontmatter}

\title{An estimate of resident time of the Oort Cloud new comets in planetary region}

\author[naoj,perc,chubu]{Takashi Ito}

\author[ksu]{Arika Higuchi}

\affiliation[naoj]{organization={Center for Computational Astrophysics, National Astronomical Observatory},
            addressline={Osawa 2--21--1}, 
            city={Mitaka},
            postcode={181--8588}, 
            state={Tokyo},
            country={Japan}}

\affiliation[ksu]{organization={Faculty of Science, Kyoto Sangyo University},
            addressline={Motoyama, Kamigamo},
            city={Kita--ku, Kyoto City},
            postcode={603--8555}, 
            state={Kyoto},
            country={Japan}}

\affiliation[perc]{organization={Planetary Exploration Research Center, Chiba Institute of Technology},
            addressline={2--17--1 Tsudanuma},
            city={Narashino},
            postcode={275--0016}, 
            state={Chiba},
            country={Japan}}

\affiliation[chubu]{organization={College of Science and Engineering, Chubu University},
            addressline={1200 Matsumoto--cho},
            city={Kasugai},
            postcode={487--8501}, 
            state={Aichi},
            country={Japan}}

\begin{abstract}
We describe the result of our numerical orbit
simulation
which traces dynamical evolution of new comets coming from the Oort Cloud.
We combine two dynamical models for this purpose.
The first one is semi-analytic, and it models an evolving comet cloud under galactic tide and encounters with nearby stars.
The second one numerically deals with planetary perturbation in the planetary region.
Although our study does not include physical effects such as fading or disintegration of comets, we found that typical dynamical resident time of the comets in the planetary region is about 
$10^8$
years.
We also found that the so-called planet barrier works when the initial orbital inclination of the comets is small.
A numerical result concerning the temporary transition of the comets into other small body populations such as transneptunian objects or Centaurs is discussed.
\end{abstract}


\end{frontmatter}

\newcommand{\orgbaselinestretch}{\baselinestretch}

 \renewcommand{\baselinestretch}{1.5}  

\section{Introduction\label{sec:intro}}
Since the historic prediction of the existence of a comet cloud that surrounds the solar system by Jan Hendrik Oort \citep{oort1950},
numerous amount of effort has been conducted both in observational and theoretical aspects of the Oort Cloud study.
Through various evidence, there is no doubt that the comet cloud exists with a shape of spherical shell stretching out to the farthest fringe of the solar system.
Discovery of a bunch of transneptunian objects (TNOs) whose aphelion distance is as large as several hundred au (e.g. (90377) Sedna, 2012 VP${}_{113}$, (541132) 2015 TG${}_{387}$) may mark the outskirt of the inner part of the inner Oort cloud \citep[e.g.][]{trujillo2014}.
Some studies have yielded an estimate that the Oort Cloud includes more than $10^{12}$ cometary objects, and its
spatial
spread is from $10^4$ to $10^5$ au centered at the Sun \citep[e.g.][]{dones2004}.
There are many studies as to how the comet cloud was created from the protoplanetary disk in the early solar system \citep[e.g.][]{duncan1987,dones2004,higuchi2006}.
There is also a hypothesis that the Oort Cloud is a product of captures of small bodies among a star cluster that
                 the Sun was supposed to belong to when it formed \citep[e.g.][]{levison2010,adams2010,wajer2024a,wajer2024b}.
Other hypothesis assumes that the cloud has undergone significant gravitational influence of the star cluster \citep[e.g.][]{fernandez1997,brasser2006,kaib2008}.
A new trend of research on the formation and evolution of comet clouds around general planetary systems and their detection is also emerging \citep[e.g][]{baxter2018,portegieszwart2021}.
Observational studies of the physical properties of comets coming from the Oort Cloud also mark a major development \citep[e.g.][]{saki2021,kwon2022,kwon2023}.
See \citet{dones2015} for a recent review of the current status of the Oort Cloud studies.

Among the many aspects of the Oort Cloud studies, our present work focuses how the comets dynamically interacts with the major planets.
In general, the study of planetary perturbations on the motion of comets requires large computational resources.
However, recent advances in computing technology have made the problem less serious, and a number of numerical results have been published along this line \citep[e.g.][]{silsbee2016,vokrouhlicky2019,fouchard2018,fouchard2023}.
Attempts are also underway to elucidate the dynamical state from the TNO disk to the inner edge of the Oort Cloud, using the distribution and the orbital motion of the scattered disk objects as a clue \citep[e.g.][]{batygin2021,nesvorny2023,hadden2024}.
Understanding of the motions and resonant states of objects with large orbital inclination in the planetary region, such as those coming from the Oort Cloud, is also improving \citep[e.g.][]{morais2017,gallardo2019a,li2021}.
In this article we particularly focus on the statistics of dynamical resident time of the new comets coming from the Oort Cloud in the planetary region (we will give our definitions of new comet and planetary region later).
We also pay attention to the comets' spatial penetration across planetary orbits.

We combine two models for our purpose.
The first model takes care of an evolving comet cloud in an analytic (and partly semi-analytic) way.
This model includes perturbations from galactic tide and stellar encounters with nearby stars.
The other model numerically deals with planetary perturbation.
Using a combination of the two models, we try to quantify dynamical characteristics that the comets show in the planetary region.
Here are our basic questions:
How long do the comets stay in the planetary region?
How efficient (or inefficient) is the so-called
planet barrier?
And, what kind of small body populations do the comets go through before they get ejected out of the solar system?
Our models are still primitive in some aspects.
For example, our models do not include any effects of physical evolution of comets such as waning and disintegration.
Also, galactic tidal force and stellar encounters are included only in the first model, and not in the second model.
However we believe our numerical result sheds some lights on the role that the Oort Cloud comets have played in the dynamical history of the small solar system populations as a whole.

Section \ref{sec:genOCNC} describes our first model and the initial state of an evolving comet cloud under galactic tide and encounters with nearby stars.
In Section \ref{sec:planetarypurt} we explain how we deal with planetary perturbation in our second model.
Section \ref{sec:result} goes to descriptions of major results obtained through our model calculations.
Section \ref{sec:summary} is devoted to a short summary of this paper.
Appendices are attached at the end of the paper for giving auxiliary information.
Note that when we refer to any small bodies in this paper, we deal with only one kind: the objects that fall from the Oort Cloud and reach the planetary region for the first time.
We may use a variety of words to describe them: comet, new comet, Oort Cloud comet, cometary object, particle, planetesimal, and so on.
But they all refer to the same type of object in this study, and there should not be any confusions about the terms.

 \section{Evolving comet cloud\label{sec:genOCNC}}
\begin{figure}[!htbp]
  \includegraphics[width=\myfigwidth]{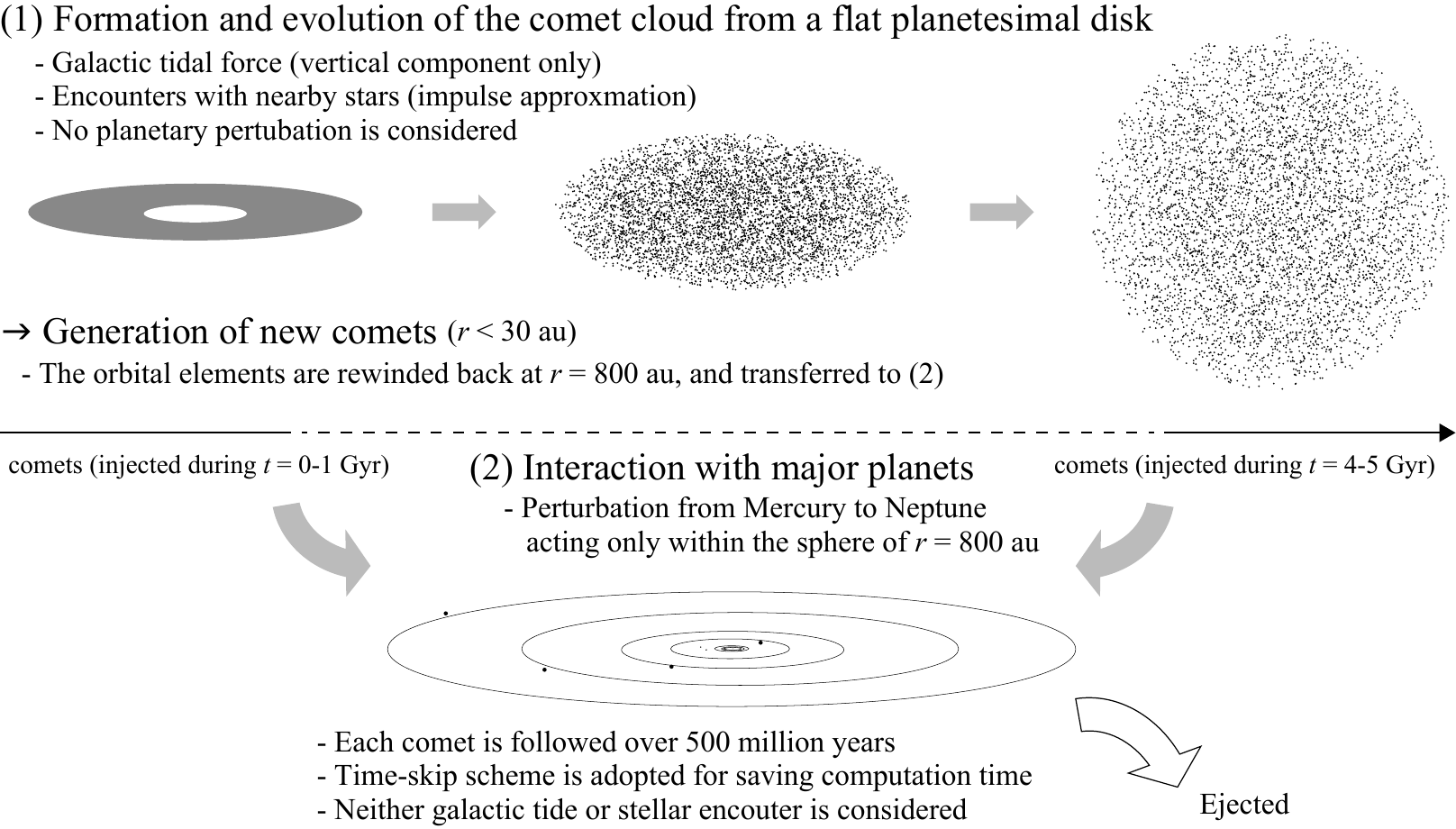}
\caption[]{%
Schematic illustration of how we model the generation of new comets and their injection into the planetary region.
The upper part of the figure illustrates the generation of the new comets in an evolving comet cloud perturbed from galactic tide and encounters with nearby stars.
This is our first model (numbered as (1)).
The orbit information of the new comets produced during the first 1 billion years and the last 1 billion years in the first model is transferred to the second model which is depicted in the lower part of the figure (numbered as (2)).
In the second model, the orbit of each comet is numerically integrated under the gravitational perturbation from the
eight
major planets up to 500 million years.
There is no consideration of the galactic tide or the stellar encounters in the second model.
Also, an approximation scheme that we call ``time-skip'' is employed for saving computation time (see Section \ref{ssec:timeskip} for the detail of the scheme).
}
\label{fig:oc-evol-schem}
\end{figure}

Our first model generates the comets in an evolving comet cloud.
Output from this model is used as input to our second model which is about numerical orbit
integration (propagation)
of the comets in the planetary region.
Transfer of objects between the first and second model is one-way: The comets handed to the second model do not return to the first model.
Figure \ref{fig:oc-evol-schem} is a schematic illustration as to how we use the two models.
In this section
we detail the first model---how we simulate the evolution of the comet cloud that generates new comets.

\subsection{Initial condition---a planetesimal disk\label{ssec:flatdisk}}
Our first model initially starts from a flat disk that consists of a swarm of massless particles.
The particles in the disk represent planetesimals that have been scattered by major planets in the early solar system.
The flat planetesimal disk later deforms into three-dimensional, nearly isotropic state particularly at its outer part in the timespan of 100 million to 1 billion years.
Galactic tidal force and encounters with nearby stars cause this.

\begin{figure}[!htbp]
 \includegraphics[width=\myfigwidth]{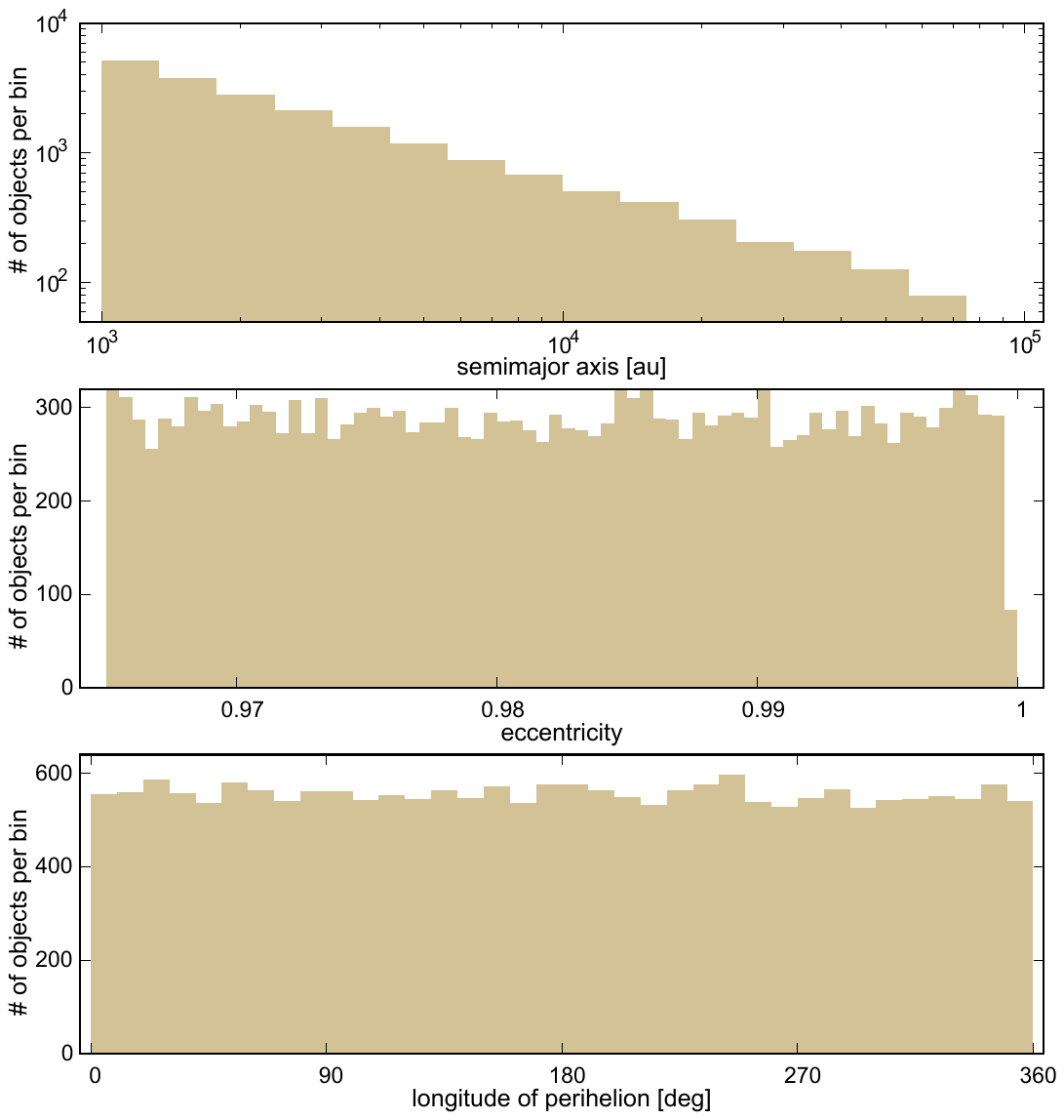}
\caption[]{%
The distribution of semimajor axis (top), eccentricity (middle), and longitude of perihelion (bottom) of the objects in the initial planetesimal disk in our study.
The number of objects that this plot includes is 20,000.
The top panel for semimajor axis $a$ indicates the relationship of
$N(a)\, \propto\, a^{-1}$.
This is equivalent to the probability distribution function
$\frac{dN(a)}{da}\, \propto\, a^{-2}$
mentioned in the main text and in \ref{appen:probability-e}.
}
\label{fig:oc-ini_disk}
\end{figure}

Initial configuration of the flat disk in this model is created through the three-body scatter of planetesimals by
massive planets \citep{higuchi2006}.
We treat all objects in the disk (and thus in the cometary cloud) as massless particles, ignoring their mutual gravity.
This treatment is justified by the low mass of the objects and their low number density \citep[e.g.][]{batygin2024a}.
The initial orbital distribution of objects in the disk is as follows:
The range of semimajor axis $a$ is between $10^3$ to $10^5$ au.
Eccentricity $e$ is uniformly distributed between 0.965 and 1.
The combination of $a$ and $e$ are chosen so that all the particles have perihelion distance $q = a(1-e) = 35$ au.
Inclination $I = 0$ (a completely flat disk), therefore longitude of ascending node is undefined in the initial condition.
Argument of perihelion (or longitude of perihelion) and mean anomaly are randomly selected between 0 and $2\pi$.
All the objects placed the initial disk are in the prograde orbits (i.e., they have positive mean motion $n > 0$), and there is no object initially orbiting in the retrograde way.
The distribution of initial semimajor axis, eccentricity, and longitude of perihelion is visualized in Figure \ref{fig:oc-ini_disk}.
Following \citet{higuchi2006,higuchi2007}, we assume that the differential number distribution of the objects follows the relationship
$\frac{dN(a)}{da} \, \propto \, a^{-2}$
where $N$ is the number of objects with semimajor axis $a$.
This relationship is seen in the top panel of Figure \ref{fig:oc-ini_disk} as
$N(a) \, \propto \, a^{-1}$.
When semimajor axis of objects has a differential number distribution (probability density function) of $a^{-2}$, and if their initial perihelion distance $q = a(1-e)$ is a common constant to all the objects (such as $q=35$ au in our model),
we find that the distribution of objects' eccentricity becomes uniform as seen in the middle panel of Figure \ref{fig:oc-ini_disk}.
We will show the reason in \ref{appen:probability-e}.
Note that the histograms in Figure \ref{fig:oc-ini_disk} represent the statistics of 20,000 objects.
Later we increase the number much more to obtain more reliable statistics.

We can justify the use of the initial perihelion distance ($q_0 = 35$ au) for the entire cometary objects (planetesimals) in our model as follows.
First, in the early solar system, long-period comets were formed as a population of planetesimals scattered by the major planets.
Here we assume that the outer edge of the planetary region is defined by the semimajor axis of the present-day Neptune, $a_\mathrm{N} = 30$ au.
Then, in the region within a heliocentric distance of $r \sim a_\mathrm{N}$, the planetesimals will always be subject to scattering from the major planets.
Eventually, this is expected to greatly reduce the number of the objects whose perihelion is located inside the planetary region.
For this reason, the initial perihelion distance $q_0$ of the long-period comets would be $q_0 > a_\mathrm{N}$.
Second, as we will describe in more detail in Section \ref{ssec:gtide}, such objects would be strongly affected by the
\citeauthor{vonzeipel1910}--\citeauthor{lidov1961}--\citeauthor{kozai1962b}
oscillation driven by the galactic tide because of the objects' large initial eccentricity.
The oscillation gradually decreases comet's eccentricity while maintaining their semimajor axis.
Consequently, the comet's perihelion moves outward, and eventually their perihelion gets detached from the planetary region.
This is the justification why we gave the initial value of $q_0 = 35$ au for the entire planetesimal disk objects.

Note that the comet's orbital inclination $I$ would slightly increase while its perihelion distance $q$ is raised from 30 au to 35 au.
However, the increase should be small, about a few degrees.
We can estimate this value by considering the conservation of the vertical component of the comet's angular momentum, $\sqrt{1-e^2} \cos I$, in the
\citeauthor{vonzeipel1910}--\citeauthor{lidov1961}--\citeauthor{kozai1962b} oscillation.
This fact justifies our assumption that the initial planetesimal disk is given $q_0 = 35$ while its inclination is all 0, i.e., the planetesimal disk is coplanar with respect to the ecliptic.

As mentioned, we assume that the comet cloud initially has a flat, planar shape without any orbital inclination $(I=0)$ with respect to the solar system invariable plane.
Here is our justification for this assumption.
In order to investigate the influence of initial conditions on the dynamical evolution of the Oort Cloud comets, generally it is desirable to prepare as simple conditions as possible.
In this model, not only the inclination but also the perihelion distance is the same for all the comets ($q=35$ au).
Therefore it would make no sense to give a very realistic distribution only to orbital inclination.
This is the reason why we set the initial inclination $I$ of all the objects to zero.

Note that all the orbital elements we use in this manuscript are osculating ones, not averaged (or secular) ones in the viewpoint of perturbation theory.
Also, we deal with the orbital elements in the heliocentric coordinates, not in the barycentric coordinates or in the hierarchical Jacobi coordinates except when we apply the time-skip scheme to the orbital motion of the comets (see Section \ref{ssec:timeskip}).
This may seem against the convention in the cometary dynamics where orbital elements are often described in the solar system barycentric coordinates, particularly in the distant region such as 250 au from the Sun \citep[e.g.][]{dybczynski2011,krolikowska2014}.
However, we believe our treatment in this manuscript does not cause practical problems because our main interest is the interaction of the Oort Cloud comets in the planetary region ($r \lesssim 30$ au) where orbital elements of objects are usually described in the heliocentric coordinates.

\subsection{Galactic tidal force\label{ssec:gtide}}
One of the main dynamical effects that work on the Oort Cloud is the galactic tide.
In the solar system neighborhood, the galactic tidal force is generally weaker than the solar gravity.
Therefore we can treat the galactic tide as a perturbation against the Keplerian motion of the Sun--comet system \citep[e.g.][]{harrington1985,byl1986,heisler1986,binney1987,fouchard2004,fouchard2005,fouchard2006}.
In particular, we consider only the vertical component of the tide whose magnitude is proportional to $-z$ where $z$ is the vertical distance of a comet from the galactic plane \citep[e.g.][see also \ref{appen:gtfunc} of this paper]{heisler1986,higuchi2007}.
This treatment is equivalent to regarding the galaxy as a homogeneous and axisymmetric plane that includes the Sun and extends infinitely.
We can interpret how this force works through an analogy to Hooke's law which explains the restoring force as a function of spring extension.
The fundamental part of this approximation is that, it makes the Sun--comet--galaxy system integrable after the procedure of canonical averaging with cometary semimajor axis being constant.
It also yields secular, time-dependent, and explicit solutions of orbital elements expressed through elliptic functions and elliptic integrals.
After the canonical averaging procedure, the vertical component of the cometary angular momentum with respect to the galactic plane becomes constant due to the axial symmetry of the galactic tidal force.
This component involves only eccentricity and inclination of the comet, and both experience a long-term oscillation with the timespan of $10^9$ years or longer.
This is nothing but a typical manifestation of the so-called \citeauthor{vonzeipel1910}--\citeauthor{lidov1961}--\citeauthor{kozai1962b} oscillation \citep{vonzeipel1910,lidov1961,kozai1962b,ito2019} which realizes a drastic change of the perihelion distance of comets \citep[e.g.][]{merritt2013}.
This is a major cause of the generation of the new comets from the Oort Cloud.

\subsection{Encounters with nearby stars\label{ssec:sencounter}}
Another major perturbation that affects the comet cloud evolution is close encounters with nearby stars.
This process randomly diffuses all the orbital elements of comets, often causing their ejection out of the cloud.
Although there are many studies being published on this subject \citep[e.g.][as a recent example]{torres2019}, in this study we follow the way \citet{higuchi2015} employed to model the stellar encounters.
\citet{higuchi2015} adopted the classical impulse approximation \citep[e.g.][]{opik1932,rickman1976,weissman1980,rickman2005} for calculating the momentum change of comets at each encounter with stars.
They presumed that each star goes through the vicinity of the solar system at a constant velocity along a straight line, and applied the impulse approximation when a star approached a comet and the Sun to the closest.
As for the distribution of stars' velocity, mass, and number density, \citet{higuchi2015} assumed those estimated from an observation of the current solar system neighborhood \citep{garcia-sanchez1999,garcia-sanchez2001}.
\citet{higuchi2015} did not consider giant molecular clouds as a source of flyby impulses, but just considered stars.
They also kept the Sun at a constant galactocentric distance throughout the calculation, and adopted the stellar parameters published in \citet[][their Table I]{rickman2008}.
\citet{rickman2008} (and hence we) adopted the total density of the galactic disk in the solar neighborhood is $\rho \sim 0.1 M_\odot \mathrm{pc}^{-3}$ where $M_\odot$ is the solar mass \citep[e.g.][]{holmberg2000}.
Consult \citet[][their Section 3.2]{higuchi2015} for the details as to how they modeled the stellar encounters.
Note that the impulse approximation and its outcome would be invalid when a very massive star passes the solar system with a small distance
such as $\ll 1,000$ au.
However, since the probability of such a very close encounter is tiny (see below),
we do not consider this kind of rare circumstance in the present study.

\begin{figure}[!htbp]
 \includegraphics[width=\myfigwidth]{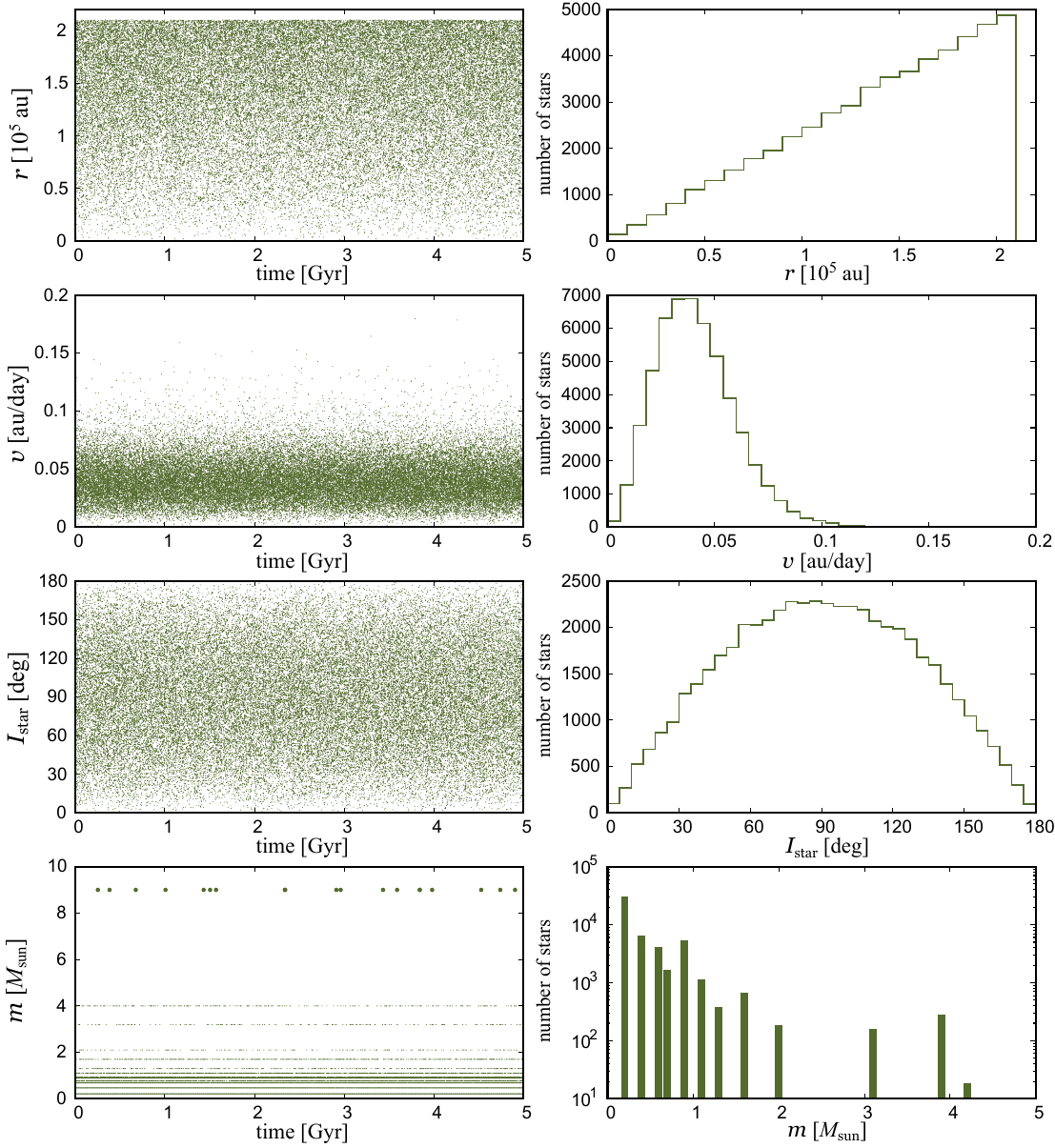}
\caption[]{%
Statistics of the stellar close encounters with the solar system occurred in the star set A.
$r$ is the closest heliocentric distance between the star and the Sun.
$v$ is the magnitude of the star's velocity relative to the Sun at that time
$I_\mathrm{star}$ is the orbital inclination of the star relative to the ecliptic plane at closest approach.
$m$ is the mass of the star.
Note that in the frequency distribution of mass $m$ in the lower right panel, the upper limit along the horizontal axis is 5 $M_\odot$.
Stellar encounters with masses larger than this are rare in our model.
We can visually resolve the occurrence of these encounters of massive stars just by looking at the time series in the lower left panel.
}
\label{fig:oc-starenc-stats_s5A}
\end{figure}

In the present study, we deal with two cases of comet cloud that are affected by two different sets of stars.
We name them the star set A and B.
Their difference is in the random number sequences used to generate the stars' mass, initial position, and initial velocity.
For avoiding clutter, we describe only the result obtained through the star set A in the main text of this paper.
We give the result obtained through the star set B in \ref{suppl:star3B}.
However, we do not see any substantial difference between the results from the two star sets.

In the star set A, our model generated 52558 close encounters between stars and the Sun over 5 billion years.
This number is 52784 in the star set B.
Figure \ref{fig:oc-starenc-stats_s5A} shows the time series and frequency distribution of some of the quantities related to the stellar encounters generated from the star set A.
In this model,
we gave the Maxwellian distribution to the star's velocity $v$.
The star's inclination $I_\mathrm{star}$ has the isotropic distribution.
As described above, the mass $m$ of the stars has a distribution along \citet[][their Table I]{rickman2008}.
As for the closest heliocentric distance $r$ between the star and the Sun, we designed our model so that the frequency (probability distribution) of the stellar encounters is simply proportional to $r$.
Because of this configuration, stellar encounters in the very close vicinity of the Sun occur only rarely.
Among the 52558 encounters happened in the star set A over 5 Gyr, we find encounters with $r < 2 \times 10^4$ au happened 507 times.
The encounters with $r < 2 \times 10^3$ au happened just 8 times.
Among the 52784 encounters happened in the star set B over 5 Gyr, we find encounters with $r < 2 \times 10^4$ au happened 486 times.
The encounters with $r < 2 \times 10^3$ au happened just 5 times.
We did not observe any encounters with $r < 1000$ au in either of the star set.

Note that in our present model, we assume that the number density and mass distribution of stars in the solar neighborhood is constant over the entire period of our numerical calculation ($t=0$--5 Gyr).
This assumption should be reasonable for the modern solar system.
However, circumstance may have been different at the time of the birth of the solar system.
If the Sun was born in a globular cluster, the stellar density in the solar neighborhood could have been much higher in the early solar system than it is today
\citep[e.g.][]{fernandez1997}.

\subsection{Evolution of the comet cloud\label{ssec:oc-evol}}

\begin{figure}[!htbp]
 \includegraphics[width=\myfigwidth]{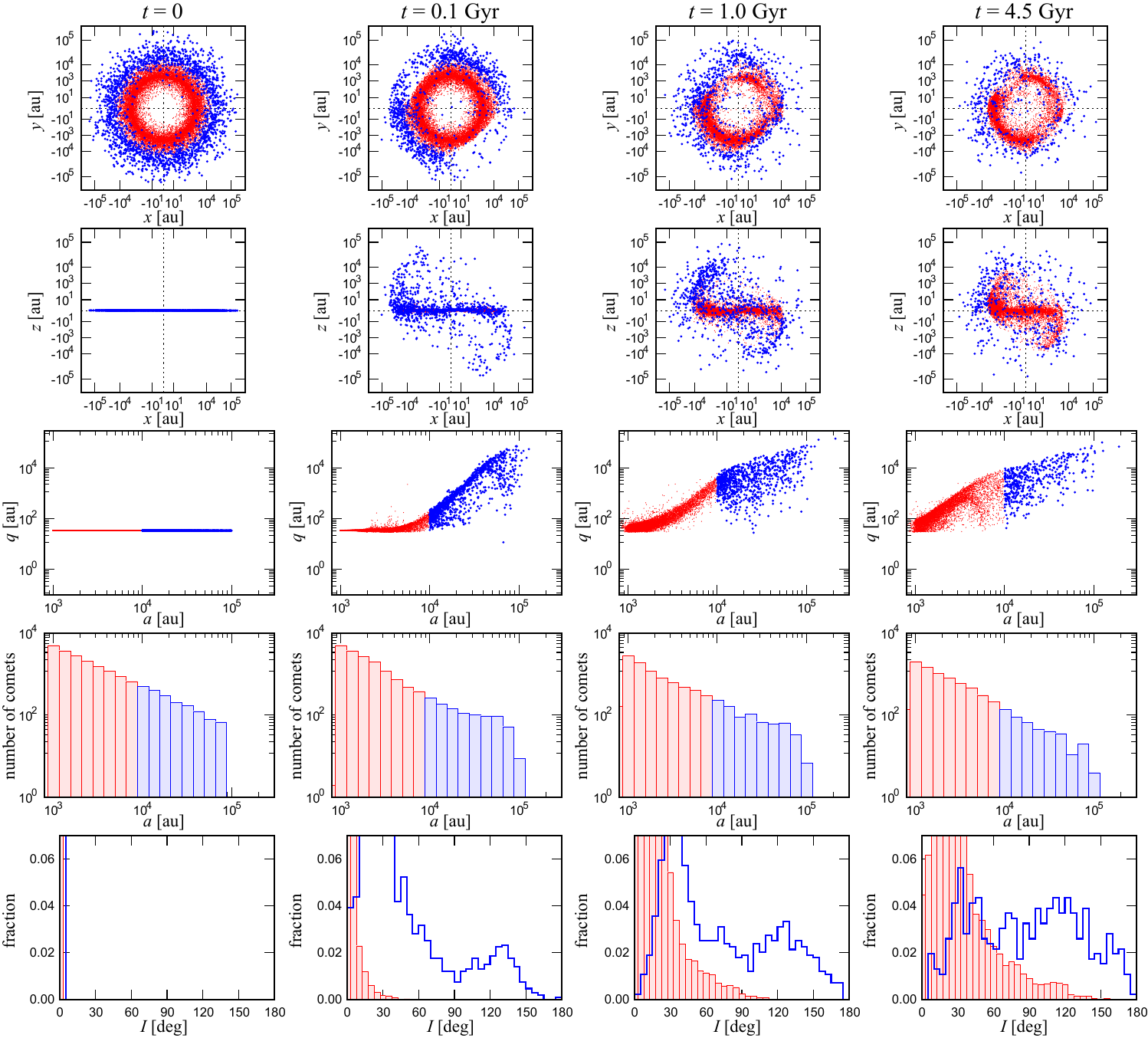}
\caption[]{%
Example snapshots of the comet cloud evolution under the start set A.
From the left, each of the four-panel column indicates the status when time $t=0$, 0.1, 1.0, and 4.5 Gyr.
Top row:
spatial distribution of the comets seen from the north (i.e. projected on the $(x,y)$ plane).
The $(x,y)$ plane corresponds to the current ecliptic, and the $x$-axis is directed toward the current vernal equinox.
The origin of the coordinate is the Sun.
The axis unit is au, but note that the distance is non-linearly normalized by a power-law, $x^{0.2}$ or $y^{0.2}$, so that we can have a better visibility of the inner part of the cloud.
The objects with semimajor axis $a < 10,000$ au are plotted in red in all the panels, while other objects ($a \geq 10,000$ au) are plotted in blue (we drew the blue dots slightly larger than the red dots).
Second top row:
spatial distribution of the comets seen from the current ecliptic (along the $(x,y)$ plane).
Third top row: scatter plots of semimajor axis $a$ and perihelion distance $q$ of the objects.
Fourth top row: absolute number distribution of semimajor axis $a$ of the comets.
Bottom row: fractional distribution of orbital inclination $I$ of the comets.
}
\label{fig:oc-evol-ah_s5A}
\end{figure}

As an illustration of how the flat planetesimal disk dynamically evolves under the perturbations, we created a set of snapshots of the evolving comet cloud under the star set A in Figure \ref{fig:oc-evol-ah_s5A}.
In this figure, we define the $y$-axis as the intersection line of the galactic plane and the ecliptic plane.
Then, the current vernal equinox seen from the Sun is roughly located along the positive $x$-direction.
However, this figure is not intended for discussing the exact locations of comet's longitude of ascending node or argument of perihelion.
Therefore it is not quite important which direction the vernal equinox is on this figure.
Note that since the inclination value between the galactic plane and the ecliptic plane is about $60^\circ$, we can regard that the galactic plane would be approximately present near the line of $z = \tan 60^\circ x \approx 1.73 x$ in the second row panels which represent the $(x,z)$ plane.

In Figure \ref{fig:oc-evol-ah_s5A}, we use two different colors depending on the semimajor axis of each object.
The red parts denote the comets with osculating heliocentric semimajor axis of $a < 10,000$ au, while other comets ($a \geq 10,000$ au) are plotted in blue.
Note that in this figure, a comet denoted in red at $t=0$ can be rendered in blue later due to the change of its semimajor axis.
Note also that in the top two rows representing the scatter plots in the $(x,y)$ and $(x,z)$ planes, the
spatial
distance is non-linearly normalized by a power-law, $x^{0.2}$ or $y^{0.2}$:
The distance of $10^1$ is scaled to $\left({10^1}\right)^{0.2} \sim  1.58$, 
the distance of $10^3$ is scaled to $\left({10^3}\right)^{0.2} \sim  3.98$, and
the distance of $10^5$ is scaled to $\left({10^5}\right)^{0.2} \sim 10.0$.
This is for getting a better visibility of  the inner part of the comet cloud.
In this figure the initial number of the objects drawn in red is 18119, and that of the blue objects is 1881.
The total number of objects $(18119 + 1881 = 20000)$ is much smaller than what we actually employ in the model calculation later, but here we did it for preventing the plots from being too busy with too many dots.
Note that the numbers of the comets in the inner (18119) and the outer (1881) part of the cloud are common to the star sets A and B, as they represent the initial condition of the cloud before any perturbation has been applied.

The panels in the leftmost column of Figure \ref{fig:oc-evol-ah_s5A} indicate the initial state of the disk $(t = 0)$.
The disk is flat, and the objects are totally confined in the $(x,y)$ plane with a uniform perihelion distance of $q = 35$ au.
In less than 100 million years from the start ($t=0.1$ Gyr), the outer part of the disk gets substantially distorted (the panels in the second left columns), while the inner part of the cloud still remains nearly flat.
The outer part of the comet cloud approaches its isotropic state over about 1 billion years as seen in the panels of the second right columns ($t=1.0$ Gyr).
The evolution of the cloud gradually slows down since this stage, and its outlook at $t = 4.5$ Gyr (the panels in the rightmost columns) is not quite different from that at $t = 1.0$ Gyr.
Note that the inner part of the cloud still concentrates around the $(x,y)$ plane (i.e. the current ecliptic) even at this point.

Readers may notice a distorted feature of the disk at $t = 0.1$ Gyr seen in the second left panel on the top row of Figure \ref{fig:oc-evol-ah_s5A}.
Here let us mention two properties of the objects in the comet cloud that we model.
First, longitudes of ascending node (denoted as $h$ throughout this paper following the notation of the Delaunay canonical elements instead of the conventional $\Omega$) of all the comets change in the same direction (i.e. they decrease) due to the galactic tide.
Second, the decrease rate of longitude of ascending node $(\frac{dh}{dt})$ of each comet is not constant, but it significantly varies in time \citep[e.g.][]{higuchi2020}.
When we look at the time variation of longitude of ascending node of many comets at the same time, these two properties can make a structure like a compression wave in the comet cloud, in particular in the early stage of its dynamical evolution such as $t \sim 0.1$ Gyr.
In our model, this structure is easily achieved because the initial orbital inclination of all the comets are zero.
As the comet cloud dynamically evolves, the comet's ascending node is scattered over all directions, and their inclination distribution approaches to the isotropic state.
Then the compression structure becomes less visible than in the early stages.
Consult \citet[][the top left panel of her Figure 3]{higuchi2020} for typical time evolution of longitude of ascending node of the comet cloud objects under the galactic tide.

Readers may also notice an apparent cut of the distribution of the comet cloud objects' perihelion distance $q$ at around $10^5$ au in the third row panels of Figure \ref{fig:oc-evol-ah_s5A}.
In other words, we do not see many comets beyond the region of $q \gtrsim 10^5$ au.
The existence of such apparent boundaries depends on the initial conditions in our model.
In the range of the energies and angular momentum (in particular its vertical component) considered in this study, it is not common for eccentricity $e$ of the comets under the influence of galactic tidal force to become so small that the object's perihelion distance is enhanced up to $q \gtrsim 10^5$ au.
A typical numerical demonstration of this kind is presented in \citet[][3rd row of her Figure 2]{higuchi2020}.
Roughly speaking, the comet's initial eccentricity is given as $e_0 \sim 1$ in our model (Figure \ref{fig:oc-ini_disk}), and we can say that it is rare that a comet with this high initial eccentricity to have significant changes to achieve a near circular orbit $(e \sim 0)$ with a large perihelion distance such as $q > 10^5$ au.

As an additional information, we can also consult a numerical calculation presented in
\citet[][see the bottom right panel of their Figure 4]{higuchi2015}.
According to this figure, function form of the frequency distribution $N(e)$ of cometary eccentricity approaches $N(e) \, \propto \, e$ as the shape of the comet cloud approaches the isotropic state.
This also indicates that not many comets with small eccentricity (hence with large perihelion distance) are produced in the comet cloud.
This is another reason why we see few comets in the region of $q \gtrsim 10^5$ au in Figure \ref{fig:oc-evol-ah_s5A}.

Note also that the comet cloud cannot become isotropic by the galactic tidal force alone like Figure \ref{fig:oc-evol-ah_s5A}.
Besides the galactic tidal force, stellar encounters acting from random directions are necessary.
\citet{higuchi2015} have suggested this fact, but as a demonstration, in \ref{appen:nostarevol} we show an example dynamical evolution of a flat planetesimal disk with only the galactic tidal force.
We will visually see that the initial planetesimal disk does not get isotropic enough in the absence of stellar encounters,

\begin{figure}[!htbp]
 \includegraphics[width=\myfigwidth]{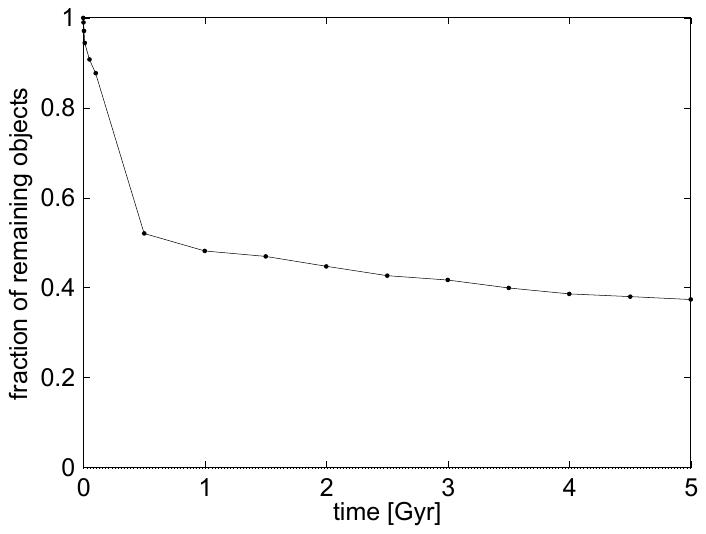}
\caption[]{%
The relative fraction of remaining objects in the comet cloud and its time variation when using the star set A.
}
\label{fig:oc-remoc_s5A}
\end{figure}

Although the galactic tidal force approximated in our model never creates objects on parabolic (eccentricity $e=1$) or hyperbolic $(e > 1)$ orbits, the encounters with stars do.
Many objects get ejected out of the cloud because of this.
In our first model we assume that a comet is removed from the cloud when its heliocentric distance $r$ becomes larger than 1 pc regardless of its eccentricity.
In other words, as long as a comet's heliocentric distance $r$ remains $r \leq 1$ pc, we do not remove it from the cloud even if its eccentricity satisfies $e \geq 1$.
Figure \ref{fig:oc-remoc_s5A} shows the relative fraction of remaining objects in the cloud when using the star set A and its time variation normalized to unity at the initial epoch, $t=0$.
We find that
$\sim 63${\%} of the objects initially in the flat disk are removed from the cloud under the star set A (this number is
$\sim 57${\%} when we use the star set B).
Note that the abrupt fractional decrease of the remaining objects in the comet cloud seen around $t \sim 0.45$ Gyr in Figure \ref{fig:oc-remoc_s5A} is due to the depletion of the comet population in the cloud.
More specifically, this decrease was caused by a strong comet shower at that time which we will mention later.

\subsection{Generation of new comets\label{ssec:genocnc}}
\begin{figure}[!htbp]
 \includegraphics[width=\myfigwidth]{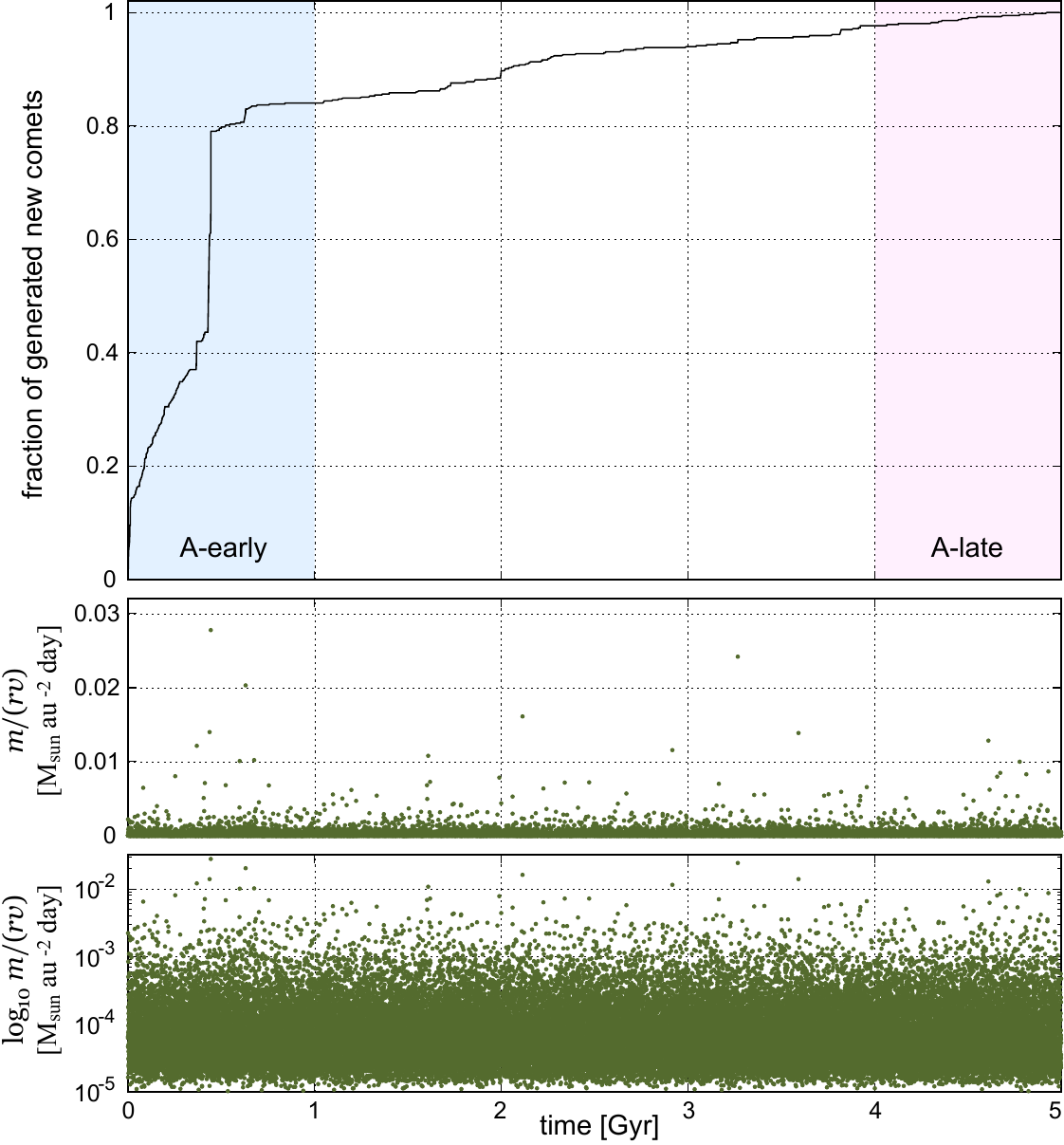}
\caption[]{%
(Top)    The cumulative fraction of the generated new comets and its time variation.
         The vertical value is normalized to unity at $t=5$ Gyr.
(Middle) The value of $\frac{m}{rv}$ in the linear scale.
(Bottom) The value of $\frac{m}{rv}$ in the logarithmic scale.
}
\label{fig:oc-genoc-starmrv_s5A}
\end{figure}

Among the objects that we initially placed in the planetesimal disk, 
$\sim 56${\%} fell into the planetary region during the entire period of five billion years under the star set A (this number is
$\sim 55${\%} when we use the star set B).
These objects become new comets.
We define that an object is recognized as a new comet when its heliocentric distance $r$ satisfies the condition, $r < 30$ au.
We show a timeline of the cumulative generation of the new comets under the star set A in the top panel of Figure \ref{fig:oc-genoc-starmrv_s5A}.
Throughout the rest of this manuscript, we pay particular attention to the two periods named ``A-early'' and ``A-late'' highlighted in this panel in colors.
A-early denotes the period of $t=0$--1 Gyr which corresponds to the early stage of the solar system history.
On the other hand,
A-late  denotes the period of $t=4$--5 Gyr which can be regarded as a proxy of late stage of the solar system history.
Similarly, we use the terms ``B-early'' and ``B-late'' for the evolution of the comet cloud under the star set B, and the result obtained in these periods are summarized in \ref{suppl:star3B}.
Note that the abrupt increase of the generation rate of the new comets seen around $t \sim 0.45$ Gyr in the top panel of Figure \ref{fig:oc-genoc-starmrv_s5A} corresponds to the abrupt decrease of the number of objects in the comet cloud that we saw in Figure \ref{fig:oc-remoc_s5A}.
This correlation supports the fact that most of the objects once out of the cometary cloud become new comets at some point.
Due to the strong comet shower that occurred $t\sim 0.45$ Gyr in this model, the production rate of the new comets became smaller after that.
This is simply because the total number of the objects remaining in the comet cloud got reduced due to the comet shower.

In the middle and bottom panels of Figure \ref{fig:oc-genoc-starmrv_s5A}, we show the values of stars' $\frac{m}{rv}$, which is a proxy of the intensity of stellar encounters responsible for the generation of new comets.
This quantity is shown in the linear scale in the middle panel, while it is shown in the logarithmic scale in the bottom panel to see its entire range.
We can intuitively and qualitatively understand the dynamical significance of the quantity $\frac{m}{rv}$ through its functional form.
Quantitatively, this quantity is roughly proportional to the velocity change that a comet undergoes when a star of mass $m$ encounters the Sun at distance $r$ and relative velocity magnitude $v$ \citep[e.g.][their Eq. (1)]{higuchi2015}.
More precisely speaking, we should have evaluated the quantity
$
  \frac{2 {\cal G} m}{v}\left(\frac{\bm{r}_\mathrm{c}}{r^2_\mathrm{c}} - \frac{\bm{r}}{r^2} \right)
$
using the position vectors $\bm{r}_\mathrm{c}$ (from the comet to the star) and $\bm{r}$ (from the Sun to the star).
However, we do not keep a record of the distance to a star ($r_\mathrm{c}$) of each comet when it encounters the star.
Since we only record the distance ($r$) of each star at its closest approach to the Sun, we approximate the quantity $\frac{m}{rv}$ as an indicator of the strength of the stellar encounter.
As we see, the period A-early is characterized by the outbreak of a strong comet shower at time $t \sim 0.45$ Gyr (the top panel of Figure \ref{fig:oc-genoc-starmrv_s5A}).
In the middle panel we find that a stellar encounter with a large value of $\frac{m}{rv} \sim 0.03$ occurred at this time.
In addition, many other stellar encounters happened with large $\frac{m}{rv}$ in A-early.
On the other hand, the number of new comets may not increase so much even if strong stellar encounters occur in later epochs.
This simply indicates that the number of comets in the comet cloud decreased in later periods, and that strong stellar encounters may not have directly led to the occurrence of new comets.

\begin{table}[!htbp]
\caption{%
Top 20 cases of stellar encounters with large $\frac{m}{rv}$ (second column from the left) that occurred when using star set A in $t=0$--5 Gyr.
They are listed in the descending order of $\frac{m}{rv}$.
The unit of $\frac{m}{rv}$ is $M_\odot \mathrm{au}^{-2} \mathrm{day}$ as in the vertical label of Figure \ref{fig:oc-genoc-starmrv_s5A}.
$I_\mathrm{star}$ is the orbital inclination of the star relative to the ecliptic plane at the time of the encounter.
We made the rightmost column (``period'') empty for the stellar encounters occurring at $1.0 < t < 4.0$ because our numerical calculation that includes planetary perturbations do not cover this period.
}
\begin{center}
\begin{tabular}{ccrccrl}
\hline
\multicolumn{1}{c}{$t$ (Gyr)} &
\multicolumn{1}{c}{$\frac{m}{rv}$} &
\multicolumn{1}{c}{$r$ (au)} &
\multicolumn{1}{c}{$v$ (au/d)} &
\multicolumn{1}{c}{$m$ ($M_\odot$)} &
\multicolumn{1}{c}{$I_\mathrm{star}$ (deg)} &
\multicolumn{1}{c}{period} \\
\hline
0.444004 & 0.02778818 &   1158.43 &  0.040384 &  1.30 &  37.242 & A-early \\
3.267966 & 0.02420527 &   4205.28 &  0.039297 &  4.00 & 112.184 &         \\
0.629985 & 0.02031706 &   1183.40 &  0.019548 &  0.47 &  21.622 & A-early \\
2.112779 & 0.01613107 &  16062.04 &  0.012351 &  3.20 &  45.377 &         \\
0.436382 & 0.01399442 &   1160.34 &  0.028944 &  0.47 &  28.292 & A-early \\
3.592014 & 0.01388100 &  66386.80 &  0.009767 &  9.00 &  95.466 &         \\
4.609899 & 0.01284837 &   5492.08 &  0.011054 &  0.78 &  48.573 & A-late  \\
0.368166 & 0.01215071 &   3180.20 &  0.028467 &  1.10 &  16.492 & A-early \\
2.916269 & 0.01154608 &  61492.30 &  0.012676 &  9.00 &  66.913 &         \\
1.607797 & 0.01078214 &  15168.66 &  0.024457 &  4.00 &  48.389 &         \\
0.675639 & 0.01021277 &  24508.44 &  0.006792 &  1.70 & 132.182 & A-early \\
0.598337 & 0.01010898 &  33858.32 &  0.004967 &  1.70 &  25.900 & A-early \\
4.777367 & 0.01001181 &   4970.94 &  0.015673 &  0.78 &  63.873 & A-late  \\
4.931365 & 0.00867046 &   4971.34 &  0.025520 &  1.10 &  46.707 & A-late  \\
4.674387 & 0.00847771 &  50258.32 &  0.009388 &  4.00 & 112.446 & A-late  \\
4.813792 & 0.00830500 &  10039.35 &  0.020389 &  1.70 &  39.343 & A-late  \\
0.253076 & 0.00805723 &  93626.89 &  0.011930 &  9.00 &  69.127 & A-early \\
4.658133 & 0.00800983 &  33863.34 &  0.011798 &  3.20 & 145.420 & A-late  \\
1.990518 & 0.00784493 &  25232.64 &  0.016166 &  3.20 &  92.377 &         \\
1.618938 & 0.00724928 &  12538.61 &  0.012102 &  1.10 & 116.870 &         \\
\hline
\end{tabular}
\end{center}
\label{tbl:starmrv-top20_s5A}
\end{table}

As more detailed numerical examples, in Table \ref{tbl:starmrv-top20_s5A} we show properties of 20 stellar encounters with the largest $\frac{m}{rv}$ that occurred in the star set A over the entire period of $t=0$--5 Gyr.
This table can be a measure of the intensity of stellar encounters that have occurred in our model calculations.
Table \ref{tbl:starmrv-top20_s5A} includes the list of orbital inclination of the encountered stars as well.
There we may find another fact that, even when $\frac{m}{rv}$ is large, the perihelion distance of comets would not change much if the star approaches from a direction nearly perpendicular to the comet's orbital plane.
Such a stellar encounter may not produce an intense cometary shower.

As a specific example of the above statement, we pay attention to the second record from the top of this table ($t = 3.267966$ Gyr).
In this event, the star approached the solar system from a direction close to perpendicular to the ecliptic plane ($I_\mathrm{star} \sim 112^\circ$), and the value of $\frac{m}{rv}$ is the second largest among the entire 5 Gyr under the star set A.
However, looking at the top panel of Figure \ref{fig:oc-genoc-starmrv_s5A}, we find that the cometary shower that occurred at this time ($t \sim 3.268$ Gyr) is not quite remarkable, even if we consider the fact that the number of objects remaining in the cloud is already much reduced by this point.
We can partly understand this by considering Gauss's form of the equations for orbital elements that express the time-derivatives of orbital elements through the perturbing force decomposed into three directions \citep[e.g.][Eq. (33) in p.~301. In particular, see the terms proportional to the force component $W$ perpendicular to the orbital plane of perturbed object]{brouwer1961}.
When orbital inclination of the star $I_\mathrm{star}$ with respect to the ecliptic plane is large, a comet orbiting near ecliptic gets a larger impulse component perpendicular to its orbit, and smaller impulse components along the orbital plane.
In order to generate a shower of comets whose orbits are along the ecliptic plane such as those in the inner Oort Cloud, their semimajor axis $a$ or eccentricity $e$ must change, not their inclination.
The change of $a$ and $e$ would need a large in-plane impulse component, not the perpendicular component.
This can explain why the stellar encounters of stars with large orbital inclination do not necessarily produce strong comet showers such as what happened at $t = 3.267966$ Gyr.

As seen in Figure \ref{fig:oc-genoc-starmrv_s5A}, nearly 90{\%} of the entire comets is produced during the period A-early, while only $\sim$1{\%} is generated in the period A-late.
In our numerical model,
we plan to prepare similar number of comets for both the period A-early and A-late for a better statistical reliability.
For this purpose we followed the orbital evolution of about 
190 million objects in our first model using the star set A, picked
about 700 thousand among them for each of the periods, and used them as the inputs into our second model.
These numbers are almost common in both the calculations using the star set A and B.
Specifically speaking, we use
710,107 objects for A-early, and
747,120 objects for A-late.
We will show the detailed statistics later as a table.

\begin{figure}[!htbp]
 \includegraphics[width=\myfigwidth]{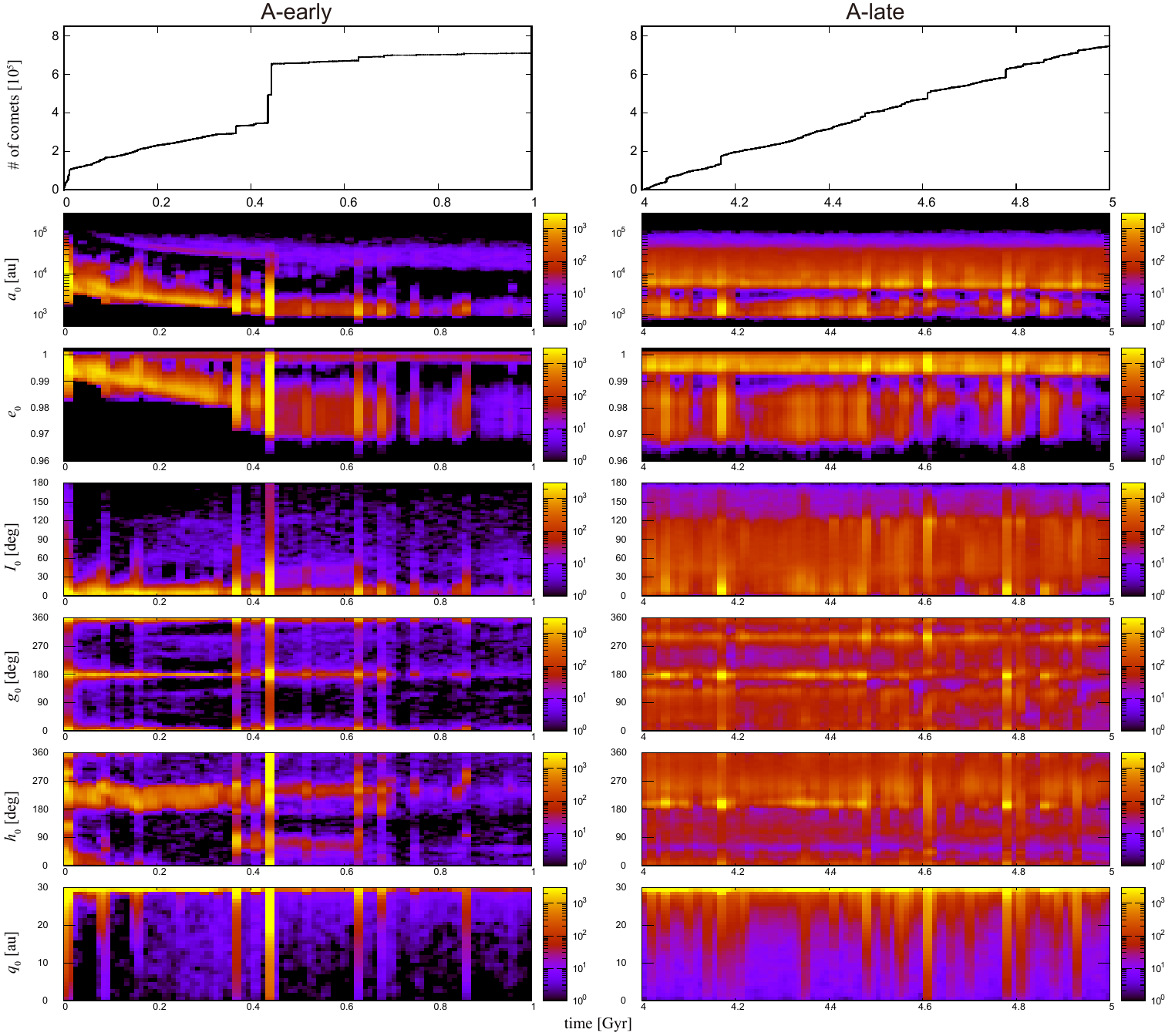}
\caption[]{%
A summary of how the new comets are generated from the comet cloud in each of the periods A-early and A-late.
The panels in the top row show the cumulative number of new comets generated in A-early and A-late.
The unit of the vertical axis is $10^5$ (comets).
The other panels are frequency maps of new comets' initial heliocentric orbital elements as a function of their generation time.
From the second top row to the bottom row,
semimajor axis (au),
eccentricity,
inclination (deg),
argument of perihelion (degree),
longitude of ascending node (degree), and
perihelion distance (au).
The color charts indicate the number of objects in the logarithmic scale.
The time interval along the horizontal axis is $0.01\, \mathrm{Gyr}$.
}
\label{fig:oc-genocnc-combined-t_s5A}
\end{figure}

Figure \ref{fig:oc-genocnc-combined-t_s5A} is a summary of how the new comets are generated from the comet cloud in each of the periods A-early and A-late.
The panels in the top row show the cumulative number of the new comets generated in each period.
The other panels in Figure \ref{fig:oc-genocnc-combined-t_s5A} are intensity maps of the new comets' initial heliocentric orbital elements in color as a function of their generation time.
In the second and the third rows, we see a decreasing trend in new comets' initial semimajor axis $a_0$ as well as in their initial eccentricity $e_0$ during the period A-early.
This trend indicates a process of relaxation of the initial conditions we gave to the planetesimal disk.
The planetesimal disk that we use as the initial state contains many objects with orbital elements similar to those of the new comets ($r < 30$ au) from the beginning.
These objects in the disk easily turn into the new comets during the first few hundred million years, undergoing perturbations from the galactic tidal force and the encounters with nearby stars.
More specific explanations are given as follows.
The galactic tidal force in our approximation does not cause any changes of semimajor axis of objects, but it changes their eccentricity.
This causes the perturbed object's perihelion distance to vary in the cloud, resulting in the generation of many new comets with the heliocentric distance of $r < 30$ au.
The timescale of this variation is shorter when the perturbed object's semimajor axis is larger \citep[e.g.][their Eq. (6)]{higuchi2007}.
This leads to the fact that the objects located in the outer part of the initial planetesimal disk (with larger semimajor axis) varies its eccentricity more quickly than those located in the inner part (with smaller semimajor axis).
Therefore the generation of the new comets (in other words, the reduction of perihelion distance of the objects in the disk below the threshold value) begins with the objects that have larger semimajor axis, and it gradually moves on to the objects that have smaller semimajor axis.
For the perihelion distance of an object with large semimajor axis to be reduced below the threshold value, its eccentricity must be high too.
This is also well illustrated in the panels for $e_0$ in Figure \ref{fig:oc-genocnc-combined-t_s5A} for the period A-early, and we see that the production of the new comets begins with the objects with higher eccentricity.
In contrast, during the period of A-late, the initial configuration that we gave to the planetesimal disk has already relaxed, resulting in a low rate of the new comet generation (Figure \ref{fig:oc-genoc-starmrv_s5A}).
This is the reason why the above mentioned decreasing trend for $a_0$ and $e_0$ which occurred in A-early is almost invisible during A-late.

In Figure \ref{fig:oc-genocnc-combined-t_s5A}, we see concentrations of the initial argument of perihelion $g_0$ of the new comets around 0 and $180^\circ$ (note that we use $g$ for argument of perihelion throughout this paper following the notation of the Delaunay canonical elements, instead of the conventional $\omega$).
The reason for this concentration is that the initial cometary cloud is perfectly flat $(I_0 = 0)$, eccentricity of the objects in it is very large $(e \sim 1)$, and the comets tend to stay near their aphelion in average.
We can better understand the reason by considering the Gauss's form of the equations of motion for the objects in the cloud.
We give more detailed explanation about it in \ref{appen:g0_180}.

Another feature that draws attention in Figure \ref{fig:oc-genocnc-combined-t_s5A} is the concentration of new comets' initial longitude of ascending node $h_0$ in a region between $180^\circ$ and $270^\circ$.
This is prominent in A-early, and also seen in A-late although weakly.
We understand that the cause of this concentration is the forced oscillation of nodal motion of the object's orbits in the planetesimal disk driven by the galactic tidal force as a secular perturbation \citep[e.g.][]{brouwer1961,murray1999}.
Actually, the ecliptic longitude of node of the galactic plane is about $270^\circ$ 
(we can confirm the value on websites such as
\href{https://ned.ipac.caltech.edu/forms/calculator.html}{NASA/IPAC Extragalactic Database} or
\href{https://lambda.gsfc.nasa.gov/toolbox/tb_coordconv.cfm}{Lambda---Tools at NASA Goddard Space Flight Center}).
Although the source of perturbation is not the galactic tide but the secular dynamical effect from Jupiter, this kind of node clustering of small bodies is also seen in the modern solar system such as in the orbital element distribution of the near-Earth asteroid populations \citep[e.g.][]{jeongahn2014} or that of the asteroid families in the main belt \citep[e.g.][]{hirayama1918,knezevic2003}.

\begin{figure}[!htbp]
 \includegraphics[width=\myfigwidth]{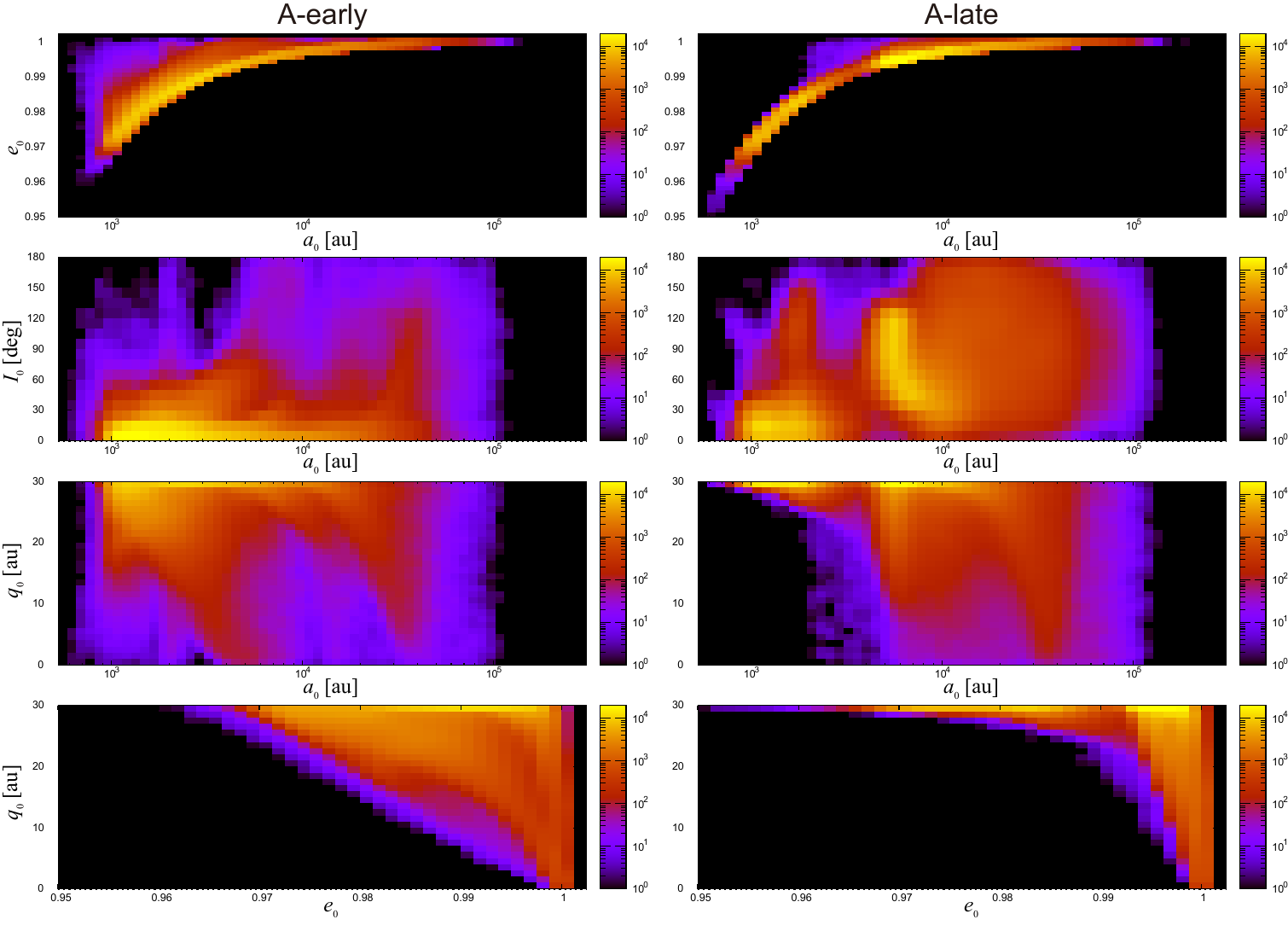}
\caption[]{%
Mutual dependence between new comets'
initial eccentricity $e_0$,
inclination $I_0$,
perihelion distance $q_0$, and
semimajor axis $a_0$.
Top:        $(a_0, e_0)$,
second top: $(a_0, I_0)$,
third top:  $(a_0, q_0)$,
bottom:     $(e_0, q_0)$.
The color charts indicate the number of objects in the logarithmic scale.
}
\label{fig:oc-aeIq0_s5A}
\end{figure}

Figure \ref{fig:oc-aeIq0_s5A} is a set of plots for mutual dependence between new comets' initial eccentricity, inclination, perihelion distance, and semimajor axis.
The $(a_0, e_0)$ panels at the top of this figure typically manifest that the new comets with large initial semimajor axis have large initial eccentricity because of their definition, $r < 30$ au.
In the period A-early, we see a prominent concentration of new comets in the small inclination region of the $(a_0, I_0)$ space.
This is because the initial planetesimal disk is confined near the ecliptic plane.
On the other hand, the color chart for $(a_0, I_0)$ in A-late shows a much weaker concentration of this kind.
The $(e_0, q_0)$ panels at the bottom show that new comets with high eccentricity tend to be generated deep into the planetary region from the start because of their small perihelion distance.

\section{Perturbation from major planets\label{sec:planetarypurt}}

\subsection{Dynamical model\label{ssec:dynamicalmodel}}
Once an object in the cloud satisfies the condition as a new comet ($r < 30$ au), we transfer it from the first model to the second model (see Figure \ref{fig:oc-evol-schem} for a recap of our model structure).
In the transfer process, we rewind the object's trajectory back to the edge of the sphere with the radius of $r = 800$ au.
The purpose of this rewind is to start the numerical orbit integration of the object under planetary perturbation from the outer edge of this sphere.
Our numerical orbit integration involves the new comets, 
eight major planets from
Mercury to Neptune, and the Sun.
Starting from their entry into the $r=800$ sphere, we follow the dynamical evolution of each comet for up to 500 million years.
We treat the comets as massless particles throughout the numerical orbit integration.

The boundary value of 800 au may sound rather arbitrary.
We chose it as a boundary within which we can regard the influences of the galactic tide and the stellar encounter small enough.
As a study that supports this threshold, we cite \citet{saillenfest2019a}.
They found that the region $a \lesssim 500$ au to be truly inert (their p. 14) in the sense that the effect of the galactic tidal force is sufficiently small.
Considering a cometary object with very large eccentricity, the region $a \lesssim 500$ au is practically equal to $r \lesssim 1000$ au.
Therefore we can presume that the region $r < 800$ au is inert, and we believe that our choice of the boundary is justified.

We adopt the current planetary masses and orbital elements for our orbit propagation.
It has been widely recognized that the planetary orbital configuration was quite different in the early solar system from what we see now \citep[e.g.][]{fernandez1984,malhotra1995}.
It is also estimated that the planetary orbits reached the current configuration through potentially violent dynamical processes such as radial migration of various types \citep[e.g.][]{gomes2005,walsh2012,lykawka2019,lykawka2023a}.
However, in this study we do not consider secular changes of major planets' orbital configuration.
We assume that the planetary orbital motion has been stable and quasi-periodic over the past $\sim 4.5$ billion years
\citep[e.g.][]{ito2002a,laskar2009}.
We do not model any non-gravitational forces such as the Yarkovsky effect \citep[e.g.][]{farinella1999,bottke2001,bottke2006} or those considered in determination of cometary orbital elements \citep[e.g.][]{marsden1973,krolikowska2014,krolikowska2020a,krolikowska2023}.
The planets and the Sun are treated as point masses without any equatorial bulges.
We consider their physical radius only when we count the number of collisions between the major bodies and the comets.

As for scheme of the numerical orbit integration, we employed the so-called Wisdom--Holman symplectic map \citep{wisdom1991,wisdom1992a} implemented as the \textsc{swift} package \citep{levison1994}.
We have modified this code and used it in our previous work \citep[e.g.][]{strom2005,ito2010,ito2016}, and we can be assured of its accuracy.
The nominal stepsize of the orbit integration is six days with a data output interval of 500 years.
The numerical integration scheme is sort of adaptive (named as the regularized mixed variable symplectic method by \citet{levison1994}), and the stepsize gets automatically shrunk together with the force center switching from being heliocentric to planetocentric when close encounters take place between a planet and a comet.
In our numerical orbit integration, we deal with many objects with large eccentricity.
To be cautious about the accuracy of results, we also ran another set of orbit integration with a smaller stepsize (two days) using about 10{\%} of the total objects used in the main orbit integration.
The statistical trend obtained from the subset integration remains unchanged even when the stepsize is reduced, and we believe this indicates that our main numerical orbit integration is overall reliable.

We carry out the orbit integration of the comets, and record their orbit changes.
Before reaching the end of the nominal integration period of 500 myr, many objects get out of the solar system by satisfying either of the following conditions (i)(ii)(iii) that we define for removal of objects:
(i)  The object's orbit turns out to be hyperbolic $(e>1)$ at the heliocentric distance $r = 800$ au.
(ii) The object's aphelion distance $Q$ becomes larger than a certain value ($Q > Q_\mathrm{max}$) at $r = 800$ au where $Q_\mathrm{max} = 2 \times 10^5$ au.
     Since $2 \times 10^5$ au is nearly equal to 1 pc, this condition implies that the object has reached a region that is under the influence of nearby stars, not that of the Sun.
(iii) The object approaches major planets or the Sun within their physical radii.
     The third condition (direct collision with planets or the Sun) rarely happened in our calculation.
We summarized frequency of the end state of the comets in Table \ref{tbl:ocncfate_s5A}.

\begin{table}[!htbp]
\caption{%
Frequency of the end state of the comets in A-early and A-late.
$e > 1$ means the comet's eccentricity exceeded unity at $r = 800$ au.
$Q > Q_\mathrm{max}$ means the comet's aphelion distance exceeded $ Q_\mathrm{max} = 2 \times 10^5$ au at $r = 800$ au.
The object names in the left column (Sun, Jupiter, Saturn, Uranus, Neptune) denotes that the comet approached within a distance of the object's physical radius and was considered lost in the collision.
``survived'' means that the comet remained on its orbit without being eliminated from the calculation until the end of the whole integration period $(5 \times 10^8\,\mathrm{years})$.
}
\begin{center}
\begin{tabular}{lrrrr} \hline
\multicolumn{1}{c}{fate}    &
\multicolumn{1}{c}{A-early} &
\multicolumn{1}{c}{A-late}  \\
\hline               
$e>1$                & 474037 & 172041 \\
$Q > Q_\mathrm{max}$ & 179245 & 124780 \\
Sun                  &     45 &      6 \\
Jupiter              &    132 &     15 \\
Saturn               &     50 &      7 \\
Uranus               &     25 &      8 \\
Neptune              &     71 &     15 \\
survived             &  56502 & 450248 \\
\hline                                 
Total                & 710107 & 747120 \\
\hline
\end{tabular}
\end{center}
\label{tbl:ocncfate_s5A}
\end{table}

\subsection{The time-skip scheme\label{ssec:timeskip}}

\begin{figure}[!htbp]
 \includegraphics[width=\myfigwidth]{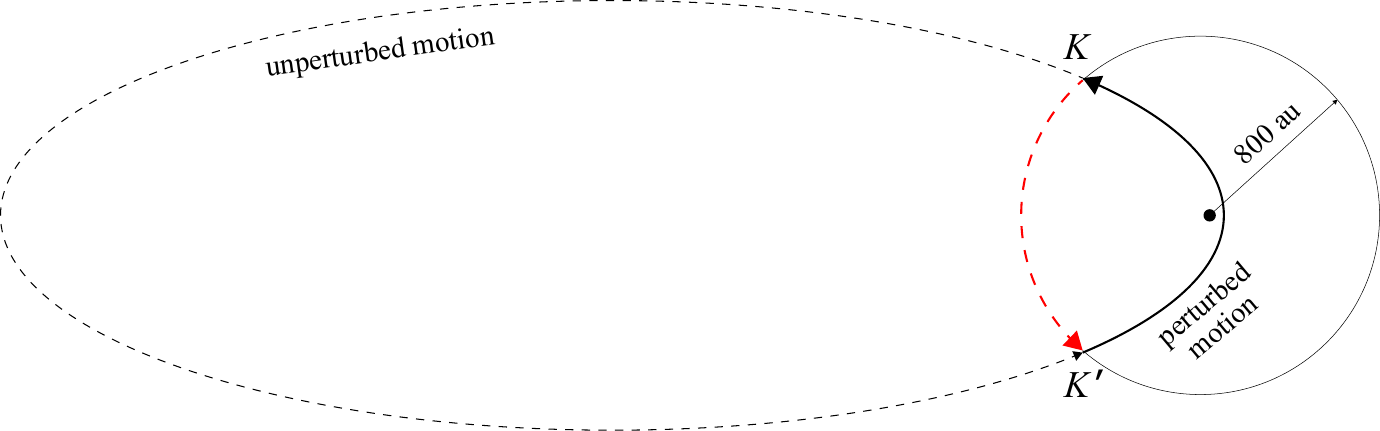}
\caption[]{%
Schematic illustration of how the time-skip scheme works in our model.
We give planetary perturbation to the orbital motion of comets only within the sphere with the radius of $r=800$ au centered at the Sun.
When a comet reaches the heliocentric distance of $r=800$ au (the point $K$ where the comet's mean anomaly is $l_K$),
we assume that the comet travels the rest of the orbit without any perturbations, and arrives back at the point $K'$ where the comet's mean anomaly is $l_K' = 2\pi - l_K$.
We do not carry out orbit integration from the point $K$ to $K'$.
}
\label{fig:oc-skipped-schem}
\end{figure}

In our second model, we give the planetary perturbation to the comets only within the sphere with the radius of $r=800$ au centered at the Sun.
However, most comets frequently go beyond the boundary of $r=800$ au, and spend most of their dynamical lifetime outside the sphere.
While directly integrating their equations of motion in the region of $r > 800$ au would take a long computation time, the influence of external perturbations there (i.e. the galactic tidal force and encounters with the nearby stars) can be limited.
Therefore we introduce an approximation scheme to reduce computational amount.
Specifically speaking, we assume that a comet that has reached the border of $r = 800$ au follows the Keplerian motion outside the sphere.
We call this approximate treatment the time-skip scheme.
See Figure \ref{fig:oc-skipped-schem} for a schematic diagram as to how this scheme works in our model.

When an outbound object has reached the border of $r=800$ au (we name this point as $K$ where the comet's mean anomaly is $l_K$) from inside the sphere, we examine if the object satisfies the above mentioned removal condition (i) or (ii).
If it does, we suspend the orbit integration for the object.
If it does not, we assume that the object travels beyond the border without any perturbations,
and that it arrives back at the point $K'$ where the object's mean anomaly is $l_K' = 2\pi - l_K$.
We assume other orbital elements than mean anomaly remain constant during this travel from $K$ to $K'$.
The position and the velocity of the object at the point $K'$ is calculated according to the new mean anomaly $l_K'$.
Using the mean motion $n_K$ at the point $K$, the actual time that the object would spend on the Keplerian orbit in the region of $r > 800$ au is calculated as
$ \Delta t_\mathrm{skip} =   n_K \left( l_K' - l_K \right)
                         = 2 n_K \left( \pi  - l_K \right) $.
We add this to the accumulated dynamical resident time of the object.
Since most objects in our model has large eccentricity, and since the skipped time $\Delta t_\mathrm{skip}$ often occupies a substantial fraction of the entire dynamical lifetime of the objects, the procedure to make an object ``skip'' from $K$ to $K'$ saves us a great amount of computational resource.
The actual computational time would be 10 to 100 times larger if we do not adopt the time-skip scheme.

Note that the time-skip scheme must be implemented in the solar system barycentric coordinates, not in the heliocentric coordinates.
This is because the objects located beyond the distance of $r > 800$ au should be regarded orbiting around the barycenter of the entire solar system, not around the Sun.
This means that, even under the time-skip scheme, the heliocentric semimajor axis of an object $a$ will be different between the points $K$ and $K'$ due to the motion of the Sun around the solar system barycenter.
Therefore the mean motion $n_K$ calculated in the heliocentric coordinates is in general different from that calculated in the barycentric coordinates, as are other orbital elements.
On the other hand, the barycentric orbital elements of the object would remain the same under the time-skip scheme except for mean anomaly.
In the actual time-skip scheme, we calculate the barycentric orbital elements of the objects that have reached $r = 800$, evaluate the values of $l_K$, $n_K$, $\Delta t_\mathrm{skip}$, and convert the orbital elements including $l_K'$ into the heliocentric coordinates again.
Note that the boundary of $r=800$ au can be defined either in the barycentric coordinates or the heliocentric coordinates; the consequence of this difference would be minor.

\begin{figure}[!htbp]
\includegraphics[width=\myfigwidth]{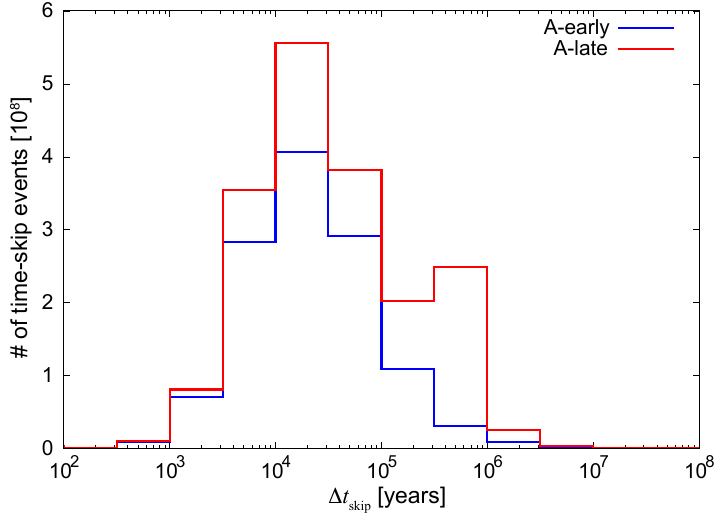}
\caption[]{%
Distribution of the skipped time $\Delta t_\mathrm{skip}$ during the periods A-early and A-late.
Unit of the vertical axis is $10^8$ events.
Total number of the time-skip events in A-early is
     1,216,843,726,
and that in A-late is
     1,875,189,946.
}
\label{fig:oc-skipped-time-lin_s5A}
\end{figure}

In Figure \ref{fig:oc-skipped-time-lin_s5A}, we summarized the distribution of the skipped time $\Delta t_\mathrm{skip}$ during each period.
Here we find that $\Delta t_\mathrm{skip}$ ranges from about $10^3$ to $10^6$ years.
The number of the time-skip events in each period is described in Figure \ref{fig:oc-skipped-time-lin_s5A}'s caption.
From the numbers of the time-skip events and the total number of the objects employed in each period, we find that each object experiences
1700--2500 time-skip events over the entire integration period in average.
In Figure \ref{fig:oc-skipped-time-lin_s5A} we see that the number of the time-skip events is much larger in A-late (denoted in red) than in A-early (denoted in blue).
This trend is common to the star set B as we show in Figure \ref{fig:oc-skipped-dist-log_s3B} in \ref{suppl:star3B}.
This trend probably reflects the difference of the average of the resident time of the comets in the planetary region that we will see in Figure \ref{fig:oc-lifetime-n1-bulk_s5A_multi}.
Later we will define this resident time as $T_\mathrm{res}$ in Section \ref{ssec:lifetime}.
The average of the resident time is longer in A-late than in A-early, and the number of the time-skip events follows this property.

Note that the averaged skipped time itself seems also longer in A-late than in the period of A-early.
This can be explained as follows.
We define the skipped time as the time required for an object to orbit in a distant region exceeding $r = 800$ au.
And, the objects with large eccentricity generally have large aphelion distance.
As we will see in Section \ref{ssec:lifetime}, the average eccentricity of the new comets in the period of A-early is slightly lower than that in A-late because of a comet shower that happened at $t \sim 0.45$ Gyr.
Therefore, the skipped time can be statistically longer in A-late where the objects' eccentricity is slightly higher in average.
This is typically expressed by the small peak of $\Delta t_\mathrm{skip} \sim 10^6$ years for the period of A-late seen in Figure \ref{fig:oc-skipped-time-lin_s5A}.

Roughly speaking, our justification to adopt the time-skip scheme in our model is as follows.
As we see in Figure \ref{fig:oc-skipped-time-lin_s5A},
the average time that most comets in our model spend in the region of $r > 800$ au is $O(10^6)$ years or shorter.
This is sufficiently short compared to the typical timescale of the action of the galactic tidal force, about $10^9$ years \citep[e.g.][]{higuchi2007}.
Therefore the effect of the galactic tidal force acting during the time when a comet is time-skipped is regarded limited.
For the same reason, the probability of a new comet undergoing fatal perturbation due to stellar encounters during this time is not high \citep[e.g.][]{higuchi2015}.
Thus the adoption of the time-skip scheme is justified.

\begin{figure}[!htbp]
\includegraphics[width=\myfigwidth]{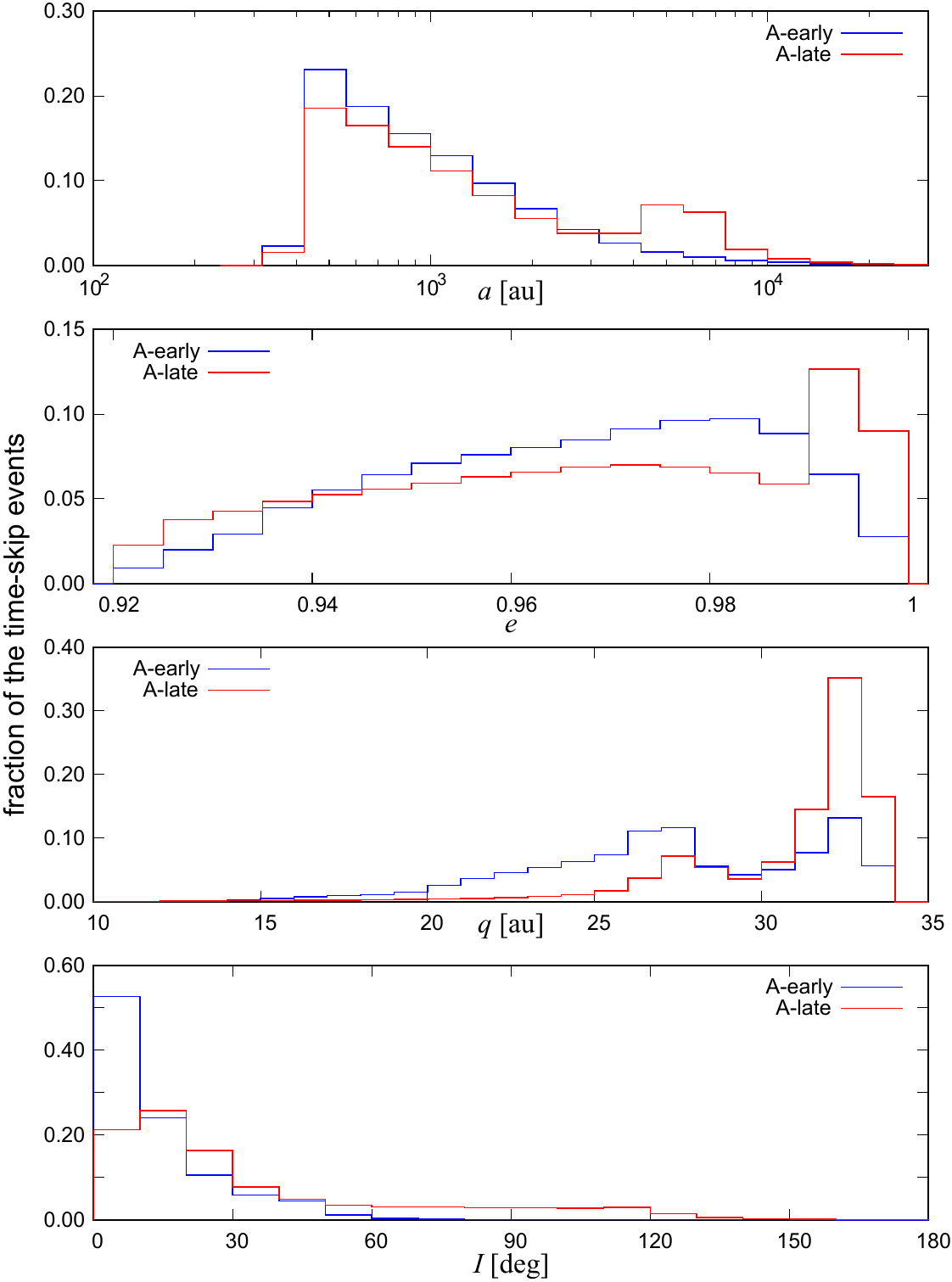}
\caption[]{%
Frequency distribution of the comets' orbital element values $(a, e, q, I)$ when each time-skip event occurred during the periods A-early and A-late.
}
\label{fig:oc-skipped-elem_s5A}
\end{figure}

We can derive a similar conclusion obtained in Figure \ref{fig:oc-skipped-time-lin_s5A} in a different way as follows.
In Figure \ref{fig:oc-skipped-elem_s5A}, we summarized the frequency distribution of orbital element values $(a, e, q, I)$ that each comet had when each time-skip event occurred during the periods A-early and A-late.
In this figure we should pay a particular attention to the distribution of semimajor axis $a$ at the time when each comet underwent the time-skip event.
As we see, the values of the semimajor axis of the comets at the time-skip events are not large.
For example, the fraction of comets whose semimajor axis was larger than $a > 10^4$ au when it was time-skipped in A-early is less than 1{\%}.
This value is almost the same in the other periods (A-late, B-early, B-late).
Thus, even if these comets were to be subjected to the galactic tidal force at a distance, their influence on the overall statistics derived from our numerical model would be limited.
This is another argument that supports the validity of using the time-skip scheme in our model.

\begin{figure}[!htbp]
 \includegraphics[width=\myfigwidth]{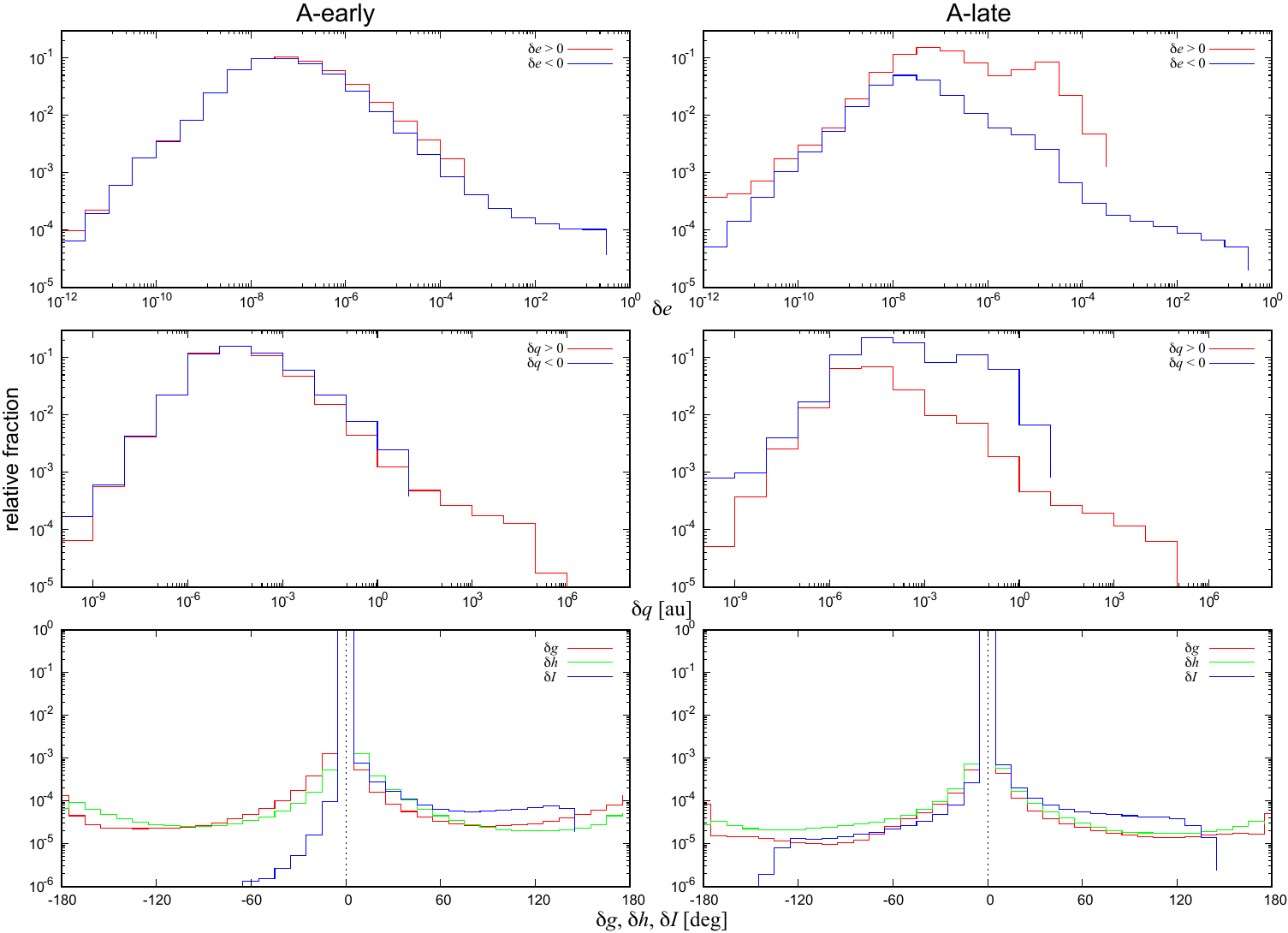}
\caption[]{%
Distribution of the comets' possible orbital element variations during the time-skip period in A-early and A-late obtained through the application of the galactic tidal force function described in \ref{appen:gtfunc}.
The vertical axis indicates relative fraction of each of the element variations, and the scale is logarithmic in all the panels.
Top:    Possible variation of eccentricity $\left| \delta e \right|$ (when $\delta e > 0$ (red) and $\delta e < 0$ (blue)).
Middle: Possible variation of perihelion distance $\left| \delta q \right|$ (when $\delta q > 0$ (red) and $\delta q < 0$ (blue)) in the unit of au.
Bottom: Possible variation of argument of perihelion $\delta g$ (red),
             that of longitude of ascending node $\delta h$ (green), and
             that of inclination $\delta I$ (blue) in the unit of degree.
}
\label{fig:oc-skipped-dist-log_s5A}
\end{figure}

In order to add more quantitative support to the justification of the use of the time-skip scheme,
we applied what we call the galactic tidal force function to the objects subject to the time-skip scheme, and quantified the influence of the galactic tide that we ignore while the time-skip scheme works.
We detail how we implement the galactic tidal force function in \ref{appen:gtfunc}.
In Figure \ref{fig:oc-skipped-dist-log_s5A} we show variations in the orbital elements of the comets resulting from the application of the galactic tidal force function.
What we present in this figure is how large change each orbital elements (such as $e$, $q$, $g$, $h$, $I$) could have if the galactic tidal force had worked during the time-skip events.
The possible changes are denoted as the variable with $\delta$ such as $\delta e$, $\delta q$, $\delta I$, and so on.
This figure tells us that the change in comets' eccentricity $e$ and perihelion distance $q$ during the time-skip event would be quite small even if the galactic tidal force acts.
The peaks of their distribution are located around $|\delta e| \sim 10^{-7}$ and $|\delta q| \sim 10^{-5}$ au, respectively.
In the approximation used in this study, the galactic tidal force does not affect the semimajor axis $a$ of comets.
Therefore the changes in comets' perihelion distance $q = a(1-e)$ seen in Figure \ref{fig:oc-skipped-dist-log_s5A} are solely due to the changes in their eccentricity $e$.
We see that the fraction of events with larger $\delta e$ (and hence $\delta q$) is somewhat greater in A-late than in A-early in this figure.
It may be based on the fact that eccentricity of the new comets in A-late is in average is higher than that in A-early, as we mentioned before and will see in Section \ref{ssec:lifetime}.

\begin{figure}[!htbp]
\includegraphics[width=\myfigwidth]{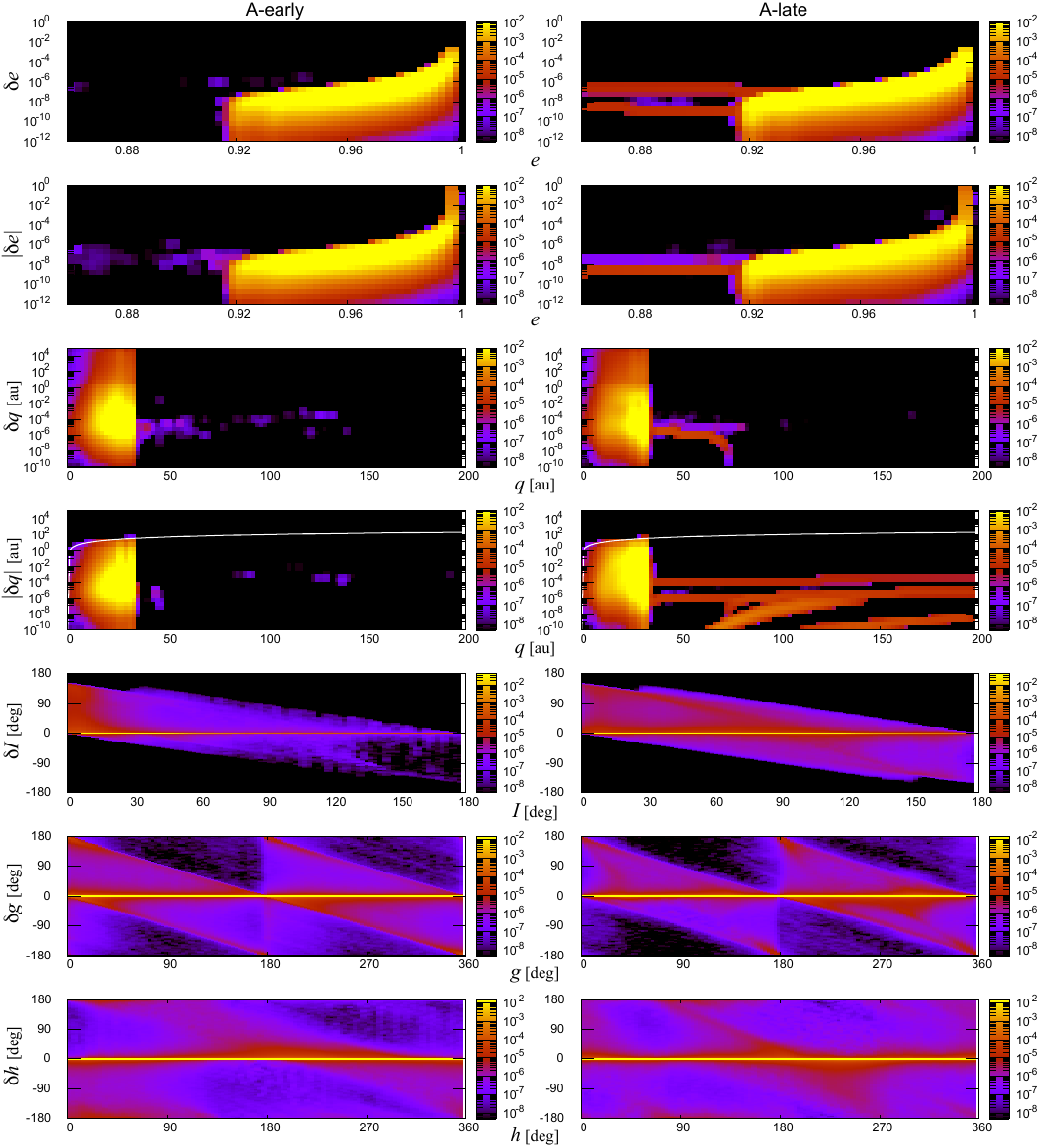}
\caption[]{%
Possible variation of comets' orbital elements due to the application of the galactic tidal force function and its dependence on the orbital elements at the beginning of the time-skip period in A-early and A-late.
The color charts indicate normalized frequency in the logarithmic scale.
From the top row,
$ \delta e$:  positive variation of eccentricity $(\delta e > 0)$.
$|\delta e|$: absolute value of negative variation of eccentricity $(\delta e < 0)$.
$ \delta q$:  positive variation of perihelion distance $(\delta q > 0)$.
$|\delta q|$: absolute value of negative variation of perihelion distance $(\delta q < 0)$ together with the white curves indicating the upper limit of the variation ($|\delta q| = q$; representing the largest change of perihelion distance when $\delta q < 0$).
$\delta I$:   variation of inclination.
$\delta g$:   variation of argument of perihelion.
$\delta h$:   variation of longitude of ascending node.
}
\label{fig:oc-skipped-diff_s5A}
\end{figure}

In order to inspect in more detail how small the potential influence of the galactic tidal force during the time-skip period is, in Figure \ref{fig:oc-skipped-diff_s5A} we present the frequency distribution of the dependence of the possible variation in the orbital elements of comets on their condition at the beginning of the time-skip event.
The horizontal axis shows the values of each element at the beginning of the time-skip period.
Similarly to Figure \ref{fig:oc-skipped-dist-log_s5A}, we again find that the possible variations of the orbital elements during the time-skip events are generally small.
We also find a periodicity of the period $\pi$ in the variation of inclination $(\delta I)$ and that of argument of perihelion $(\delta g)$.
The direct cause of this as we understand is that the disturbing function representing the galactic tidal potential largely depends on the components proportional to $\cos 2I$ and $\cos 2g$ when it is expressed through the heliocentric orbital elements \citep[e.g.][Eq. (A.5)]{saillenfest2019a}.
In Figure \ref{fig:oc-skipped-diff_s5A} the variation of cometary longitude of ascending node $(\delta h)$ also seems to have the periodicity of $\pi$, but it is less clear than those of $\delta I$ or $\delta g$.
This is probably an outcome of the fact that the disturbing function has both the components of $\cos h$ and $\cos 2h$, and the contribution of the $\cos 2h$ terms (with the period of $\pi$) may not be quite dominant.

In Figure \ref{fig:oc-skipped-diff_s5A}, we may find rather clear boundaries in the plot of $\delta e$ at $e \sim 0.92$ as well as in the plot of $\delta q$ at $q \sim 35$ au.
The existence of these boundaries can be explained as follows.
As we will see in the next section (Section \ref{sec:result}), the perihelion distance of the comets tend to retain the influence of their initial conditions for a large part of the entire numerical integration period.
Specifically speaking, their perihelion remains to be located inside or just around Neptune's orbit, $q \lesssim 35$ au.
This value forms the boundary seen in the panel for $\delta q$ in Figure \ref{fig:oc-skipped-diff_s5A}.
On the other hand, for the time-skip scheme to be triggered, semimajor axis of an object must have a value greater than 400 au (i.e. $2a \geq 800$ au).
Through the definition of perihelion distance ($q=a(1-e)$), we have $e = 1 - \frac{q}{a}$.
Applying the values of $q \lesssim 35$ au and $a \geq 400$ au, we get $e \gtrsim 1 - \frac{35}{400} \sim 0.9125$.
This value forms the boundary seen in the panel for $\delta e$ in Figure \ref{fig:oc-skipped-diff_s5A}.

Let us note a few more points that we need to pay attention to in Figure \ref{fig:oc-skipped-diff_s5A}.
Perihelion distance $q$ of a small fraction of comets can significantly change due to the galactic tidal force during the time-skip period, and the amount of the variation could reach over $10^4$ au.
For these comets, the stellar perturbation can have a non-negligible effect during the time-skipped period.
Although the fraction of such objects is as small as about $10^{-3}$ or less, we should be aware of these facts because the existence of these potentially high-$q$ objects indicates a limitation of the time-skip scheme.
In addition, we should recall that the function form of the galactic tidal force that we consider is the averaged one which is supposed to be valid on time scales longer than the orbital period of the comets (see \ref{appen:gtfunc} for details).
Therefore it may not be obvious whether we can safely apply the function to the time-skip period which is shorter than orbital period of new comets.
In addition, we may need to make a quantitative assessment of the consequences of ignoring close encounters of nearby stars that could happen during the time-skip period.

\clearpage
\thispagestyle{empty}

\clearpage

\section{Results\label{sec:result}}
In what follows we describe four main results of our numerical orbit integration of the comets under the planetary perturbation.
They are:
dynamical resident time of the comets (Section \ref{ssec:lifetime}),
spatial penetration of the new comets and planet barrier (Section \ref{ssec:barrier}),
transition into other small body populations (Section \ref{ssec:otherpops}), and
survivors over 500 million years (Section \ref{ssec:survivors}).

\begin{figure}[!htbp]
\includegraphics[width=\myfigwidth]{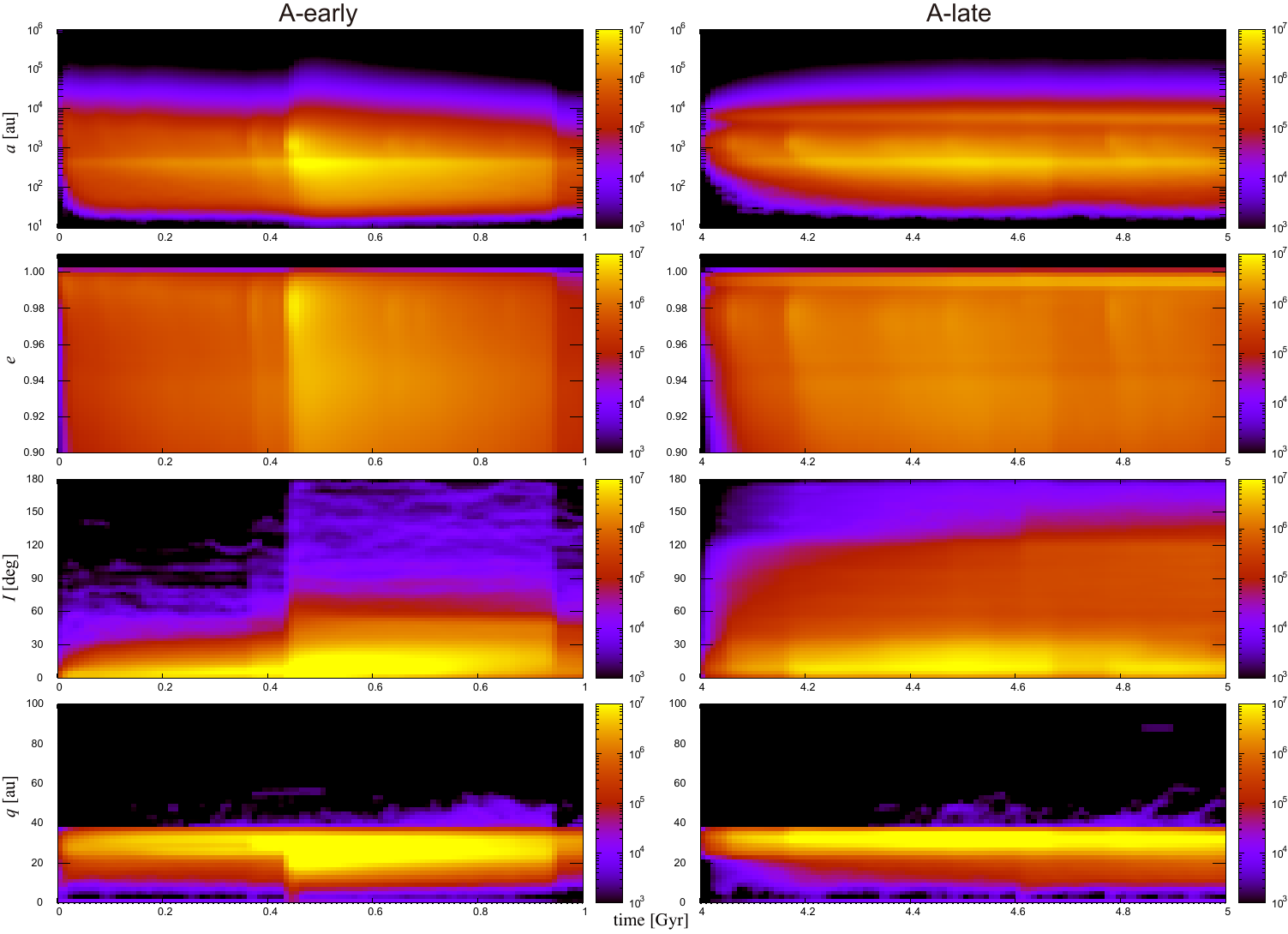}
\caption[]{%
Time variation of the frequency distribution of the orbital elements of the comets in A-early and A-late.
The color charts indicate the number of objects in the logarithmic scale.
The time interval along the horizontal axis is $0.002\, \mathrm{Gyr}$.
}
\label{fig:oc-t-aeIqQ_s5A}
\end{figure}

\subsection{Dynamical resident time of the comets\label{ssec:lifetime}}
In Figure \ref{fig:oc-t-aeIqQ_s5A} we draw time variation of the frequency distribution of the orbital elements of the comets in A-early and A-late.
In other words, these plots show time series of new comets' generation, orbital evolution, and annihilation.
This figure can be contrasted to Figure \ref{fig:oc-genocnc-combined-t_s5A} which is just about new comets' generation.
In Figure \ref{fig:oc-t-aeIqQ_s5A} we find sharp increases of cometary flux at several occasions.
They represent strong comet showers invoked by close encounters of nearby stars.
In particular, the comet shower that happened at time $t \sim 0.45$ Gyr in A-early is remarkable.
The average initial eccentricity of the new comets in the period of A-early is slightly lower than that in A-late, because this shower generated many new comets with relatively lower initial eccentricity, $e_0 \lesssim 0.96$.

\begin{figure}[!htbp]
 \includegraphics[width=\myfigwidth]{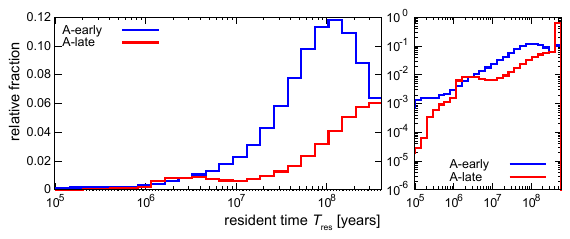}
\caption[]{%
Distribution of the resident time $T_\mathrm{res}$ of the comets.
The vertical axis of the left panel is on a linear scale, while that of the right panel is on a logarithmic scale.
The vertical axis is normalized in both the panels so that the total value becomes unity for each period.
}
 \label{fig:oc-lifetime-n1-bulk_s5A_multi}
\end{figure}

\label{pg:ejectcondition}
Figure \ref{fig:oc-lifetime-n1-bulk_s5A_multi} shows the distribution of dynamical resident time of the comets since each object's generation (i.e. injection into the planetary region) until its elimination from the system.
We denote this time $T_\mathrm{res}$ throughout this paper.
In Figure \ref{fig:oc-lifetime-n1-bulk_s5A_multi} we find a clear difference of the distributions of $T_\mathrm{res}$ in A-early (blue) and in A-late (red).
The distribution of $T_\mathrm{res}$ in A-early has a peak at around $10^8$ years, while there seems no peak in the distribution of $T_\mathrm{res}$ in A-late.
This means that $T_\mathrm{res}$ of many comets is longer than the the maximum length of the numerical integration of each comet ($5 \times 10^8$ years) in A-late.
We can qualitatively interpret the difference as follows.
The period A-early models the early stage of the solar system when the Oort Cloud objects are concentrated along the ecliptic plane.
In this period, frequency of close encounters of the comets with major planets that orbit near along the ecliptic plane is larger.
This enhances the likelihood that the incoming comets are scattered away by the planets and ejected out of the system.
In contrast, the period A-late models the modern solar system, and the comets are nearly isotropically distributed, in particular in the outer part of the Oort Cloud.
In other words, fraction of the comets coming along the ecliptic plane is smaller in A-late than in A-early, and the relative number of comets encountering the major planets is lower in A-late.
This statistically enforces the dynamical stability of the cometary orbits in A-late, leading to the longer $T_\mathrm{res}$ in this period than in A-early.
Note that regarding the distribution $T_\mathrm{res}$, the result obtained from the star set B is almost similar to that obtained from the star set A (compare Figure \ref{fig:oc-lifetime-n1-bulk_s5A_multi} and Figure \ref{fig:oc-lifetime-n1-bulk_s3B_multi}).

In the right panel of Figures \ref{fig:oc-lifetime-n1-bulk_s5A_multi}, the $T_\mathrm{res}$ distribution for A-late seems to have a strong peak at the right edge of the panel.
This is because more than half of the comets have survived the entire integration period in this period, resulting in $T_\mathrm{res} > 5 \times 10^8$ years.
However, the $T_\mathrm{res}$ distribution in the left panel suggests that the distribution approaches its peak at $T_\mathrm{res} \sim 4 \times 10^8$ years in A-late.
Empirically, the shape of the $T_\mathrm{res}$ distribution of the comets in our model is somewhat close to the log-normal distribution.
The reason for this is unspecified yet and is a subject for our future work.
But from this, we guess that the $T_\mathrm{res}$ distribution in A-late has a peak value around several $10^8$ years.
To identify the true peak of the $T_\mathrm{res}$ distribution in A-late we would have to extend the period of numerical integration to $10^9$ years or longer.
Due to the limitation of computational resources available to us, currently we are unable to get this extension realized.

\begin{figure}[!htbp]
 \includegraphics[width=\myfigwidth]{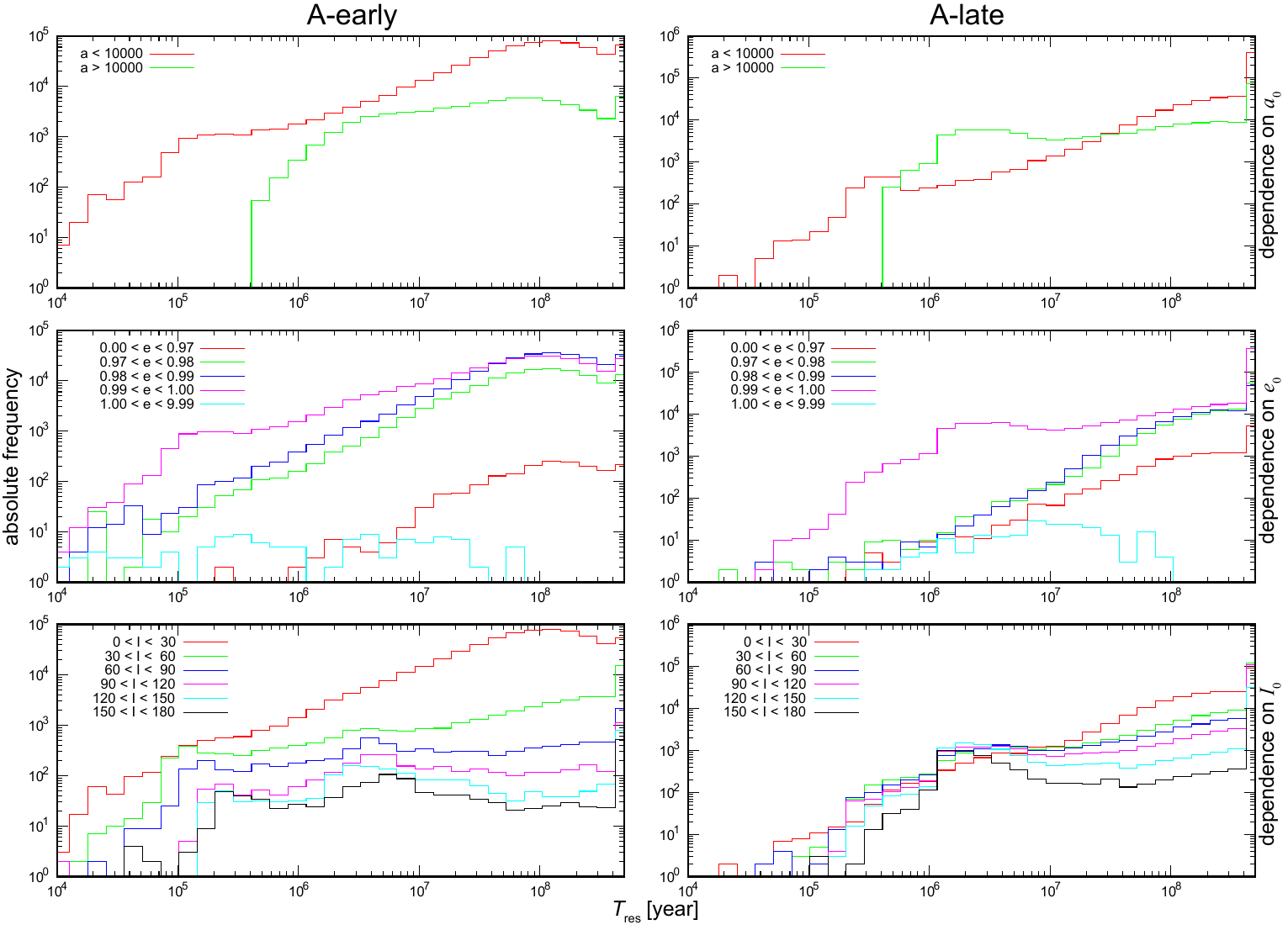}
\caption[]{%
Dependence of the resident time $T_\mathrm{res}$ on comets' initial orbital elements during A-early and A-late.
The vertical axis in the panels is not normalized, and its scale is logarithmic.
The subscript 0 is omitted in the panel legends for avoiding clutter, such as $a_0 \to a$ and $e_0 \to e$.
}
\label{fig:oc-lifetime-depend-n0-log-multi_s5A}
\end{figure}

Figure \ref{fig:oc-lifetime-depend-n0-log-multi_s5A} shows more detailed dependence of the resident time $T_\mathrm{res}$ on new comets' initial orbital elements, $a_0$, $e_0$, and $I_0$.
Note that the vertical axis in the panels is not normalized, and its scale is logarithmic.
What we can observe in this figure is as follows:
\begin{itemize}
  \item The new comets that have large initial semimajor axis ($a_0 > 10,000$ au) tend to have relatively shorter $T_\mathrm{res}$ than those that have smaller initial semimajor axis ($a_0 < 10,000$ au): the distribution peak for the comets with $a_0 > 10,000$ au seems less sharply focused in the large $T_\mathrm{res}$ region than the other group ($a_0 < 10,000$ au).
        This is because the new comets with large initial semimajor axis tend to have large eccentricity in our model by our definition, and they are close to the state of ejection on a hyperbolic orbit from the beginning.
  \item The new comets with initially on very eccentric orbits $(e_0 \gtrsim 0.99)$ tend to have less sharp peaks in the large $T_\mathrm{res}$ region probably by the same reason mentioned above.
  \item The dependence of the $T_\mathrm{res}$ distribution on the initial inclination seems unclear or weak except for the low-inclination comets ($0 < I_0 < 30^\circ$) occurred in A-early that have the maximum around $T_\mathrm{res} = 10^8$ years.
As for this point, we should recall that the low-inclination objects are the primary component of the new comets that are generated in this period.
Their resident time distribution has a peak at $T_\mathrm{res} \sim 10^8$ years as shown in Figures \ref{fig:oc-lifetime-n1-bulk_s5A_multi}, and other higher-inclination comets do not occur as frequently in A-early.
  \item
Except for the property of the low-inclination comets ($0 < I_0 < 30^\circ$) occurred in A-early mentioned above, we do not see a remarkable difference between the distributions of $T_\mathrm{res}$ of the prograde objects ($I_0 < 90^\circ$) and the retrograde objects ($I_0 > 90^\circ$).
Considering the general fact that the interaction between the retrograde objects and the major planets is weaker than that between the prograde objects and the planets, this result seems rather unexpected.
One of the reasons for this is that the initial orbital element distribution is different between the prograde comets and the retrograde comets in our numerical model.
As we saw in the panels in the second top row in Figure \ref{fig:oc-aeIq0_s5A}, the distribution of the initial orbital inclination $I_0$ and initial semimajor axis $a_0$ of the comets is not symmetric with respect to the boundary between the prograde and retrograde orbital motion, i.e. $I_0 = 90^\circ$.
Later we will find some difference of the dynamical property between the prograde comets and the retrograde comets in Section \ref{ssec:barrier} where we discuss the smallest perihelion distance of each comet.
Consult the discussions about Figure \ref{fig:oc-qmin-depend-n0-log-init-multi_s5A} and Figure \ref{fig:oc-eIqminTe_s5A} in that section.
  \item
The absolute frequency of occurrence of the retrograde comets is much less than that of the prograde comets.
This seems reasonable if we remember the initial configuration of the flat planetesimal disk that contains no retrograde objects.
  \item
Regarding the dependence of the resident time $T_\mathrm{res}$ on new comets' initial orbital elements, the result obtained from the star set B does not seem quite different from that obtained from the star set A. Compare
Figure \ref{fig:oc-lifetime-depend-n0-log-multi_s5A} and
Figure \ref{fig:oc-lifetime-depend-n0-log-multi_s3B}.
\end{itemize}

\begin{figure}[!htbp]
  \includegraphics[width=\myfigwidth]{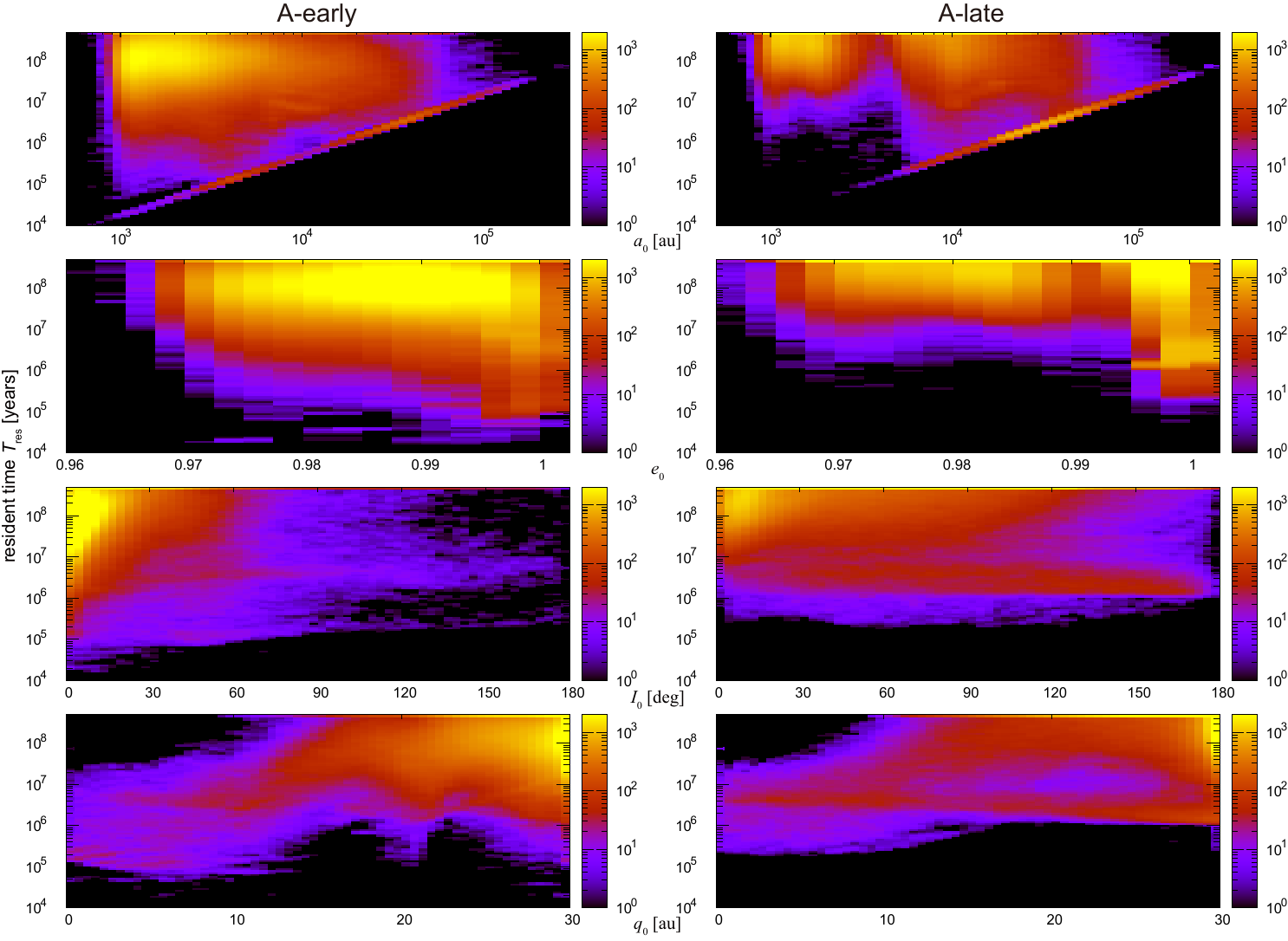}
  \caption[]{%
Frequency distribution of $T_\mathrm{res}$ and its dependence on the comets' initial orbital elements in A-early and A-late.
Top:        $(a_0, T_\mathrm{res})$,
second top: $(e_0, T_\mathrm{res})$,
third  top: $(I_0, T_\mathrm{res})$,
bottom top: $(q_0, T_\mathrm{res})$.
The color charts indicate the number of objects in the logarithmic scale.
}
\label{fig:oc-aeIqQ-Te-log_s5A}
\end{figure}

Figure \ref{fig:oc-aeIqQ-Te-log_s5A} is the frequency distribution of $T_\mathrm{res}$ and its dependence on the comets' initial orbital elements, $a_0$, $e_0$, $I_0$, and $q_0$.
In the top right panel for $(a_0,T_\mathrm{res})$ in A-late, we see two concentrations of the frequency around $a_0 \sim 10^3$ au and $a_0 \sim10^4$ au.
This reflects the initial orbital distribution of the comets shown in Figure \ref{fig:oc-aeIq0_s5A} where we find two concentrations of the initial semimajor axis of the new comets around $10^3$ au and $10^4$ au.
A concentration seen around $e_0 \gtrsim 0.99$ in the second top right panel of $(e_0,T_\mathrm{res})$ in the period A-late also reflects the condition shown in Figure \ref{fig:oc-genocnc-combined-t_s5A} where we find a nearly constant production of new comets with $e_0 \sim 0.99$--1.0 throughout the period ($t=4$--5 Gyr).
In the panels of $(I_0,T_\mathrm{res})$ for A-early, we find a concentration around $I_0 \sim 0$, but the corresponding concentration around $I_0 \sim 0$ in the panel for A-late is less dense.
This is a reflection of the change of the comet cloud shape from a two-dimensional disk in A-early to a three-dimensional, isotropic form in A-late.

Note that in the panels of $(a_0,T_\mathrm{res})$, we find linearly-shaped concentrations from the bottom left to the top right.
They are caused by how we calculate $T_\mathrm{res}$.
We reckon $T_\mathrm{res}$ from the aphelion of the initial orbit of each new comet.
A new comet's $T_\mathrm{res}$ would be very short if it gets ejected out of the system just after its first apparition.
In this case we have $T_\mathrm{res} \sim \frac{P_0}{2} \, \propto \, a_0^{1.5}$ where $P_0$ is the orbital period of the comets on their initial orbits, and this forms the sharp boundary seen in the top-row panels for $(a_0, T_\mathrm{res})$.
Since the initial eccentricity of the comet cloud objects is close to unity, non-negligible fraction of the new comets get ejected out of the system just after their first apparition, and they are plotted on this border.
The specific fractions of these objects in each period are:
1.38{\%} in A-early,
3.92{\%} in A-late,
1.85{\%} in B-early, and
2.05{\%} in B-late.
Although this may be regarded as a kind of model artifact, the major conclusion about the new comets' dynamical resident time in the planetary region remains intact even with this artifact.

\subsection{Spatial penetration of comets and planet barrier\label{ssec:barrier}}

\begin{figure}[!htbp]
 \includegraphics[width=\myfigwidth]{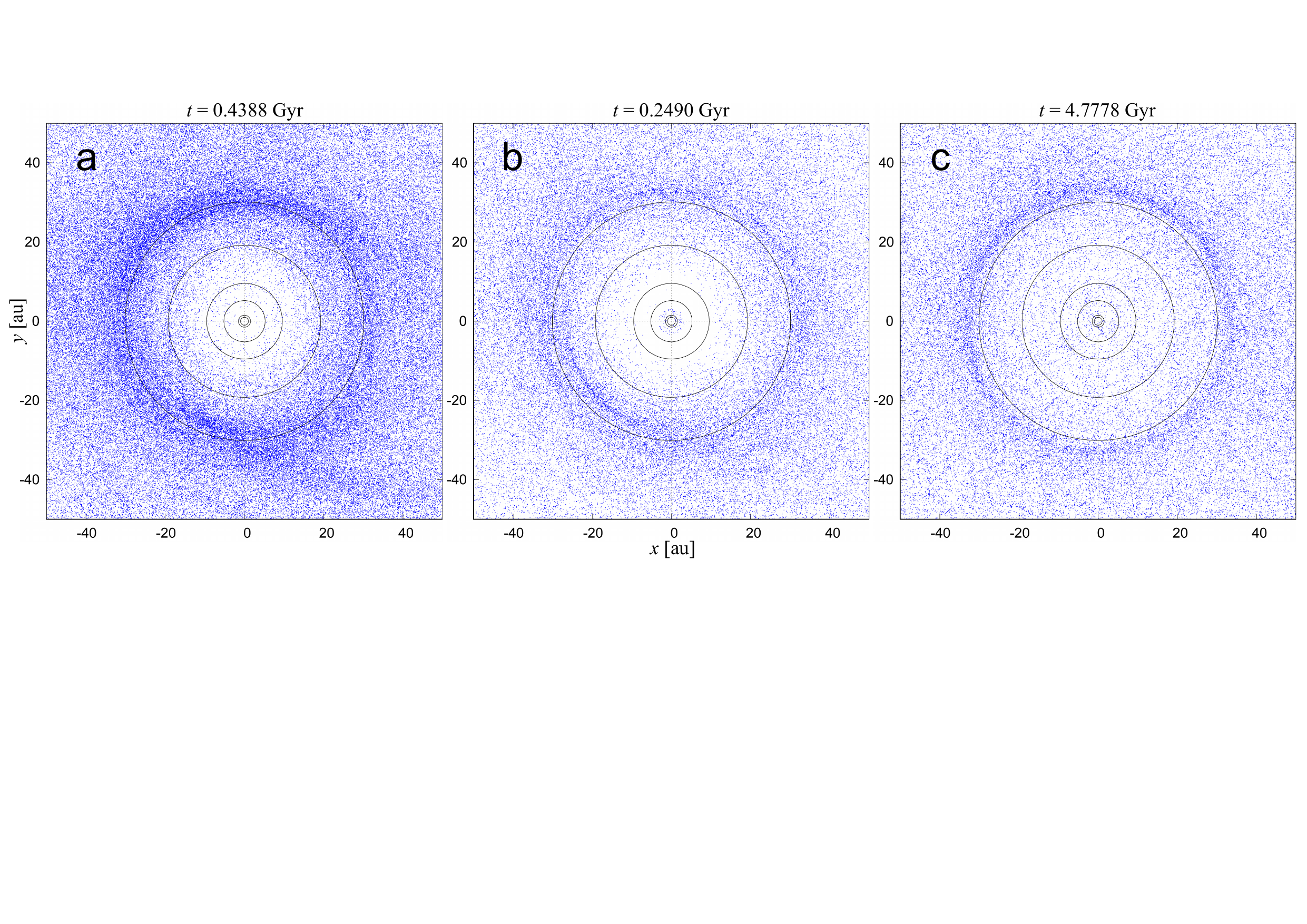}
\caption[]{%
Snapshots that show how the planet barrier works viewed from the north of the solar system.
The $x$-axis in each panel is directed toward the current vernal equinox seen from the Sun (J2000.0), and the $z$-axis is directed toward the ecliptic north pole.
The snapshots \textsf{a} and \textsf{b} were observed in the period A-early, and \textsf{c} was observed in A-late.
The blue dots indicate comets' positions.
1,000 positions of each comet over $5 \times 10^5$ years are marked in each panel (note that the average interval of data output is 500 years except for the time-skipped period).
The black concentric circles indicate approximate locations of major planetary orbits:
From inside, Earth, Mars, Jupiter, Saturn, Uranus, and Neptune.
The orbits of Mercury and Venus are omitted here for avoiding clutter.
The time when each snapshot was recorded is described at the top of each panel.}
\label{fig:oc-jsb_new}
\end{figure}

A certain number of comets originating from the Oort Cloud reach the Earth's vicinity.
However, the flux could be reduced if the comets get scattered by major planets.
This conjecture gave birth to the idea called the Jupiter barrier where giant planets protect the Earth from getting the impacts of small bodies coming from the outer part of the solar system \citep[e.g.][]{everhart1973,wetherill1994,horner2008}.
Our numerical result partially confirms this hypothesis, showing that this kind of barrier (hereafter referred to as the planet barrier) actually works when incoming cometary flux is nearly two-dimensional such as in A-early when the comet cloud is not yet isotropic enough.
The major component of the planet barrier is
Uranus and Neptune.
Once the comet cloud has become isotropic such as in the period A-late, the incoming cometary flux arrives at the planetary region from almost any directions, and the barrier is no longer quite effective.

In Figure \ref{fig:oc-jsb_new} we show three visual examples of how the planet barrier takes place in our numerical orbit integration.
The panel \textsf{a} shows a snapshot in A-early where the barrier is working efficiently.
We see relatively lower comet density inside Uranus' orbit compared with its outside.
The panel \textsf{b} shows a snapshot in A-early where the planet barrier is working and confining some comets inside Jupiter's orbit.
This status continued for $10^5$ to $10^6$ years in our model.
In other words, Jupiter is composing the planet barrier in this case.
The panel \textsf{c} shows a snapshot where the barrier is not working as efficiently as seen in the panel \textsf{a}.
The status shown in the panel \textsf{c} was observed in A-late when the new comets approach the planetary region from almost any directions.

\begin{figure}[!htbp]\centering
\includegraphics[width=\myfigwidth]{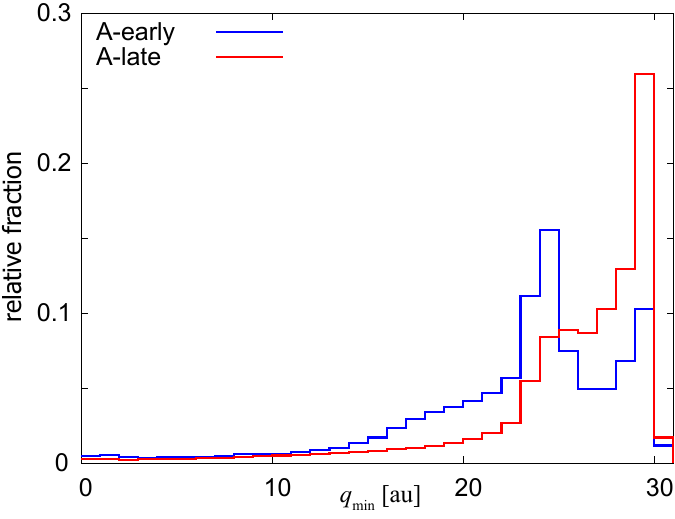}
\caption[]{%
Frequency distribution of the minimum perihelion distance $q_\mathrm{min}$ of the comets occurred during A-early and A-late.
The vertical values are normalized so that the integrated frequency becomes 1 for each period.}
\label{fig:oc-qmin-n1-bulk_s5A}
\end{figure}

Figure \ref{fig:oc-qmin-n1-bulk_s5A} shows the frequency distribution of the smallest perihelion distance $q_\mathrm{min}$ of each comet during its whole integration period.
The vertical values are normalized by the total number of the comets produced in each period.
In the $q_\mathrm{min}$ distribution obtained during A-early (blue), we see a local maximum between the orbits of Uranus and Neptune ($20 \lesssim q_\mathrm{min} \lesssim 25$ au).
This reflects the effect of the planet barrier.
Another local maximum in A-early seen around $q_\mathrm{min} \sim 30$ au probably reflects the initial condition for the new comets, $r < 30$ au.

In contrast to the period A-early, the distribution of $q_\mathrm{min}$ in Figure \ref{fig:oc-qmin-n1-bulk_s5A} just shows a strong peak around 30 au (which must be derived from our initial setting), and we do not see any local maximum within.
In other words, the planetary barrier is not quite effective during this period.
This contrast is more pronounced when we use the other star set, B.
We find that the relative fraction of $q_\mathrm{min}$ inside Uranus' orbit in B-early is much less than that in B-late (Figure \ref{fig:oc-qmin-n1-bulk_s3B}).

\begin{figure}[!htbp]\centering
  \includegraphics[width=\myfigwidth]{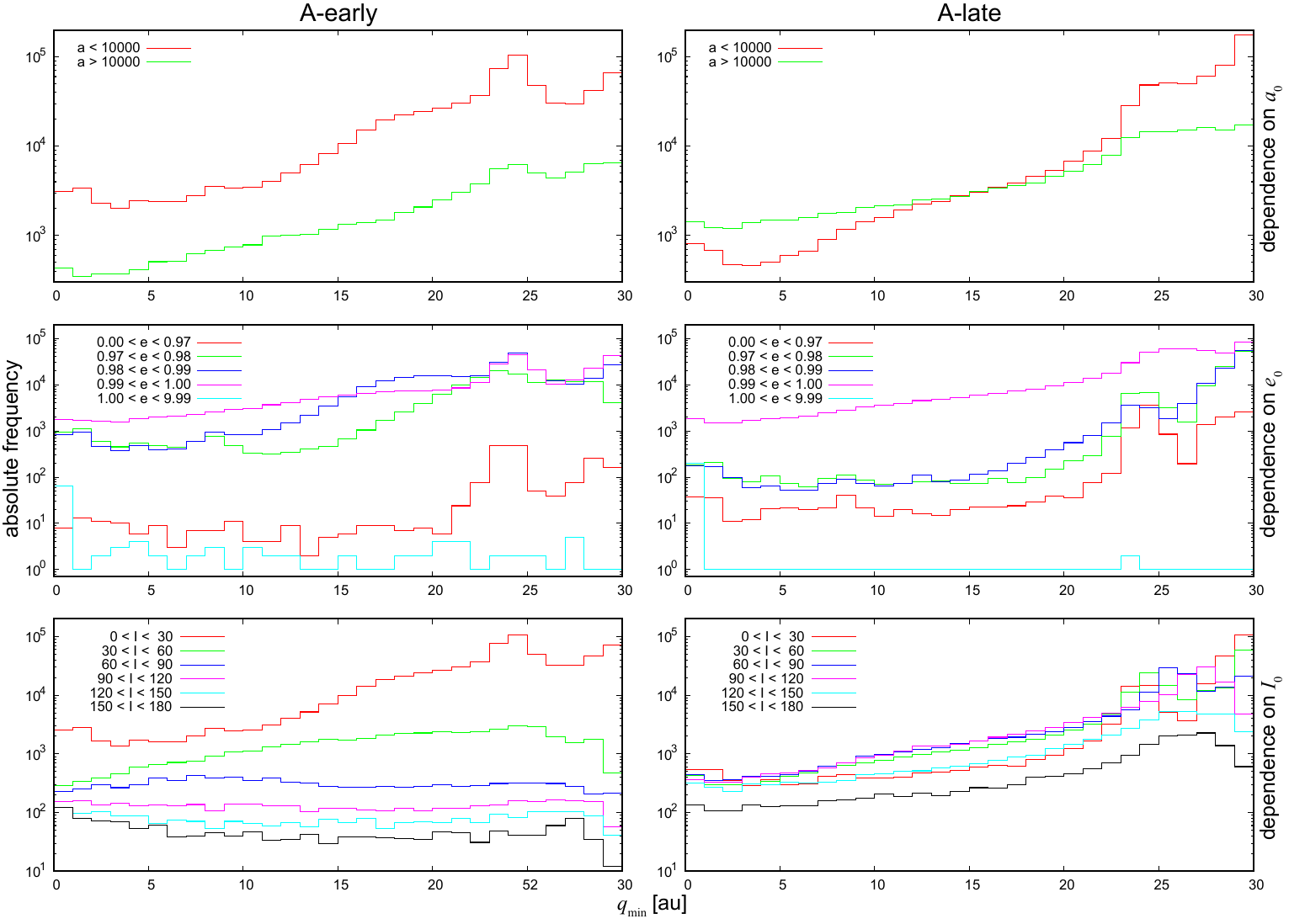}
  \caption[]{%
Dependence of the minimum perihelion distance $q_\mathrm{min}$ on comets' initial orbital elements $a_0$, $e_0$, and $I_0$ during A-early and A-late.
The vertical axis in the panels is not normalized, and its scale is logarithmic.
The subscript 0 is omitted in the panel legends for avoiding clutter, such as $a_0 \to a$ and $e_0 \to e$.
}
  \label{fig:oc-qmin-depend-n0-log-init-multi_s5A}
\end{figure}

In Figure \ref{fig:oc-qmin-depend-n0-log-init-multi_s5A} we show more detailed dependence of $q_\mathrm{min}$ on comets' initial semimajor axis $(a_0)$, initial eccentricity $(e_0)$, and initial inclination $(I_0)$.
Note that the vertical axis in the panels is not normalized, and its scale is logarithmic.
What we observe in this figure is as follows.
\begin{itemize}
\item 
In the panels in the top row, we find that the frequency of $q_\mathrm{min}$ of the comets coming from the inner part of the comet cloud ($a_0 < 10,000$ au, denoted in red) shows a more rapid decrease inside 20--25 au.
This tendency seems clear if we compare it with that of the comets coming from the outer part of the cloud ($a_0 > 10,000$ au, denoted in green) whose decrease in that region is weaker.
We interpret it as a consequence of the difference of their eccentricities.
This trend is prominent in A-early.
\item
In the panels in the bottom row, we find a trend that the retrograde objects $(I_0 > 90^\circ)$ seem to get into the deep planetary region more easily than the prograde objects $(I_0 < 90^\circ)$ do.
More specifically, the slope for the curves of $I_0 > 90^\circ$ seems rather shallower than that of the curves of $I_0 > 90^\circ$, in particular in A-early.
This may be related to the general fact that the interaction between the retrograde objects and the major planets is weaker than that of the prograde objects.
In general, retrograde objects are considered to be dynamically more stable than prograde objects when their perihelion distance is of the same order of magnitude, as shown in studies of the three-body problem \citep[e.g.][]{harrington1972,harrington1975,donnison1994}.
However, this does not mean that the resident time $T_\mathrm{res}$ of the retrograde comets in our model is always longer than that of the prograde objects.
When retrograde objects cross the planetary barrier and enter the terrestrial planetary region, frequent interaction with planets there will shorten the timescale of their orbital stability.
Then the stability characteristics of retrograde orbits becomes indiscernible in our numerical result, as we saw in the bottom row panels of Figure \ref{fig:oc-lifetime-depend-n0-log-multi_s5A} and mentioned in Section \ref{ssec:lifetime}.
\item
The dependence of the $q_\mathrm{min}$ frequency on the initial orbital inclination $I_0$ is weaker in A-late (the right bottom panel) than in A-early (the left bottom panel) except for the comets with low inclination $(0 < I_0 < 30^\circ)$ that are easily subject to the planet barrier.
This is probably because the initial eccentricity $e_0$ of the new comets generated during A-late is overall larger than that in A-early (see the panels for $e_0$ in Figure \ref{fig:oc-genocnc-combined-t_s5A}, or the panels $(a_0, e_0)$ in Figure \ref{fig:oc-aeIq0_s5A}).
For this reason, many new comets generated in A-late breach the planetary barrier without much regard to their initial orbital inclination, resulting in the less remarkable peaks in the $q_\mathrm{min}$ distribution around $q_\mathrm{min} \sim 25$ au.
\item
We see contribution from the objects initially on hyperbolic objects $(e_0 > 1)$ in Figure \ref{fig:oc-qmin-depend-n0-log-init-multi_s5A} (the panels in the middle row) although their frequency is low.
It seems that most of them pass through the barrier without leaving any local maxima in the $q_\mathrm{min}$ distribution.
The large relative velocity of the initially hyperbolic objects to the planets keeps the interaction between the objects and the planets weak.
\end{itemize}

Note that not only Uranus and Neptune but also Jupiter and Saturn could act as barriers, naturally.
However, the incoming comets seem barriered by the Uranus--Neptune zone before they reach the Jupiter--Saturn zone in our model.
As a result, the role of the inner two giant planets as a barrier seems apparently smaller in this case.
However, which type of planetary barrier is effective will depend on the initial orbital condition of the incoming comets.
In our numerical simulations using the star set A, the Neptune--Uranus barrier appears to work strongly as described above.
But in our simulations using the star set B, we find the Saturn--Jupiter barrier works more effectively than it does in the simulation using the star set A (consult Figures \ref{fig:oc-qmin-n1-bulk_s3B} and \ref{fig:oc-qmin-depend-n0-log-init-multi_s3B}).

\begin{figure}[!htbp]\centering
  \includegraphics[width=\myfigwidth]{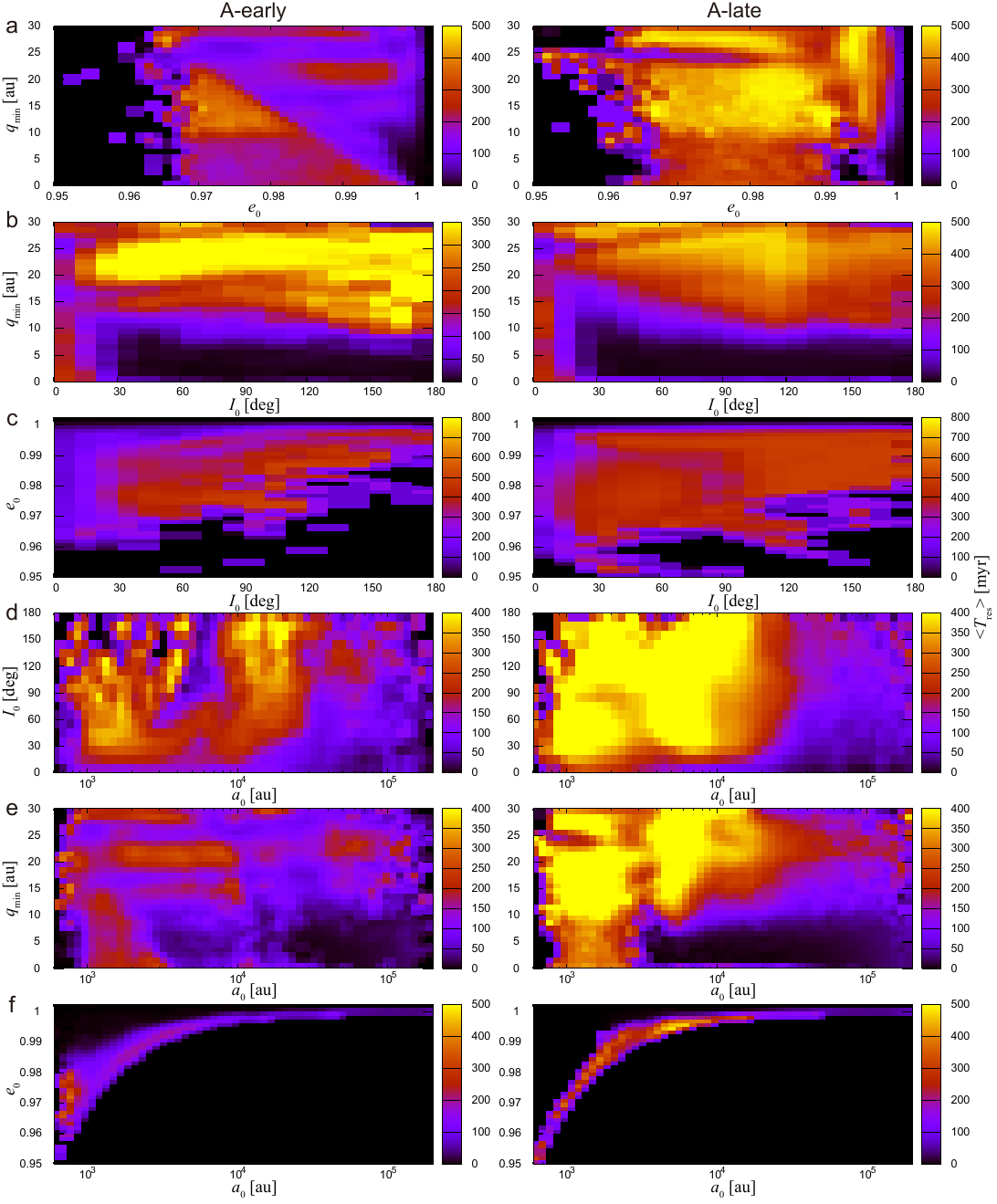}
  \caption[]{%
Distribution of $\left< T_\mathrm{res} \right>$ in phase space during A-early and A-late and projected onto various planes.
From the top to bottom, the panels are drawn on the
\textsf{a}: $(e_0, q_\mathrm{min})$,
\textsf{b}: $(I_0, q_\mathrm{min})$,
\textsf{c}: $(I_0, e_0)$,
\textsf{d}: $(a_0, I_0)$,
\textsf{e}: $(a_0, q_\mathrm{min})$,
\textsf{f}: $(a_0, e_0)$ planes, respectively.
The color charts indicate the value of $\left< T_\mathrm{res} \right>$ in the linear scale, and their unit is million years.
}
  \label{fig:oc-eIqminTe_s5A}
\end{figure}

At the end of this subsection, let us visualize the relation between $q_\mathrm{min}$ and comets' resident time $T_\mathrm{res}$.
The result is summarized in Figure \ref{fig:oc-eIqminTe_s5A}.
For this figure we calculated the distribution of average $T_\mathrm{res}$ (hereafter denoted as $\left< T_\mathrm{res} \right>$) in phase space that $q_\mathrm{min}$ and other initial orbital elements compose.
$\left< T_\mathrm{res} \right>$ is defined as a simple summed average of $T_\mathrm{res}$ of all comets in a phase space cell, i.e.
$\left< T_\mathrm{res} \right> = \sum_{i=1}^n T_{\mathrm{res},i} / n$
where $n$ is the number of comets in the considered cell.
Figure \ref{fig:oc-eIqminTe_s5A} shows the resulting distribution of $\left< T_\mathrm{res} \right>$ plotted onto the
$(e_0, q_\mathrm{min})$,
$(I_0, q_\mathrm{min})$,
$(I_0, e_0)$,
$(a_0, I_0)$,
$(a_0, q_\mathrm{min})$, and
$(a_0, e_0)$ planes.
We may find the following characteristics in these panels:
\begin{itemize}
\item %
In the        top row panels (\textsf{a}) that show the distribution of $\left< T_\mathrm{res} \right>$ projected on the $(e_0, q_\mathrm{min})$ plane, we see an apparent horizontal boundary with shorter $\left< T_\mathrm{res} \right>$ around $q_\mathrm{min} \sim 25$ au (the dark regions), particularly in the panel for A-early (left).
This reflects the existence of the
planet barrier.
The boundary also seems to exist in the panel for A-late (right),
Also, we may see that $\left< T_\mathrm{res} \right>$ in the region $q_\mathrm{min} \lesssim 10$ au is shorter in this panel than in other regions, possibly indicating that the Jupiter--Saturn barrier working to some extent.
This trend may be seen in the fifth top panels (\textsf{e}) that show the $\left< T_\mathrm{res} \right>$ distribution on the $(a_0, q_\mathrm{min})$ plane as well.
\item
There is nearly a linear boundary from the top left to the bottom right across the panel \textsf{a} for A-early in the top row.
We presume that this boundary somewhat reflects the initial configuration of the comet's eccentricity $(e_0)$ and perihelion distance $(q_0)$ that we showed at the bottom of Figure \ref{fig:oc-aeIq0_s5A}.
\item
In the second top panels (\textsf{b}) that show the $\left< T_\mathrm{res} \right>$ distribution on the $(I_0, q_\mathrm{min})$ plane,
we may find that the initially retrograde objects ($I_0 > 90^\circ$) appear to have a slightly longer $\left< T_\mathrm{res} \right>$ when $q_\mathrm{min} \lesssim 20$ au than the prograde objects ($I_0 < 90^\circ$):
the dark region that indicates shorter $\left< T_\mathrm{res} \right>$ near the bottom of each panel is somewhat narrower when $I_0 > 90^\circ$.
As we mentioned before, retrograde objects generally possess stronger stability than prograde ones, but the difference is subtle here.
\item
In the fourth top panels (\textsf{d}) that show the $\left< T_\mathrm{res} \right>$ distribution on the $(a_0, I_0)$ plane, particularly in the right panel for A-late, we can confirm what we already explained before:
when comets' initial semimajor axis $a_0$ is larger, $\left< T_\mathrm{res} \right>$ tends to be shorter regardless of the initial inclination $I_0$.
\end{itemize}

We already mentioned that the period A-late in our model is a proxy of the modern solar system.
During this period, the comets tend to fly over to the terrestrial planetary region nearly isotropically, and the planet barrier works less efficiently than in A-early.
However, even in the present day, if a star encounters the solar system in close proximity, and if a large number of new comets fall along the ecliptic plane as a shower, the planet barrier would work efficiently to protect the terrestrial planets from the bombardment of comets.
In fact, comet showers in our model generally occur along the ecliptic plane.
We think the reason for this is simple.
A comet shower occurs when a strong perturbation is applied to a region with a high number density of comets.
This region corresponds to the inner part of the comet cloud in our model, and most comets there are concentrated near the ecliptic plane even after the age of the solar system (see Figure \ref{fig:oc-evol-ah_s5A}).
Therefore, the comet showers due to stellar encounters in our model tend to occur along the ecliptic plane, and the planet barrier can efficiently work against them.

 \subsection{Transition into other small body populations\label{ssec:otherpops}}
It is not rare for the small solar system bodies to make transitions from one population to another over a long-term, such as from Centaurs to Jupiter-family comets \citep[e.g.][]{sarid2019,steckloff2020}.
It would be reasonable to assume that such a transition chain includes the Oort Cloud comets.
This kind of transition is also seen in our numerical result, and the comets experience various dynamical states during their stay in the planetary region.
They sometimes turn into different small body populations such as TNOs, Centaurs, Jupiter-family comets, and even near-Earth objects.
First, we define six small body populations for the following discussion.
In what follows we use the symbol $\land$ for logical conjunction (equivalent to logical \textsf{AND}):
\begin{itemize}
  \item near-Earth comets: ($q < 1.3$ au) $\land$ ($P < 200$ years)
  \item near-Earth asteroids: ($q < 1.3$ au) $\land$ ($a < 3.5$ au)
  \item main belt asteroids: ($2.1 < a < 3.5$ au) $\land$ ($e < 0.35$)
  \item Jupiter-family comets: ($2.0 < T_\mathrm{J} < 3.0$) $\land$ ($a \leq 10$ au)
  \item Jupiter Trojans: ($5.05 < a < 5.35$ au) $\land$ ($e < 0.2$) $\land$ ($I < 40^\circ$)
  \item Centaurs: ($a < a_\mathrm{N}$) $\land$ ($q > a_\mathrm{J}$)
\end{itemize}
Here
  $P$ denotes orbital period,
  $T_\mathrm{J}$ is the Tisserand parameter with respect to Jupiter,
  $a_\mathrm{J}$ is Jupiter's semimajor axis ($=5.2$ au), and
  $a_\mathrm{N}$ is Neptune's semimajor axis ($=30$  au).
$T_\mathrm{J}$ is defined as
  $T_\mathrm{J} = a_\mathrm{J}/a + 2 \sqrt{\left(a/a_\mathrm{J}\right) \left(1-e^2\right)} \cos I$
\cite[e.g.][]{danby1992}.
As for the dynamical definition of Centaurs, we adopt the categorization given in the IAU Minor Planet Center (MPC: \url{https://www.minorplanetcenter.net/}) or in \citet{jewitt2009}.

\begin{figure}[!htbp]\centering
  \includegraphics[width=\myfigwidth]{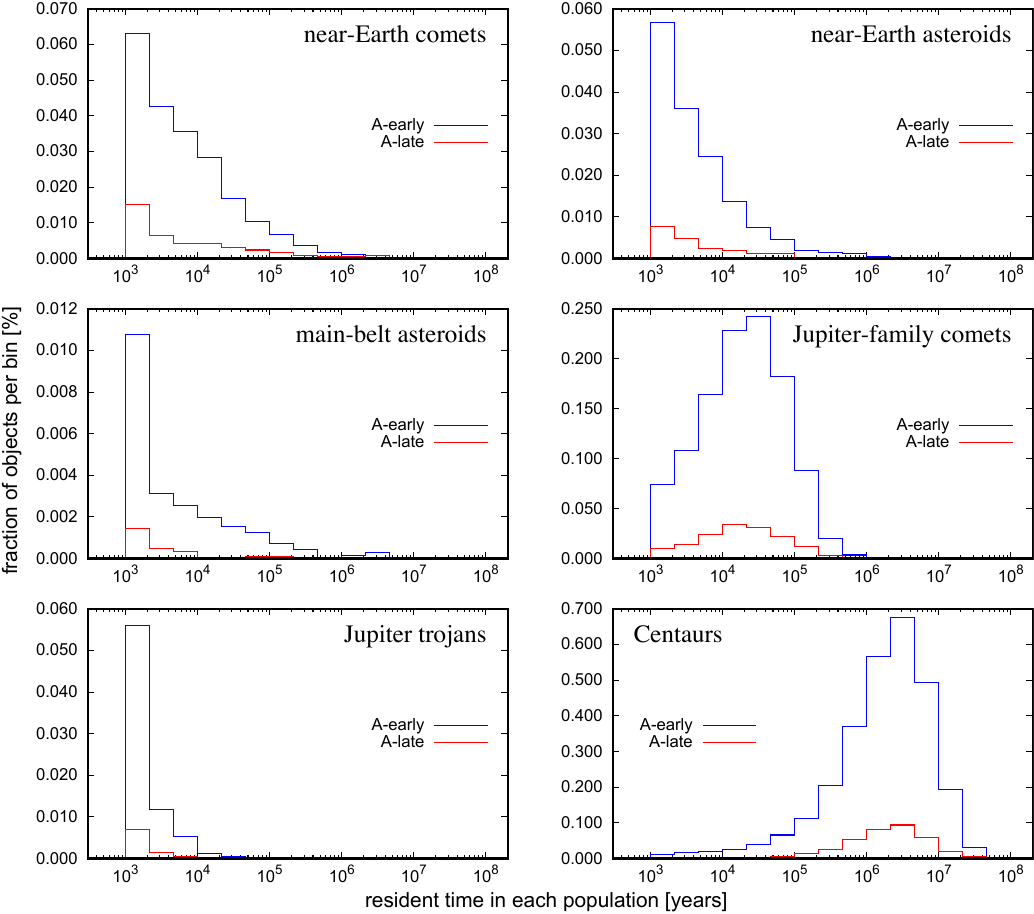}
  \caption[]{Distribution of the resident time of comets in each of the six small solar system body populations.
The values on the vertical axis are normalized, and multiplied by a factor of 100.
}
  \label{fig:oc-group-rt-1_s5A}
\end{figure}

In Figure \ref{fig:oc-group-rt-1_s5A} we show histograms for the resident time distribution of the new comets in the orbital space of the six populations defined as above.
The values on the vertical axis are normalized by the total number of the objects generated in the comet cloud in each period, and also multiplied by a factor of 100 so that the values are expressed in per cent.
For example, the vertical value of 0.01{\%} in a bin means that the actual fraction of objects in the bin is $0.01/100 = 0.0001$.
Since the total number of the new comets generated in each period in our model is between
710 and 747 thousand (see Table \ref{tbl:ocncfate_s5A}), the fractional value 0.0001 would be equivalent to having
71--75 objects in the bin.
Note also that the apparent peaks seen at the leftmost bins in some panels (where the resident time is $\sim 10^3$ years) reflect the contribution from the objects that were ejected out of the system just after the first apparition.
This may be considered as a model artifact, and we already mentioned it (see our discussion given for Figure \ref{fig:oc-aeIqQ-Te-log_s5A}).

As expected, few comets experience the state of the main belt asteroids (the middle left panel) or the Jupiter Trojans (the bottom left panel) which have moderate to small eccentricity.
As we see in the top row panels in Figure \ref{fig:oc-group-rt-1_s5A}, a small fraction of objects experiences to be the near-Earth comets or the near-Earth asteroids, but the time spent in these populations is about $10^5$ years or less.
This is an order of magnitude shorter than the typical dynamical lifetime of the near-Earth asteroids observed in the current solar system \citep[e.g.][]{gladman1997,ito2006c,granvik2016}.
The shorter resident time is probably caused by the fact that the eccentricity of the objects of this type in our model is larger than that of typical near-Earth asteroids in the current solar system.

We also find that the number of objects experiencing the near-Earth comets or the near-Earth asteroids is notably higher in the period A-early than in A-late.
Since orbital inclination is not included in our definitions of the near-Earth comets or the near-Earth asteroids, we cannot attribute this difference to the difference in the distribution of initial orbital inclination of the objects in A-early and A-late.
We rather interpret this as simply reflecting the occurrence of a very intensive cometary shower at time $\sim 0.45$ Gyr in A-early which brought the perihelion of many new comets into the vicinity of Earth's orbit.
Incidentally, let us mention that we did not observe any Atens ($a < 1$ au $\land$ $Q > 0.983$ au) or Atiras ($a < 1$ au $\land$ $Q < 0.983$ au) in our numerical orbit simulation.

While our definitions of the near-Earth asteroids or the main-belt asteroids in this study do not depend on orbital inclination, those of the Jupiter-family comets, the Jupiter Trojans, and the Centaurs do.
We find the consequence of these dependencies in Figure \ref{fig:oc-group-rt-1_s5A}.
Specifically speaking, in the period A-early when many new comets with small orbital inclinations occur, the frequency of the objects that experience the state of the Jupiter-family comets, Jupiter Trojans, and Centaurs is higher than in A-late.
In A-early, the integrated frequency (sum of the values in all bins) of the objects experiencing the Jupiter family comet population reaches $\sim 1${\%}, and that of the objects experiencing the Centaur population reaches several {\%}.

Although not as frequently as in the period A-early, transitions to other small body population also occur in the period A-late.
As we have mentioned several times, the period A-late is a proxy of the modern solar system.
In the panel for the Jupiter-family comets in Figure \ref{fig:oc-group-rt-1_s5A} (in the middle right panel), if we add up the numbers in all the bins
for the period of A-late (denoted in red), we roughly find that a few $\times 0.1${\%} of all the comets we deal with during one billion year in the period A-late have transitioned into the Jupiter-family comets, although temporarily.
In fraction (not in {\%}), this value is equivalent to a few $\times 10^{-3}$.
These numbers did not change much either when we used another star set in the period of B-late (see Figure \ref{fig:oc-group-rt-1_s3B}).
Also, they are roughly consistent to preceding studies such as \citet[][Table 1 in their p.~212]{biryukov2007} that discusses the capture probability of the Oort Cloud comets into the Jupiter-family orbits in the modern solar system.

Although Figure \ref{fig:oc-group-rt-1_s5A} does not include specific information on orbital inclination of the objects in this population, let us note that the objects that temporarily became the Jupiter-family comets have a broad distribution of orbital inclination.
Its time-average value $\left< I_\mathrm{JFC} \right>$ is,
in A-early,        $\left< I_\mathrm{JFC} \right> = 24.5^\circ \pm 12.6^\circ$ where the $\pm$ error is one standard deviation from the distribution of inclination.
In A-late,   it is $\left< I_\mathrm{JFC} \right> = 21.9^\circ \pm 11.3^\circ$.

In the bottom right panel for Centaurs in our Figure \ref{fig:oc-group-rt-1_s5A}, we find that the sum of all the bin values is 3 to 4 times larger than that of the Jupiter-family comets (the middle right panel).
In the period of A-late (denoted in red), the summed value is a few $\times 0.1${\%}.
In fraction (not in {\%}), this range is equivalent to several $\times 10^{-3}$ or close to $10^{-2}$.
This number is not inconsistent with preceding studies such as \citet[][their p.~1347]{emelyanenko2005} that discussed the transition of small bodies from the Oort Cloud into Centaurs and the Jupiter-family comets.
Consult also \citet{levison2001} which worked on the origin of the Halley-type comets and its relation to the Oort Cloud.

\begin{figure}[!htbp]\centering
  \includegraphics[width=\myfigwidth]{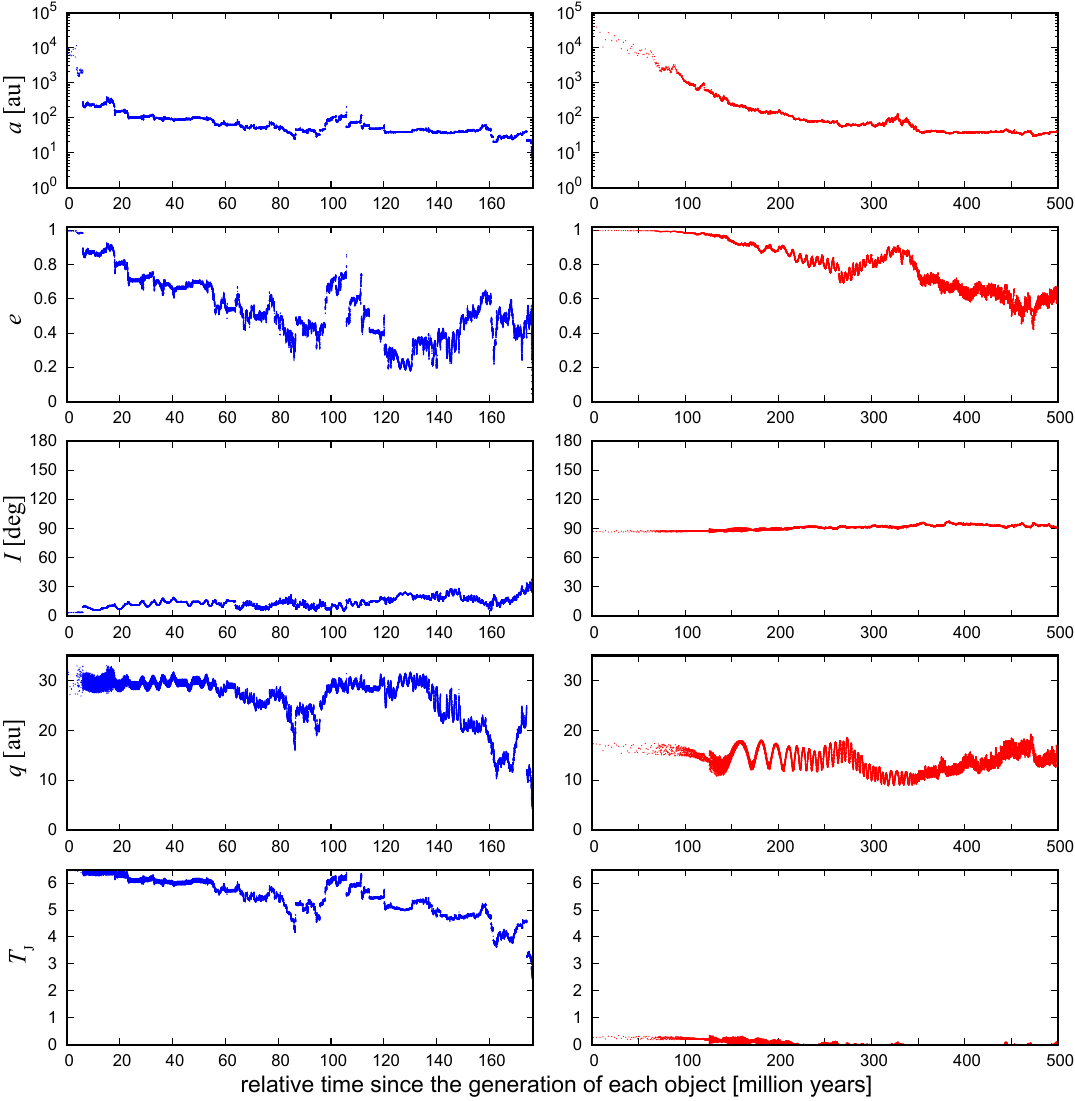}
  \caption[]{%
A pair of examples of the transition of objects among different small body populations.
Left  (plotted in blue): an object generated in A-early at $t = 9.7117221 \times 10^6$ years.
Right (plotted in red):  an object generated in A-late  at $t = 4.4774344 \times 10^9$ years.
From the top, time variation of
semimajor axis $a$,
eccentricity $e$,
inclination $I$,
perihelion distance $q$, and
the Tisserand parameter with respect to Jupiter $T_\mathrm{J}$.
The horizontal axis denotes the relative time since the generation of each object in the comet cloud.
}
  \label{fig:typorbit-2021}
\end{figure}

Figure \ref{fig:typorbit-2021} displays a pair of examples of the transition of objects among different small body populations in our model.
The object in the left column (plotted in blue) was generated in A-early.
This object remains on an orbit along the ecliptic with small inclination $I$, and it gradually reduces its eccentricity.
On the other hand, its semimajor axis was quickly reduced from the initial value of $a \sim 10^3$ au to $10^2$ au or less within the first $\sim 10$ million years, and it keeps decreasing after that too.
As a result, its perihelion distance $q$ became lower than $\sim 20$ au once at the relative time $\sim 80$ million years, and then again at $\sim 150$ million years.
In other words, this object penetrated the Neptune--Uranus barrier at these points.
However, its $q$ did not get lower than 10 au until the end, and we can say that this object was practically blocked by the Saturn--Jupiter barrier.
On the course of the dynamical evolution, this object experienced
the status of Centaur,
and finally got ejected out of the system by a close encounter with Jupiter.

On the other hand, the object in the right panel (plotted in red) was generated in A-late.
Its semimajor axis changes from $a \sim 10^4$ au to $10^2$ au or less, while it remains on a highly eccentric orbit $(e > 0.95)$ for the first $\sim 100$ million years.
Then its eccentricity starts a slow decrease.
A remarkable point about this object is that its orbital inclination remains high $(I \sim 90^\circ)$ throughout the entire integration period.
In other words, this object is a typical polar corridor object (see below), and its perihelion distance stays between the orbits of Uranus and Saturn ($10 \lesssim q \lesssim 20$ au).
This object survived the entire integration period of 500 million years, but it also experienced both the status of Centaurs and the Jupiter-family comets during the period.

So far we have described the cases where the comets experience the small body populations inside  the Centaur orbits.
         Next we discuss the cases where the comets become            the TNO populations outside the Centaur orbits.
The populations we consider here are the Classical TNOs and the Detached TNOs.
We define each of them as follows, largely adopting the definitions that \citet{gladman2008} give.
In addition, we also pay attention to the objects approaching the inner part of the solar system through the so-called polar corridor.
This is a dynamical path that connects the outer part of the planetary region and the Oort Cloud with large inclination, possibly a favorable location for the Centaurs that are once captured by Jupiter and then left orbit \citep{namouni2018b,namouni2020}:
\begin{itemize}
  \item 
        Classical TNOs: ($q > a_\mathrm{N}$) $\land$ ($39.4 < a  < 47.8$ au) $\land$ ($e < 0.24$) $\land$ ($I < 35^\circ$)
  \item 
        Detached TNOs:  ($q > q_\mathrm{DT}$) $\land$ ($a < 2000$ au) $\land$ ($e > 0.24$)
  \item Polar corridor objects: ($30 < a < 1000$ au) $\land$ ($60^\circ < I < 120^\circ$)
\end{itemize}

Note that our definition of the detached TNOs is rather an extended one including (90377) Sedna.
Considering the ambiguity of the definition of the detached TNOs, in this study we tested three values of the smallest perihelion distance of the detached TNOs ($q_\mathrm{DT} = 35$, 38, 40 au) and see how different the result is.

\begin{figure}[!htbp]\centering
  \includegraphics[width=\myfigwidth]{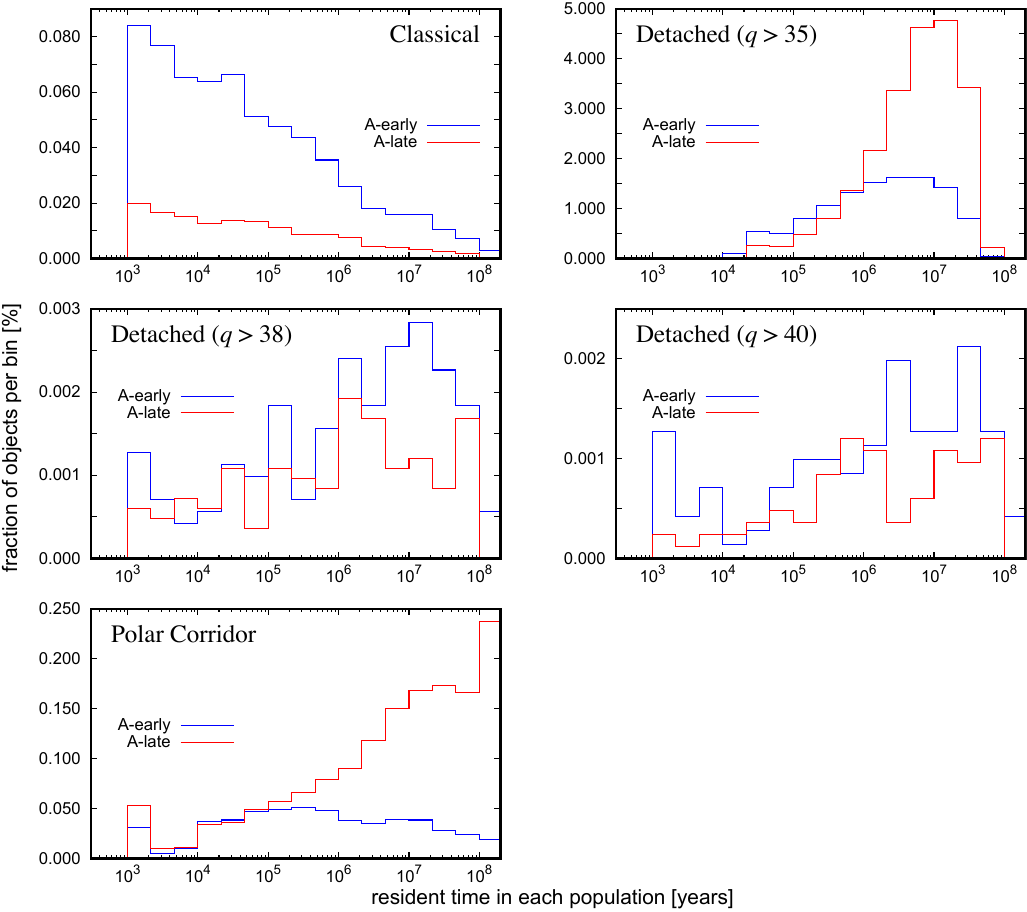}
  \caption[]{Distribution of the resident time of comets as
the classical TNOs,
the detached TNOs with $q_\mathrm{DT} = 35, 38, 40$ au, and
the polar corridor objects.
The values on the vertical axis are normalized, and multiplied by a factor of 100.
}
  \label{fig:oc-group-rt-2_s5A}
\end{figure}

Figure \ref{fig:oc-group-rt-2_s5A} shows distribution of the resident time of the comets in the orbital space of the populations (classical TNOs, detached TNOs with $q_\mathrm{DT} = 35$, 38, 40 au, and the polar corridor objects defined as above).
The values on the vertical axis are normalized by the total number of the comets generated in each period, and also multiplied by a factor of 100 so that the values are expressed in per cent.
From this figure we can tell that the number of the comets experiencing the classical TNO population is not large.
The main reason for this is that classical TNOs are defined as objects with moderate eccentricity $(e < 0.24)$, and only a small fraction of the comets in our model satisfy this condition.
Also, the definition includes the condition $T_\mathrm{J} \geq 3.05$ which generally prefers the objects with smaller inclination. 
This brings an outcome that the number of the comets that experience the classical TNOs is even more limited in the period A-late.
On the other hand, the fraction of the comets that experience the detached TNOs is large.
The integrated fraction of the new comets experiencing the detached TNOs (when $q_\mathrm{DT} = 35$ au) is 10{\%} or so, indicating that a significant fraction of new comets once become detached TNOs.
However, the fraction of the detached TNOs becomes much smaller when we adopt $q_\mathrm{DT} = 38$ or 40 au, at most some 0.01{\%} in total.
This indicates that if the detached TNO population contains any comets originated from the Oort Cloud, their perihelion distance is possibly $q \lesssim 35$ au.
It should be noted, however, that these probabilities in our result may strongly depend on our definition of the new comets ($r < 30$ au).
If we change this definition to a different one such as $r < 35$ au, the probability distribution for the comets to experience the detached TNOs with $q_\mathrm{DT} \gtrsim 38$ au can be substantially larger than what is shown in Figure \ref{fig:oc-group-rt-2_s5A}.

As is depicted in \citet{namouni2018b}, the polar corridor occupies a region with large orbital inclination in phase space.
Thus, a large number of objects in our model pass through this region during the period A-late when many new comets travel toward the planetary region nearly isotropically.
In other words, many of the new comets observed in the modern solar system possibly came through this corridor.
This is not the case in A-early when the inclination of the new comets is generally small.

Since the spatial region of the polar corridor is quite large, it is possible that a large fraction of high-inclination Centaurs and some TNOs once go through this region before dynamically evolving into different population.
The source of high-inclination Centaurs (with small $T_J$) that later evolve into Halley-type comets such as what \citet{fernandez2018} discussed may have something to do with the polar corridor.
This statement can be particularly the case in the late stage (such as in A-late or B-late) when the comet cloud has dynamically evolved into the three-dimensional shape with high inclination.
Several TNOs are already detected and recognized in the orbital region of the polar corridor \citep[e.g.][their Figure 4 and the box in p.~233]{gladman2021}.
See also \citet{namouni2022,namouni2024} about the relevance of the polar corridor to the planet-crossing asteroids.

In \ref{suppl:transient}, we attached a series of the frequency distribution map of the orbital elements $a$, $e$, $I$ and $q$ of these transitioning objects that we have described in this section (Figures \ref{fig:oc-group-elem-ae_s5A}--\ref{fig:oc-group-elem-aq-gladman_s5A}).

\begin{figure}[!htbp]\centering
  \includegraphics[width=\myfigwidth]{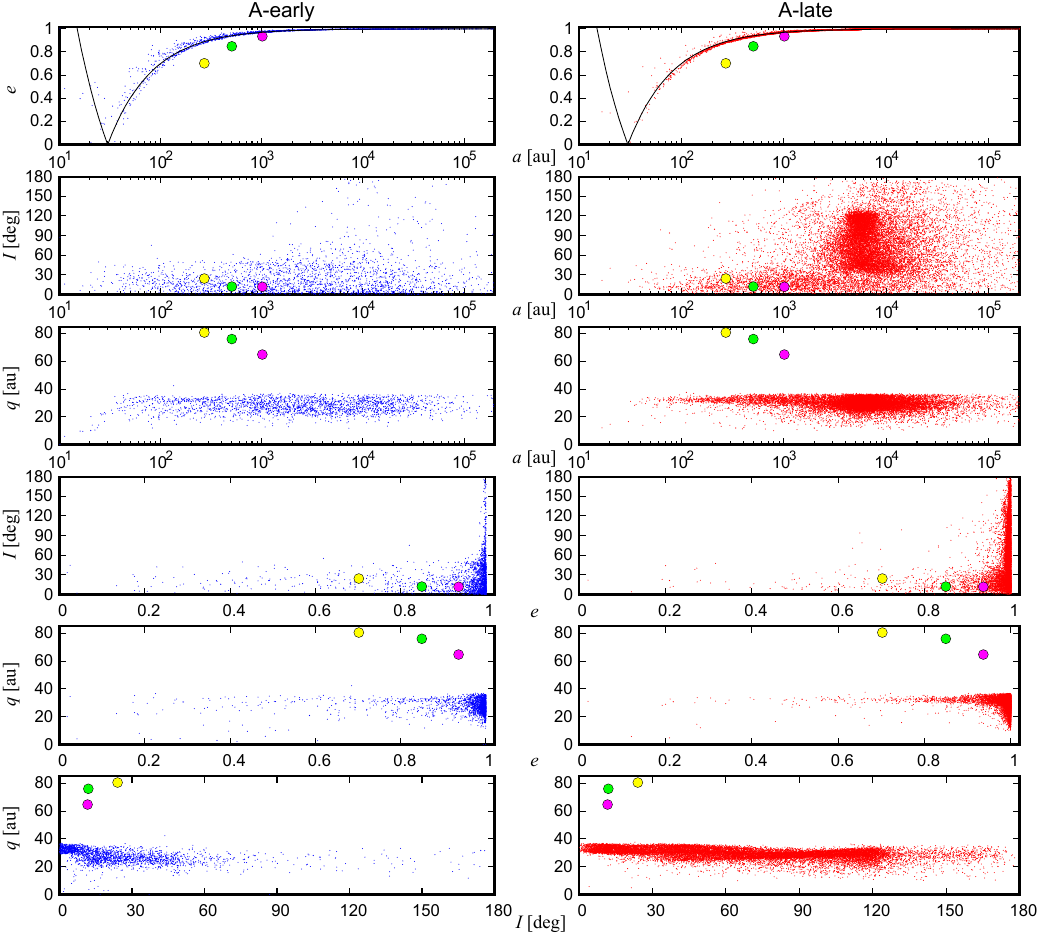}
  \caption[]{%
Scatter plots of the survivors' orbital elements at their end state in A-early and in A-late.
We also plotted the current osculating orbital elements of three peculiar TNOs for comparison:
(90377)  Sedna        (2003 VB${}_{12}$)  as the large green   circles,
                       2012 VP${}_{113}$  as the large yellow  circles, and
(541132) Leleakuhonua (2015 TG${}_{387}$) as the large magenta circles.
}
  \label{fig:oc-survivors_s5A}
\end{figure}

\subsection{Survivors over 500 Myr\label{ssec:survivors}}

In our numerical orbit integration,
some fraction of objects survive over the entire integration period of 500 million years since their generation in the comet cloud (Table \ref{tbl:ocncfate_s5A}).
Figure \ref{fig:oc-survivors_s5A} collects scatter plots of orbital elements $a$, $e$, $I$, $q$ of the survivors at their final state together with the orbital location of three peculiar TNOs:
(90377)  Sedna        (2003 VB${}_{12}$),
                       2012 VP${}_{113}$, and
(541132) Leleakuhonua (2015 TG${}_{387}$).
In the panels of $(a,e)$ on the top row, we find that the distribution of many survivors' perihelion distance $q$ follows the curve representing $q = 30$ au.
This indicates that they survived with their perihelion just outside Neptune.
In the panels of $(a,I)$ and $(e,I)$, we find that survivors' inclination distribution largely reflecting the initial state:
   the survivors in A-early tend to have smaller inclination,
while      those in A-late  tend to have larger  inclination.
In particular, a cluster of objects in the panels of $(a,I)$ with high inclination in $a = 10^3$--$10^4$ au is noticeable.
This should be compared with Figures \ref{fig:oc-genocnc-combined-t_s5A} and \ref{fig:oc-aeIq0_s5A} that depict the initial orbital distribution of the comets.

Dynamical origin of the detached TNOs with unique orbits such as Sedna is not well understood \citep[e.g.][]{schwamb2010,soares2013}.
Our numerical simulations did not reproduce the high perihelion distances of these three objects either.
Our result about this point may somewhat vary depending on the initial conditions of the comets we set.
However, it seems difficult to explain the orbital distribution of these objects quantitatively enough using only the perturbations from the major planets that we incorporated in our numerical model.
Thus, we have to agree with the statement that the dynamical origin of these extremely detached objects still remains mystery.

\section{Summary and conclusion\label{sec:summary}}

We employed a pair of dynamical models to estimate the resident time of the Oort Cloud comets in the planetary region.
The initial state of the modeled comet cloud is a two-dimensional planetesimal disk.
This disk then dynamically evolves into three-dimensional, nearly isotropic state due to the galactic tidal force and stellar encounters with nearby stars.
During the dynamical evolution of the comet cloud, many objects in the cloud fall into planetary region and become new comets.
Our main results are as follows:
\begin{itemize}
\item
Typical dynamical resident time of the comets in the planetary region $T_\mathrm{res}$ is about
$10^8$ years.
The value of $T_\mathrm{res}$ becomes longer in the later part of the solar system history when comet clouds approach the isotropic shape.
\item 
When the initial orbital inclination of the comets is small, the so-called planetary barrier gets in effect.
The barrier efficiently prevents the comets from penetrating into the terrestrial planetary region, particularly when the comet cloud is still nearly flat.
It also works during the periods when strong comet showers occur along the ecliptic plane due to close approaches of stars.
\item
During their stay in the planetary region, the comets experience transitions into other small body populations such as transneptunian objects.
Many of them pass through the so-called polar corridor.
Also, a non-negligible amount of objects temporarily become Jupiter-family comets or Centaurs.
\end{itemize}

Our study does not consider any kind of physical effects such as fading or disintegration of comets.
Also, our dynamical model is incomplete or inaccurate in some aspects, as we have described in previous sections.
However, we believe that this study has provided some useful insights into the duration of the Oort Cloud comets' stay in planetary region and their spatial penetration.
In addition, our numerical data, particularly that presented in Section \ref{ssec:otherpops}, combined with theories of physical and thermal evolution of comets, may allow us to theoretically model their dynamical transfer and physical evolution over a vast spatial region from the Oort Cloud through the Kuiper Belt to the inner solar system.
This could provide constraints on the origin of short-period comets as well as long-period comets from various source regions.
Historically speaking, we can presume such attempts began in \citet{fernandez1980}.
Now that the techniques for numerical simulation and the accuracy of observational data have both been developed and refined to a great deal, realization of such comprehensive modeling has become a reality.

In this study, we estimated the frequency distribution of the resident time that the Oort Cloud comets transition to and stay in various populations of the small solar system bodies in the planetary region (Section \ref{ssec:otherpops}).
Generally, objects in the Oort Cloud are considered to have the properties of comets.
Recently, however, the interrelationships between the Oort Cloud comets and asteroids have been the focus of much attention \citep[e.g.][]{weissman1997,shannon2015,shannon2019}.
The transition between comets and asteroids is known to occur in the inner region of the solar system \citep[e.g.][]{jewitt2012,ito2018a}, and physical models to realize it have been proposed \citep[e.g.][]{miura2022}.
Therefore, as our numerical calculations indicate, we cannot rule out the possibility that objects from the Oort Cloud are present among the near-Earth objects population as asteroids, rather than comets, in numbers that cannot be ignored.
Including this conjecture, the validity of the results of our model calculation should be verified by large-scale observational surveys in the near future \citep[e.g.][]{aihara2018a,ivezic2019,yoshida2011a,yoshida2024,fraser2024}.
For example, if a wide-field survey of distant comets has detected many (probably faint) comets with perihelion located beyond Uranus' orbit, and if a local maximum in their perihelion distance distribution has been confirmed between Uranus' orbit and Neptune's orbit (or between Jupiter's orbit and Saturn's orbit), it would be a vindication that the planetary barrier that we have studied in this article is at work.

\section*{CRediT authorship contribution statement}
Takashi Ito:   Conceptualization, Methodology, Software, Validation, Investigation, Resources, Writing {\textemdash} Original Draft, Writing {\textemdash} Review {\&} Editing, Visualization, Funding acquisition.
Arika Higuchi: Conceptualization, Methodology, Software,             Investigation,                                                  Writing {\textemdash} Review {\&} Editing.

\section*{Declaration of competing interest}
The authors declare no competing interests.

\section*{Data availability}
Numerical data supporting this study's findings are available from the authors upon reasonable request.

\section*{Acknowledgments}
We wish to greatly thank Julio A. Fernandez and an anonymous reviewer for providing us with detailed and critical reviews.
These suggested directions that made better the presentation and quality of the paper to a great deal.
We started this study in June 2006.
With the help of   Marc Fouchard who provided helpful comments on the direction of this research in its early stages, the numerical simulation including planetary perturbation was completed in mid 2023.
The idea to test three kinds of boundary values $q_\mathrm{DT}$ for the definition of the detached TNOs in Figures \ref{fig:oc-group-rt-2_s5A} and \ref{fig:oc-group-rt-2_s3B}, and that to see how many new comets go through the polar corridor, came from a question that we received from Brett Gladman at the 52th DPS/AAS meeting, October 2020.
The numerical orbit integration carried out for this study was largely performed at Center for Computational Astrophysics (CfCA), National Astronomical Observatory of Japan.
This study is also supported by the JSPS Kakenhi Grant (JP25400458/2013--2016, JP16K05546/2016--2018, JP18K03730/2018--2021, JP20K04054/2020--2022), and the JSPS bilateral open partnership joint research project (2014--2015).
This study has made use of NASA's Astrophysics Data System (ADS) Bibliographic Services.

\clearpage

\appendix

\section{Probability density function of $e$ and $a$\label{appen:probability-e}}
As we wrote in Section \ref{sec:genOCNC}, when semimajor axis of objects has a differential number distribution (probability density function) proportional to $a^{-2}$, and if their perihelion distance $q = a(1-e)$ is a common constant to all the objects, the distribution of objects' eccentricity becomes uniform.
Here we show the process of the conversion.

First, we consider semimajor axis $a$ is the random variable.
Let us write that the probability density function $f$ of $a$ as follows:
\begin{equation}
  f(a) = \kappa a^{-2} ,
\label{eqn:a-pdf-a}
\end{equation}
where $\kappa$ is a constant coefficient.
In our model, we assume that initial perihelion distance $q_0 = a(1-e)$ is a common constant to all the objects.
Therefore we have the following relation:
\begin{equation}
  a = \frac{q_0}{1-e}.
\label{eqn:a-expressed-by-e}
\end{equation}

Now we regard the eccentricity $e$ as the new random variable.
The derivative function between the old $(a)$ and the new $(e)$ random variables is as follows:
\begin{equation}
  \frac{da}{de} = q_0 \left( 1-e \right)^{-2} .
\end{equation}

Therefore the probability density function $\gamma$ of the new random variable $(e)$ is derived as follows:
\begin{equation}
\begin{aligned}
  \gamma (e) &= q_0 \frac{da}{de} f(a) \\
             &= q_0 \left( 1-e \right)^{-2} f \left( \frac{q_0}{1-e} \right) \\
             &= q_0 \left( 1-e \right)^{-2} \left( \kappa \frac{q_0}{1-e} \right)^{-2} \\
             &= \kappa^{-2} q_0^{-1} .
\end{aligned}
\label{eqn:a-pdf-gamma-e}
\end{equation}

Eq. \eqref{eqn:a-pdf-gamma-e} means that the probability density function of $e$ has a constant value.
This brings us the conclusion that the probability distribution of $e$ is uniform.

In a similar manner, we can derive the probability density function form of semimajor axis $a$ as being proportional to $a^{-2}$, starting from the constant probability density function presented as Eq. \eqref{eqn:a-pdf-gamma-e}.
Now, consider eccentricity $e$ is the original random variable.
Its probability density function is given in Eq. \eqref{eqn:a-pdf-gamma-e}, and we write it as
\begin{equation}
  \gamma (e) = \widetilde{\kappa}^{-2} ,
\end{equation}
where we regard $\widetilde{\kappa}^{-2} = \kappa^{-2} q_0$ is a constant parameter.
We already have the relationship between $e$ and $a$ as Eq. \eqref{eqn:a-expressed-by-e}.
Then we regard the semimajor axis $a$ as the new random variable.
The derivative function between the old $(e)$ and the new $(a)$ variables is as follows:
\begin{equation}
  \frac{de}{da} = q_0 a^{-2} .
\end{equation}

Therefore the probability density function $\widetilde{f}$ of the new variable $(a)$ becomes as follows:
\begin{equation}
\begin{aligned}
  \widetilde{f}(a) &= \frac{de}{da} \gamma(e) \\
                   &= q_0 a^{-2} \cdot \widetilde{\kappa}^{-2} \\
                   &= \widehat{\kappa} a^{-2} ,
\end{aligned}
\label{eqn:a-pdf-rev-a}
\end{equation}
where $\widehat{\kappa} = q_0 \widetilde{\kappa}^{-2}$ is a constant parameter.
The probability density function $\widetilde{f}(a)$ in Eq. \eqref{eqn:a-pdf-rev-a} has the same form to $f(a)$ in Eq. \eqref{eqn:a-pdf-a}.

The function form seen in Eq. \eqref{eqn:a-pdf-rev-a} or in Eq. \eqref{eqn:a-pdf-a} is equivalent to the differential differential number distribution of the semimajor axis of the objects in our comet cloud model (denoted as $\frac{dN(a)}{da} \, \propto \, a^{-2}$) that we mentioned in Section \ref{sec:genOCNC}.

\clearpage
 \section{Distribution of $g_0$ of new comets\label{appen:g0_180}}

In this appendix we show our understanding as to why we see concentrations of the initial argument of perihelion $g_0$ of the new comets around 0 and $180^\circ (= \pi)$ in Figure \ref{fig:oc-genocnc-combined-t_s5A}.

First, let us pick two relevant equations of motion in Gauss's form from \citet[][Eq. (33) in their p. 301]{brouwer1961} as follows:
\begin{equation}
\begin{aligned}
   \frac{dI}{dt} &= \frac{\mathrm{W}}{na\sqrt{1-e^2}} \frac{r}{a}       \cos \left(g + f\right),           \\
   \frac{dh}{dt} &= \frac{\mathrm{W}}{na\sqrt{1-e^2}} \frac{r}{a} \frac{\sin \left(g + f\right)}{\sin I} ,
\end{aligned}
\label{eqn:gauss-I-h}
\end{equation}
where $t$ is time, $f$ is true anomaly of the object, $r$ is its heliocentric distance, and $\mathrm{W}$ is the force component perpendicular to the object's orbital plane.
We consider $\mathrm{W}$ is positive in the direction along which the orbital motion appears counterclockwise (i.e. the right-hand rule).
Any kind of perturbations including the galactic tidal force and encounters with nearby stars can invoke $\mathrm{W}$.
Let us also add one more equation of motion about argument of pericenter $g$ in Gauss's form:
\begin{equation}
\begin{aligned}
  \frac{dg}{dt} &= \frac{\sqrt{1-e^2}}{nae}
    \left[ - \mathrm{R} \cos f + \mathrm{S} \left( 1 + \frac{r}{a\left(1-e^2\right)}  \right) \sin f \right] \\
                & \quad
                 - \frac{\mathrm{W}}{n a \sqrt{1-e^2}} \frac{r}{a} \frac{\sin \left(g+f\right)}{\tan I} ,
\end{aligned}
\label{eqn:gauss-g}
\end{equation}
where $\mathrm{R}$ is the force component in the direction of the radius vector of the object, and $\mathrm{S}$ is the component perpendicular to $\mathrm{R}$ in the orbital plane.
$\mathrm{R}$ is positive in the direction that the radius vector of the object extends along it, and
$\mathrm{S}$ takes positive values in the direction of increasing longitude of the object orbiting along it.

Among the pair of equations in Eqs. \eqref{eqn:gauss-I-h},
the first equation about $\frac{dI}{dt}$ tells us that $\mathrm{W}$ works most efficiently with respect to the change of inclination $I$ when $\cos \left( g+f \right) = \pm 1$.
This condition is equivalent to $g+f = 0$ or $\pi$, hence the comet must be at its ascending node $(g+f = 0)$ or its descending node $(g+f = \pi)$.
In the initial comet cloud that we consider, eccentricity of the objects is very large, $e \sim 1$ (see Figure \ref{fig:oc-ini_disk}).
Therefore, in average, we can expect that many objects stay near their apocenter ($f \sim \pi$).
Then, it is obvious that an object's orbital inclination is efficiently excited
when $g \sim -\pi$ (at its ascending  node where $g+f=0$)   or
when $g \sim    0$ (at its descending node where $g+f=\pi$).

The above discussion implies that the variation rate of comet's inclination $\frac{dI}{dt}$ remains small unless $g \sim -\pi$ or $g \sim 0$ (assuming $f \sim \pi$).
Meanwhile, the second equation of Eqs. \eqref{eqn:gauss-I-h} tells us that the time derivative of longitude of ascending node $\frac{dh}{dt}$ is proportional to $\sin (g+f) / \sin I$.
This means that $\frac{dh}{dt}$ becomes small when the object is around its ascending node ($g+f=0$) or its descending node $(g+f=\pi)$ because of the factor $\sin (g+f)$.
However, another factor ($1/\sin I$) enhances $\frac{dh}{dt}$ when the comet's inclination $I$ is small, and the comet's nodes move fast in this case.

Having the above properties in mind, let us consider the motion of an object in the cloud that initially has $I = 0$ and $e \sim 1$.
\begin{itemize}
\item For an object initially with $I=0$, no ascending node is defined.
However, if any perturbing forces kick in, its ascending node is defined in a short time and its inclination starts to rise.
\item Whatever the source of the perturbing force is,
the first equation of Eq. \eqref{eqn:gauss-I-h} tells us that
it is the objects located near its ascending node ($g+f=0$) or descending node ($g+f=\pi$) whose inclination gets most efficiently enhanced.
We should also remember the fact that most of the objects we consider here stick around their apocenter with $f \sim \pi$ due to their large eccentricity.
Therefore, the objects are enhanced their inclination most efficiently when their argument of pericenter is $g \sim -\pi$ or $g \sim 0$.
This means that the object's perihelion is around its descending node $(g \sim -\pi)$ or ascending node $(g \sim 0)$.
\item Once the inclination $I$ of an object increases, motion of its nodes slow down due to the factor $1/\sin I$ in the right-hand side of the second equation of Eq. \eqref{eqn:gauss-I-h}.
\item Due to the above mentioned reason, the inclination $I$ of an object that is not in the vicinity of its ascending node or descending node is less efficiently enhanced.
However, once the object approaches its ascending node $(g+f=0)$ or descending node $(g+f=\pi)$, the increase rate of its inclination (invoked by $\mathrm{W}$) gets higher.
Then, its ascending node begins to move more slowly as explained above, and the condition $g \sim -\pi$ or $g \sim 0$ is likely to remain fulfilled.
Note that the condition $g \sim -\pi$ is practically equivalent to $g \sim \pi$.
\end{itemize}

Let us say one more thing about the motion of object's argument of pericenter $g$ from the form of Eq. \eqref{eqn:gauss-g}:
\begin{itemize}
  \item
Eccentricity of the objects considered here is so large $(e \sim 1)$ that the effect of the first and second terms on the right-hand side of Eq. \eqref{eqn:gauss-g} can be suppressed (see the coefficient $\sqrt{1-e^2}$ multiplied by both terms).
The second term also has a coefficient $\sin f$ with $f \sim \pi$, which turns into $\sin f \sim 0$.
Therefore, in what follows we consider only the third term.
The effect of the third term depends on the sign of $\mathrm{W}$ as well as on argument of pericenter $g$ itself, and can be divided into the following two cases.
Here we assume $\tan I > 0$ (objects in prograde orbit):
  \begin{itemize}
    \item When $\mathrm{W} > 0$ : The third term is approximately proportional to $-\sin (g + \pi)$.
      Thus, if $0   < g <  \pi$, then $\frac{dg}{dt} > 0$.
            If $\pi < g < 2\pi$, then $\frac{dg}{dt} < 0$.
      In both cases, argument of pericenter $g$ changes toward $g = \pi$.
    \item When $\mathrm{W} < 0$ : The third term is proportional to $+\sin (g + \pi)$.
      Thus, if $0   < g <  \pi$, then $\frac{dg}{dt} < 0$.
            If $\pi < g < 2\pi$, then $\frac{dg}{dt} > 0$.
      In both cases, argument of pericenter $g$ changes toward $g = 0$ (or $g = 2\pi$).
  \end{itemize}
Therefore in the considered system, the force component $\mathrm{W}$ is responsible for keeping argument of pericenter of the objects in the vicinity of $g=0$ or $g = \pi$.
  \item The above conclusion remains the same even if we assume $\tan I < 0$ (objects in retrograde orbit). More specifically, when $\tan I < 0$:
  \begin{itemize}
    \item When $\mathrm{W} > 0$ : The third term is proportional to $+\sin (g + \pi)$.
      Thus, if $0   < g <  \pi$, then $\frac{dg}{dt} < 0$.
            If $\pi < g < 2\pi$, then $\frac{dg}{dt} > 0$.
      In both cases, argument of pericenter $g$ changes toward $g = 0$ (or $g = 2\pi$).
    \item When $\mathrm{W} < 0$ : The third term is proportional to $-\sin (g + \pi)$.
      Thus, if $0   < g <  \pi$, then $\frac{dg}{dt} > 0$.
            If $\pi < g < 2\pi$, then $\frac{dg}{dt} < 0$.
      In both cases, argument of pericenter $g$ changes toward $g = \pi$.
  \end{itemize}
\end{itemize}

\begin{figure}[!htbp]
 \includegraphics[width=\myfigwidth]{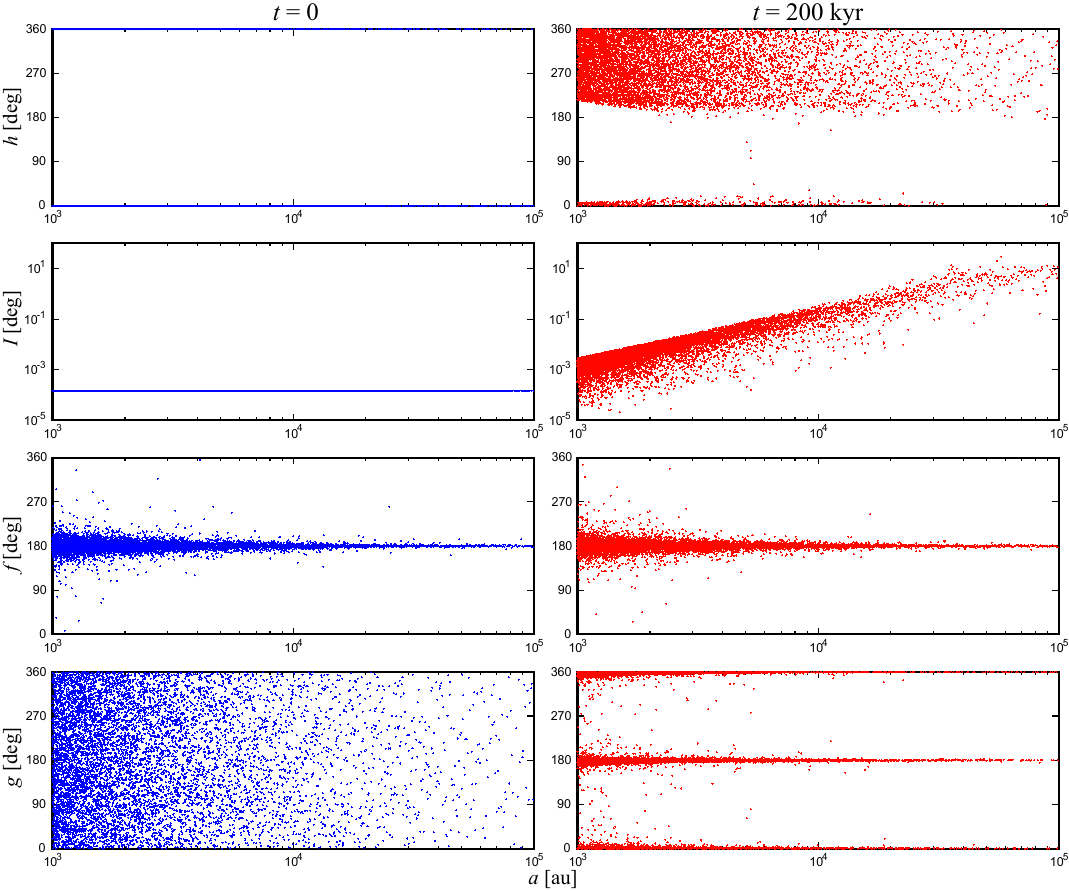}
\caption[]{%
(Left column)
Scatter plots of the four heliocentric variables against semimajor axis $a$ of 8000 objects placed in the comet cloud under the star set A at the initial state, $t=0$.
(Right column) The same plots of the same variables after a short interval, $t = 200\; \mathrm{kyr}$.
The horizontal scale is logarithmic.
The rows are
(top) longitude of ascending node $h$,
(second top) inclination $I$,
(third top) true anomaly $f$, and
(bottom) argument of perihelion $g$.
Note that longitude of ascending node $h$ seems to be clustered at $h=0$ as well as $h \sim 360^\circ$ at $t=0$ due to a technical reason on coding.
}
\label{fig:oc-g0-arikadisk}
\end{figure}

As an example for illustration, we extracted 8,000 objects from the modeled comet cloud under the star set A, and plotted in Figure \ref{fig:oc-g0-arikadisk} how the distributions of their heliocentric variables $h$, $I$, $f$, and $g$ change within a short time interval from the initial state.
In our model, comet's longitude of ascending node is formally assumed to be $h=0$ at $t=0$, and the orbital inclination $I$ is zero
(actually it is set to about $10^{-4}$ degrees for a technical reason on coding).
Comets' initial mean anomaly is randomly distributed between $l=0$ and $360^\circ$ in the cloud.
But when this is converted to true anomaly $f$, the distribution concentrates on $f \sim 180^\circ (= \pi)$ as shown in the left panel of the third row of Figure \ref{fig:oc-g0-arikadisk}.
As for the initial argument of perihelion, we uniformly distribute it using random numbers as in the left bottom panel.
After a short time interval $(t = 200\; \mathrm{kyr})$, comet's longitude of ascending node $h$ is getting clearly defined as the orbital inclination begins to be excited.
Most objects still remain near aphelion $(f \sim 180^\circ)$ at this point, but their argument of perihelion $g$ is concentrated at $180^\circ$ and 0 (the right bottom panel).
It is this situation that is reflected in the distribution of $g_0$ of the new comets in Figure \ref{fig:oc-genocnc-combined-t_s5A}.
Note that in the right top panel of Figure \ref{fig:oc-g0-arikadisk}, longitude of ascending node $h$ of comets is clustered around $h = 270^\circ$.
This reflects the fact that longitude of ascending node of the galactic plane relative to the ecliptic plane is about $270^\circ$ (see Section \ref{ssec:genocnc}).

\clearpage
\section{Evolution of comet cloud without stellar encounters\label{appen:nostarevol}}
\begin{figure}[!htbp]
  \includegraphics[width=\myfigwidth]{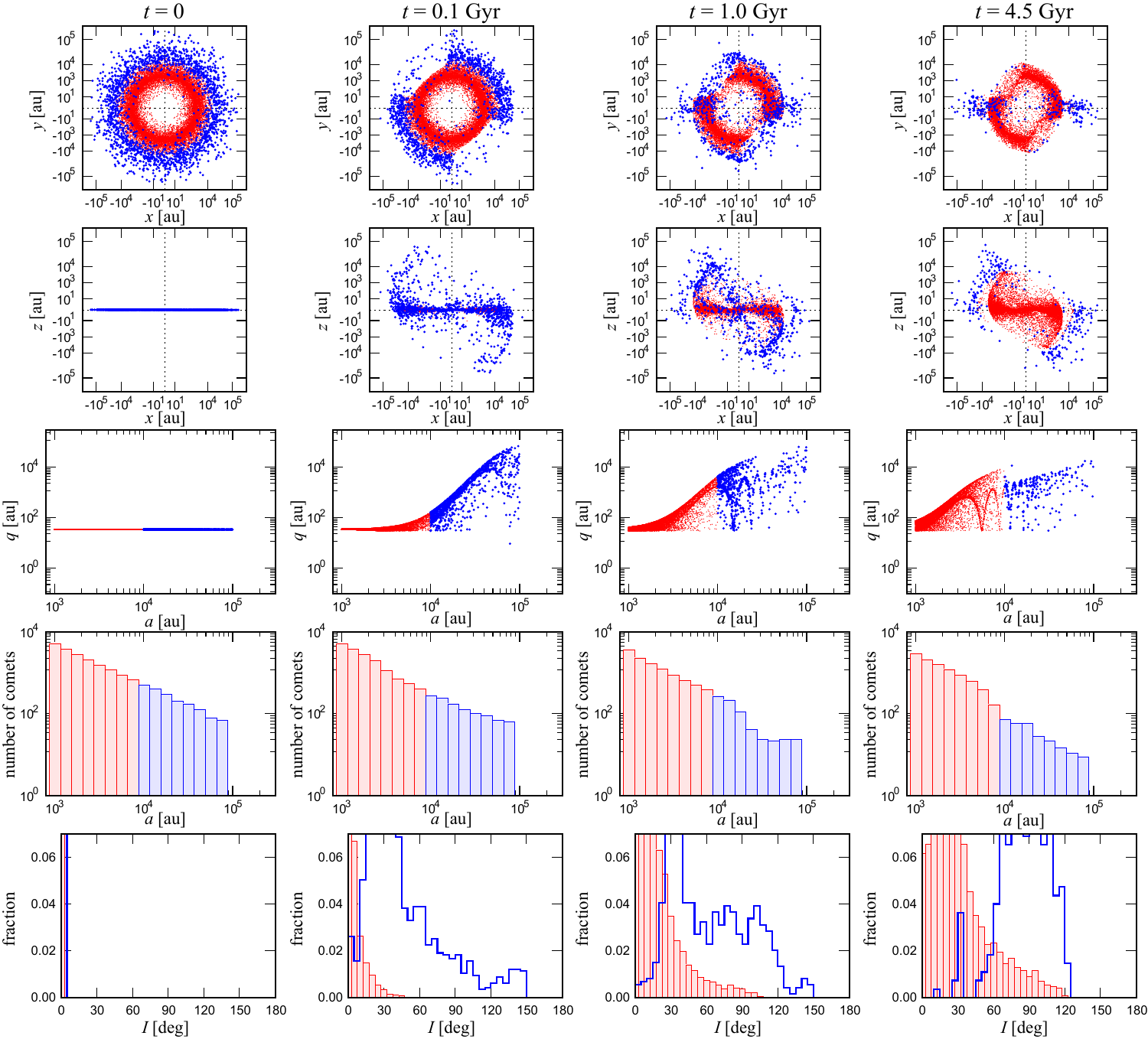}
\caption[]{%
Example snapshots of the comet cloud evolution considering only the galactic tidal force and not the stellar encounters.
This figure should be compared to Figure \ref{fig:oc-evol-ah_s5A} and Figure \ref{fig:oc-evol-ah_s3B} where stellar encounters are considered.
Consult Figure \ref{fig:oc-evol-ah_s5A}'s caption for details of the color, axis, and coordinates used in the panels.
}
\label{fig:oc-evol-nostar-ah_s5A}
\end{figure}

As we mentioned in Section \ref{ssec:genocnc}, comet cloud would not become substantially isotropic without stellar encounters from random directions.
One of the reasons for this is that the galactic tidal force is fundamentally a periodic oscillation with its period generally as long as $O(10^9)$ years.
It does not act to realizing a steady isotropic state of the cloud.
As an example of this fact, in Figure \ref{fig:oc-evol-nostar-ah_s5A} we show snapshots of the dynamical evolution of a two-dimensional planetesimal disk identical to Figure \ref{fig:oc-evol-ah_s5A} but with only the galactic tidal forces applied.
This is an example of the evolution of the comet cloud without stellar encounters, and it should be compared with our numerical simulation carried out with stellar encounter (Figure \ref{fig:oc-evol-ah_s5A}).

The initial conditions ($t=0$) we use are common to both Figures \ref{fig:oc-evol-nostar-ah_s5A} and \ref{fig:oc-evol-ah_s5A}.
We see no noticeable difference up to time $t=0.1$ Gyr.
However, after $t=1$ Gyr, we find a difference.
In particular, in Figure \ref{fig:oc-evol-nostar-ah_s5A} without stellar encounters, we find the outer part of the comet cloud (denoted by the blue points) is not isotropically distributed at $t=4.5$ Gyr.
Rather, in the projections to the $(x,y)$ and $(x,z)$ planes in the top two rows of the right column, the distribution of the outer comet cloud objects appears to stand perpendicular to the ecliptic plane.
The orbital inclination distribution in the bottom right panel shows that many of the objects in the outer part of the cloud have large inclination concentrated between $60^\circ$ and $120^\circ$, and they are not in the isotropic state.
In what follows let us try explaining what causes these differences.
\begin{itemize}
\item
As we mentioned in Section \ref{ssec:gtide} and will mention in \ref{appen:gtfunc}, the vertical component of the comet's angular momentum relative to the galactic plane is conserved in the dynamical model that we used.
Under this constraint, each comet tends to keep its perihelion distance near its maximum value for relatively long time \citep[][the third top panels of her Figure 2]{higuchi2020}.
During this state, the comet's orbital inclination with respect to the galactic plane takes a value close to $90^\circ$ \citep[][the bottom panels of her Figure 2]{higuchi2020}.
These are related to the \citeauthor{vonzeipel1910}--\citeauthor{lidov1961-en}--\citeauthor{kozai1962b} oscillation \citep[][her Figure 3]{higuchi2020}.
\item
Similarly to our Figures \ref{fig:oc-evol-ah_s5A} and \ref{fig:oc-evol-ah_s3B}, in Figure \ref{fig:oc-evol-nostar-ah_s5A} we define the $y$-axis as the intersection line of the galactic plane and the ecliptic plane.
Note that since the inclination value between the galactic plane and the ecliptic plane is about $60^\circ$, we can regard that the galactic plane would be approximately present near the line of $z = \tan 60^\circ x \approx 1.73 x$ in the second row panels which represent the $(x,z)$ plane.
\item
There is no particular concentration in the distribution of comet's ascending node with respect to the galactic plane.
However, many of the comets with such high orbital inclination have their argument of perihelion librating around $\pm 90^\circ$ due to the \citeauthor{vonzeipel1910}--\citeauthor{lidov1961-en}--\citeauthor{kozai1962b} oscillation \citep[][her Figure 3]{higuchi2020}.
There are also comets that are on highly inclined orbits with their argument of perihelion circulating instead of librating, but they also spend the longest time in the state where their orbital inclination to the galactic plane is close to $90^\circ$ \citep[][the black curves in the bottom panels of her Figure 2]{higuchi2020}.
\item
Also, such comets with large eccentricity tend to stay around its aphelion ($f \sim \pi$) for most of their orbital period.
\item
Above series of facts realizes a concentration of the comets along the direction of galactic north pole or the south pole, near the line of $z = \tan (60^\circ +90^\circ) x \approx -0.58 x$ in the second row panels of Figure \ref{fig:oc-evol-nostar-ah_s5A} which represent the $(x,z)$ plane.
This is particularly true for the outer part of the cloud where the galactic tidal force is strong.
\item
This concentration is projected on to the $(x,y)$ plane in the top row panels of Figure \ref{fig:oc-evol-nostar-ah_s5A} where we find many objects along the $x$-axis.
\end{itemize}

Thus we can conclude that the distribution of the comets in the outer part of the cloud (blue in Figure \ref{fig:oc-evol-nostar-ah_s5A}) is not isotropic.
This is one of the evidence that the evolution of comet clouds solely by the galactic tidal force does not realize the isotropic state.

Note that in the third and fourth panels from the top of Figure \ref{fig:oc-evol-nostar-ah_s5A}, we find the perihelion distance $q$ of the outer comet cloud objects at time $t=4.5$ Gyr is smaller than before ($t = 0.1$ or 1.0 Gyr).
This is because, in systems where the \citeauthor{vonzeipel1910}--\citeauthor{lidov1961-en}--\citeauthor{kozai1962b} oscillation is at work, the closer the perturbed object is to the perturbing source, the shorter the period of the oscillation becomes.
\citet[][her Eq. (20)]{higuchi2007} shows that the period $(P)$ of the oscillation is inversely proportional to the orbital period of the comet, i.e. $P \, \propto \, a^{-\frac{3}{2}}$ (the readers find a more rigorous treatment in \citet[][her Eq. (89)]{higuchi2020}).
Thus, the objects in the outer comet cloud with larger semimajor axis $a$ have shorter oscillation period of eccentricity than those in the inner part of the cloud, and hence they have shorter oscillation period of perihelion distance.
Therefore, we can interpret the decrease in the perihelion distance $q$ of the outer cloud objects seen in Figure \ref{fig:oc-evol-nostar-ah_s5A} at time $t=4.5$ Gyr as the propagation of the effect of the galactic tidal force starting from the outer toward the inner part of the cloud through time.
On the other hand, in Figures \ref{fig:oc-evol-ah_s5A} and \ref{fig:oc-evol-ah_s3B} which show the dynamical evolution of the comet cloud under both the influence of the galactic tide and the stellar encounters, perihelion distance $q$ of the outer cloud objects can also vary according to the stellar encounters.
Therefore, the change in perihelion distance $q$ will no longer have the systematic nature seen in Figure \ref{fig:oc-evol-nostar-ah_s5A}.

Incidentally, note that the above trend is typically seen in the classical inner \citeauthor{vonzeipel1910}--\citeauthor{lidov1961-en}--\citeauthor{kozai1962b} oscillation \citep[e.g.][]{antognini2015,naoz2016} as well.
Let $a$ be the semimajor axis of the inner perturbed object (e.g. an asteroid), and $a' \, (>a)$ be the semimajor axis of the outer perturbing object (e.g. Jupiter).
Then, the period of the classical inner \citeauthor{vonzeipel1910}--\citeauthor{lidov1961-en}--\citeauthor{kozai1962b} oscillation $(P_\mathrm{i})$ has an approximately relation of $P_\mathrm{i} \, \propto \, a'^3 a^{-\frac{2}{3}}$ \citep[e.g.][]{antognini2015,naoz2016}.
Although the function form of the dependence of the period on $a$ is slightly different, here we again find that the closer the perturbed body is to the perturbing body (i.e. the larger $a$ is), the shorter the period of the oscillation in these systems.

Let us note a little more about the contribution of the stellar encounters on the evolution of the Oort Cloud.
In the dynamical evolution process of the new comets slipping out of the Oort Cloud and approaching the inner solar system, it is known that an intermediate region called the tidally active zone (TAZ) plays a major role
\citep[e.g.][]{fouchard2011a,fouchard2011b}.
The definition of TAZ is rather abstract, but it is defined as the region in phase space where the galactic tidal force can generate observable comets on its own (without help from the stellar encounters).
The definition of the observable comets is those whose perihelion distance $q<5$ au.
The number density of the comets inside TAZ determines the incoming flux of the observable comets 
\citep[e.g.][]{fouchard2017}.
When there are some objects in TAZ, the galactic tidal force is capable of reducing their perihelion distance and bringing them into the inner solar system.
However, it is not the action of the galactic tidal force that brings the objects in the Oort Cloud into TAZ.
It is the stellar encounters that enhances the injection of cometary objects into TAZ.
In other words, although current comet injection from the Oort Cloud to the inner part of the solar system may heavily rely on the galactic tide, the stellar encounters also play a fundamental role in replenish the depleted TAZ 
\citep[e.g.][]{rickman2014}.

\clearpage
\section{Galactic tidal force function\label{appen:gtfunc}}
Here we briefly summarize how we implement the galactic tidal force function and apply it to the orbital motion of the new comets at each of the time-skip event described in Section \ref{ssec:timeskip}.
Our formalization basically follows that of \citet{higuchi2007} and \citet{higuchi2020} who deal with the galactic tidal force in an integrable approximation.

In the approximation method that we described in Section \ref{ssec:timeskip}, it is the vertical component of the galactic tidal force that matters for our model \citep[e.g.][his Section 9.4]{tremaine2023}.
In our galaxy, it is known that the variation of the galactic disk density within the galactic plane is much smaller (and slower) than the vertical variation.
In other words, we can presume the galactic disk is thin, and its density is just a function of the vertical distance $z'$.
Then we consider the environment around the Sun, placing the Sun at the coordinate origin.
Although the galactic disk is thin around the Sun (several 100 pc), it is still much thicker than the radius of the comet cloud ($< 1$ pc).
So, expanding the gravitational potential that the galactic disk creates in the Taylor series around the Sun (i.e. at $z'=0$) is justified.
As a result, we get an approximate formula for the galactic potential whose leading-order term is proportional to $z'^2$.
Naturally, we get an expression of the vertical component of the galactic tidal force $\mathbf{f}$ working on the unit mass of
an object like a comet in the rotating coordinates centered at the Sun through the epicyclic approximation as follows \citep[e.g.][]{binney1987}:
\begin{equation}
  \mathbf{f} = - \nu_0^2 \mathbf{z}' ,
  \label{eqn:galactictide-z}
\end{equation}
where
$\mathbf{z}'$ is the vertical position of the comet measured from the galactic plane that includes the Sun.
The quantity $\nu_0 = \sqrt{4 \pi \kappa^2 \rho}$ is the vertical frequency of the tidal force in the epicyclic approximation with the average star mass density in the solar neighborhood, $\rho \sim 0.1 M_\odot$ $\mathrm{pc}^{-3}$ \citep{holmberg2000}.
$\kappa^2$ denotes the gravitational constant.
The tidal force expressed as Eq. \eqref{eqn:galactictide-z} serves as a perturbation against the Keplerian motion of comets around the Sun.
Under the approximation, Hamiltonian $F$ of a comet becomes as follows:
\begin{equation}
  F = -\frac{\mu}{2 a} + \frac{\nu_0^2}{2} a^2 \left( 1-e^2 \right)^2
       \sin^2 I_{\cal G} \frac{\sin^2 (f + g_{\cal G})}{\left( 1 + e\cos f\right)^2},
  \label{eqn:hamiltonian-full}
\end{equation}
where $\mu = \kappa^2 M_\odot$.
The orbital elements $(a, e, I_{\cal G}, g_{\cal G}, f)$ are those defined in the galactocentric frame.
Since we are talking about secular orbital evolution of a comet, we canonically average the Hamiltonian $F$ in Eq. \eqref{eqn:hamiltonian-full} over the orbital period of the comet around the Sun.
Making a definite integral of the second term of Eq. \eqref{eqn:hamiltonian-full} by comet's mean anomaly $l$, we get the average Hamiltonian $F^\ast$ as
\begin{equation}
  F^\ast = -\frac{\mu}{2 a^\ast} +\frac{\nu_0^2}{4} {a^\ast}^2 \sin^2 I^\ast_{\cal G} \left( 1 - {e^\ast}^2 + 5{e^\ast}^2 \sin^2 g^\ast_{\cal G} \right) ,
  \label{eqn:galactichamiltonian-averaged}
\end{equation}
where the superscript $\ast$ means that the variable has been canonically averaged.
If we rewrite the new Hamiltonian $F^\ast$ in Eq. \eqref{eqn:galactichamiltonian-averaged} using the Delaunay elements,
it would have a function form of $F^\ast (L^\ast, G^\ast, g^\ast_{\cal G})$ where
$L^\ast = \sqrt{\mu a^\ast}$ and
$G^\ast = L^\ast \sqrt{1 - {e^\ast}^2}$
(note that the canonical momenta $L^\ast$ and $G^\ast$ are common in the heliocentric frame and the galactocentric frame by their definitions).
Since $F^\ast$ does not include mean anomaly, its conjugate momentum $L^\ast$ is constant, and $F^\ast$ has a form of $F^\ast (G^\ast, g^\ast_{\cal G})$.
Also, since the Hamiltonian $F^\ast$ itself is constant because it is originated by a conservative, central force field.
Therefore the system that $F^\ast$ describes is integrable.

We should also pay attention to the fact that the Hamiltonian $F$ or $F^\ast$ does not depend on longitude of ascending node either.
This means its conjugate momentum $H^\ast_{\cal G} = L^\ast \sqrt{1 - {e^\ast}^2} \cos I^\ast_{\cal G}$ is conserved, which yields the conservation of the vertical component of comet's angular momentum per unit mass, $\sqrt{1 - {e^\ast}^2} \cos I^\ast_{\cal G}$.
We denote this fact as follows:
\begin{equation}
  j = \sqrt{1-{e^\ast}^2} \cos I^\ast_{\cal G} = \mathrm{constant}.
  \label{eqn:constant-j}
\end{equation}
We can say that the relationship \eqref{eqn:constant-j} comes from the axial symmetry of the galactic disturbing potential expressed as the second term of the Hamiltonian $F$ in Eq. \eqref{eqn:hamiltonian-full}.

Now we know both the averaged Hamiltonian $F^\ast$ and the averaged semimajor axis $a^\ast$ are constant (through the fact that $L^\ast = \sqrt{\mu a^\ast}$ is a constant).
The unperturbed part of $F^\ast$ in Eq. \eqref{eqn:galactichamiltonian-averaged}, $-\frac{\mu}{2 a^\ast}$ which dominates the Keplerian motion of the comet around the Sun, is constant because $a^\ast$ is a constant.
This means that the perturbed part of $F^\ast$ in Eq. \eqref{eqn:galactichamiltonian-averaged} is also constant.
Here let us define a constant of motion $C$ derived from the perturbed part of $F^\ast$ as follows:
\begin{equation}
   C = \frac{1}{2}\left( 1 - \frac{j^2}{1-{e^\ast}^2} \right)
                  \left( 1 - {e^\ast}^2 + 5{e^\ast}^2 \sin^2 g^\ast_{\cal G} \right) ,
  \label{eqn:constant-C-10}
\end{equation}
in other words,
\begin{equation}
  F^\ast = -\frac{\mu}{2 a^\ast} +\frac{\nu_0^2}{4} {a^\ast}^2 C .
  \label{eqn:Fast-expressed-by-C}
\end{equation}
Let us define a variable $\chi$ as
\begin{equation}
  \chi = 1 - {e^\ast}^2 ,
  \label{eqn:def-chi}
\end{equation}
and $C$ in Eq. \eqref{eqn:constant-C-10} can be expressed as follows:
\begin{equation}
  C = \left( 1 - \frac{j^2}{\chi} \right)
      \left( \chi + 5 \left( 1 - \chi \right) \sin^2 g^\ast_{\cal G} \right) ,
\label{eqn:constant-C-A1-newarika}
\end{equation}
and an inversion of Eq. \eqref{eqn:constant-C-A1-newarika} yields
\begin{equation}
  \sin^2 g^\ast_{\cal G} = \frac{\chi \left( C + j^2 - \chi \right)}
                       {5 \left( 1-\chi \right) \left( \chi -j^2 \right)} .
\label{eqn:constant-sin2g-newarika}
\end{equation}

In what follows we try to express the time-dependent analytic solution of $e^\ast$, $I^\ast_{\cal G}$, $g^\ast_{\cal G}$, and $h^\ast_{\cal G}$ (longitude of ascending node of the comet described in the galactocentric frame) at this approximation level.
We begin with eccentricity $e^\ast$, but here we pay a stronger attention to the variable $\chi$ that has a closer affinity to canonical variables.
The canonical equations of motion of the averaged system that the Hamiltonian $F^\ast (G^\ast, g^\ast_{\cal G})$ with one degree of freedom governs are as follows:
\begin{equation}
  \frac{dG^\ast         }{dt^\ast} = -\frac{\partial F^\ast}{\partial g^\ast_{\cal G}}, \quad
  \frac{dg^\ast_{\cal G}}{dt^\ast} =  \frac{\partial F^\ast}{\partial G^\ast}.
\label{eqn:eom-general-FGg}
\end{equation}
where 
$t^\ast$ is the parametric (or ``stretched'') time in the averaged (i.e. canonically transformed) system \citep[e.g.][]{hori1966,yuasa1973,ito2012}.

Since $L^\ast$ is a constant, the time derivative of $G^\ast$ becomes as follows:
\begin{equation}
  \frac{d G^\ast}{d t^\ast} = \frac{L^\ast}{2 \sqrt{\chi}} \frac{d \chi}{d t\ast} .
\label{eqn:dGdt}
\end{equation}
Using Eq. \eqref{eqn:dGdt}, the first equation in Eqs. \eqref{eqn:eom-general-FGg} about $G^\ast$ can be rewritten as follows:
\begin{equation}
  \frac{d \chi}{d t^\ast} = -\frac{2 \sqrt{\chi}}{L^\ast}
                             \frac{\partial F^\ast}{\partial g^\ast_{\cal G}} .
\label{eqn:dchidt-1}
\end{equation}
If we write the Hamiltonian $F^\ast$ in Eq. \eqref{eqn:Fast-expressed-by-C} as
\begin{equation}
  F^\ast = F_0^\ast + F_1^\ast,
\label{eqn:def-F=F0F1}
\end{equation}
with the unperturbed part $F_0^\ast$ and the perturbed part $F_1^\ast$ as
\begin{equation}
  F_0^\ast  = -\frac{\mu}{2 a^\ast} = -\frac{\mu^2}{2{L^\ast}^2}, \quad
  F_1^\ast  =  \frac{\nu_0^2}{4} {a^\ast}^2 C ,
\label{eqn:def-F0F1}
\end{equation}
then we find that the unperturbed part $F_0^\ast$ does not explicitly depend on $g^\ast_{\cal G}$, therefore
$\frac{\partial F_0^\ast}{\partial g^\ast_{\cal G}} = 0$.
And we have
\begin{equation}
  \frac{\partial F_1^\ast}{\partial g^\ast_{\cal G}}
= \frac{5\nu_0}{2} \frac{L^\ast}{n^\ast} \frac{\chi-j^2}{\chi} \sin g^\ast_{\cal G} \cos g^\ast_{\cal G} ,
\label{eqn:dpF1dpg}
\end{equation}
where $n^\ast$ is the mean motion of the perturbed body in this system.

Applying Eq. \eqref{eqn:dpF1dpg} to Eq. \eqref{eqn:dchidt-1}, we get
\begin{equation}
\begin{aligned}
  \frac{d \chi}{d t^\ast}
  &= -\frac{2 \sqrt{\chi}}{L^\ast}
      \frac{\partial F_1^\ast}{\partial g^\ast_{\cal G}} \\
  &= 
-\frac{2 \sqrt{\chi}}{L^\ast}
 \frac{5\nu_0}{2} \frac{L^\ast}{n^\ast} \frac{\chi-j^2}{\chi} \sin g^\ast_{\cal G} \cos g^\ast_{\cal G} .
\end{aligned}
\label{eqn:dchidt-2}
\end{equation}

Now let us try to express $\sin g^\ast_{\cal G} \cos g^\ast_{\cal G}$ in the right-hand side of Eq. \eqref{eqn:dchidt-2} using $\chi$.
This quantity can be positive and negative depending on the value of $g^\ast_{\cal G}$, but its square is uniquely expressed as follows.
Using Eq. \eqref{eqn:constant-sin2g-newarika}, we have
\begin{equation}
   \sin^2 g^\ast_{\cal G} \cos^2 g^\ast_{\cal G}
  = \frac{4 \chi \left( \chi_0^\ast -\chi \right)
                 \left( \chi_1^\ast -\chi \right)
                 \left( \chi_2^\ast -\chi \right)}
         {\left( 5 \left( 1-\chi \right) \left( \chi-j^2 \right) \right)^2 } ,
\label{eqn:sin2gcos2g-1}
\end{equation}
where 
\begin{equation}
  \chi_0^\ast = C + j^2 ,
  \label{eqn:def-chi0}
\end{equation}
and $\chi_1^\ast$, $\chi_2^\ast$ are the two solutions of the quadratic equation of $\chi$
\begin{equation}
  5 \left( 1-\chi \right) \left( \chi-j^2 \right)
- \chi \left( C + j^2 - \chi \right) = 0 ,
  \label{eqn:equation-chi1chi2}
\end{equation}
where we assume $\chi_1^\ast < \chi_2^\ast$.
Since $\sin g^\ast_{\cal G} \cos g^\ast_{\cal G} = \frac{\sin 2g^\ast_{\cal G}}{2}$,
the square root of $\sin^2 g^\ast_{\cal G} \cos^2 g^\ast_{\cal G}$
becomes positive when $\sin 2g^\ast_{\cal G} > 0$, and it
becomes negative when $\sin 2g^\ast_{\cal G} < 0$.
More specifically writing, from Eq. \eqref{eqn:sin2gcos2g-1} we have
\begin{equation}
  \sin g^\ast_{\cal G} \cos g^\ast_{\cal G}=
\left\{
\begin{aligned}
& +\sqrt{
    \frac{4 \chi \left( \chi_0^\ast -\chi \right)
                 \left( \chi_1^\ast -\chi \right)
                 \left( \chi_2^\ast -\chi \right)}
         {\left( 5 \left( 1-\chi \right) \left( \chi-j^2 \right) \right)^2 }
                                  }
& \; \left(\sin 2g^\ast_{\cal G} > 0 \right), \\
& -\sqrt{
    \frac{4 \chi \left( \chi_0^\ast -\chi \right)
                 \left( \chi_1^\ast -\chi \right)
                 \left( \chi_2^\ast -\chi \right)}
         {\left( 5 \left( 1-\chi \right) \left( \chi-j^2 \right) \right)^2 }
                                  }
& \; \left(\sin 2g^\ast_{\cal G} < 0 \right),
\end{aligned}
\right.
\label{eqn:sin2gcos2g-2-pm}
\end{equation}

Applying Eq. \eqref{eqn:sin2gcos2g-2-pm} to the expression of $\frac{d\chi}{dt^\ast}$ of Eq. \eqref{eqn:dchidt-2}, we have:
\begin{equation}
  \frac{d \chi}{d t^\ast}
  = -\frac{2 \sqrt{\chi}}{L^\ast}
      \frac{\partial F_1^\ast}{\partial g^\ast_{\cal G}}
  = \left\{
\begin{aligned}
& -\frac{2 \nu_0^2}{n^\ast}
       \sqrt{ \left( \chi - \chi_0^\ast \right)
              \left( \chi - \chi_1^\ast \right)
              \left( \chi - \chi_2^\ast \right) }
& \; \left(\sin 2g^\ast_{\cal G} > 0 \right), \\
& +\frac{2 \nu_0^2}{n^\ast}
       \sqrt{ \left( \chi - \chi_0^\ast \right)
              \left( \chi - \chi_1^\ast \right)
              \left( \chi - \chi_2^\ast \right) }
& \; \left(\sin 2g^\ast_{\cal G} < 0 \right),
\end{aligned}
\right.
\label{eqn:dchidt-3}
\end{equation}

Among the three solutions $\chi^\ast_0$, $\chi^\ast_1$, $\chi^\ast_2$ of $\frac{d\chi}{dt^\ast} = 0$ in Eq. \eqref{eqn:dchidt-3}, let us denote
the        smallest one as $\alpha_0$,
the second smallest one as $\alpha_1$, and
the        largest  one as $\alpha_2$.
\citet[][their Appendix, p. 1705]{higuchi2007} gave the time-dependent solution of an ordinary differential equation of $\chi$ (their Eq. (A11), which is equivalent to our Eq. \eqref{eqn:dchidt-3}) by employing Jacobi elliptic function $\mathrm{cn}$ as follows:
\begin{equation}
  \chi = \alpha_1 + \left( \alpha_0 - \alpha_1 \right) \mathrm{cn}^2 \left( \theta, k\right) ,
  \label{eqn:sol-A11}
\end{equation}
or equivalently,
\begin{equation}
  \chi = \alpha_0 - \left( \alpha_0 - \alpha_1 \right) \mathrm{sn}^2 \left( \theta, k\right) .
  \label{eqn:sol-A11-sn}
\end{equation}
$\theta$ is defined as
\begin{equation}
  \theta = \frac{2K(k)}{\pi} \left( \widehat{g}_{\cal G} + \frac{\pi}{2} \right),
  \label{eqn:arika-theta}
\end{equation}
where $K(k)$ is the complete elliptic integral of the first kind with the modulus $k$ \citep[e.g.][]{byrd1971}.
$k^2$ is defined as
\begin{equation}
  k^2 = \frac{\alpha_1 - \alpha_0}{\alpha_2 - \alpha_0} .
  \label{eqn:arika-k2}
\end{equation}
$\widehat{g}_{\cal G}$ in Eq. \eqref{eqn:arika-theta} is defined as
\begin{equation}
  \widehat{g}_{\cal G} = n_{\widehat{g}_{\cal G}} t^\ast + \widehat{g}_{{\cal G},t^\ast=0},
  \label{eqn:arika-omega_star}
\end{equation}
with a constant of integration $\widehat{g}_{{\cal G},t^\ast=0}$ expressed as follows:
\begin{equation}
  \widehat{g}_{{\cal G},t^\ast=0} = \frac{\pi}{2} \left(\frac{    F(\varphi_{t^\ast=0},k)}{K(k)} - 1\right),
  \label{eqn:arika-omega_i}
\end{equation}
where $F(\varphi_{t^\ast=0},k)$ is the incomplete elliptic integral of the first kind with the modulus $k$ and amplitude $\varphi_{t^\ast=0}$ defined as
\begin{equation}
  \sin \varphi_{t^\ast=0} = \sqrt{\frac{\chi_{t^\ast=0} - \alpha_0}{\alpha_1 - \alpha_0}} .
  \label{eqn:arika-sinphi-t=0}
\end{equation}
Here $\chi_{t^\ast=0}$ denotes the initial value of $\chi$ at $t^\ast = 0$.
At general time $t$, the amplitude $\varphi$ is defined as follows:
\begin{equation}
  \sin \varphi = \sqrt{\frac{\chi - \alpha_0}{\alpha_1 - \alpha_0}} .
  \label{eqn:arika-sinphi-general}
\end{equation}

Note that we presume $\sin \varphi \geq 0$ and $\mathrm{sn}\; \theta \geq 0$ from Eq. \eqref{eqn:arika-theta} through Eq. \eqref{eqn:arika-sinphi-general}.
This assumption is justified from the symmetry of the considered system with respect to the variable $2g$ that the perturbed part of the Hamiltonian $F_1^\ast$ in Eq. \eqref{eqn:def-F0F1} governs.
Extension to other quadrants (where $\sin \varphi < 0$ or $\mathrm{sn}\; \theta < 0$ is straightforward, which would change the sign of $F(\varphi_{t^\ast=0},k)$ in Eq. \eqref{eqn:arika-omega_i}.
See \citet{higuchi2020} for more rigorous discussions.
Also, see \citet{matese1989} for an earlier study of the time-dependent solution expressed as Eq. \eqref{eqn:sol-A11} or Eq. \eqref{eqn:sol-A11-sn}.

We can regard $n_{\widehat{g}_{\cal G}}$ in Eq. \eqref{eqn:arika-omega_star} as the ``mean motion'' of $\widehat{g}_{\cal G}$ in \eqref{eqn:arika-theta} which has the form of
\begin{equation}
  n_{\widehat{g}_{\cal G}} = \frac{2 \pi \nu_0^2}{4 n^\ast K(k)} \sqrt{\alpha_2 - \alpha_0} .
  \label{eqn:arika-n-omega-star}
\end{equation}

From Eqs. \eqref{eqn:sol-A11}\eqref{eqn:arika-theta}\eqref{eqn:arika-omega_star}, we find that the variable $n_{\widehat{g}_{\cal G}}$ in Eq. \eqref{eqn:arika-n-omega-star} dominates the oscillations of $\chi$, $e^\ast$, and $I_{\cal G}^\ast$.
Eq. \eqref{eqn:arika-n-omega-star} also tells us that $n_{\widehat{g}_{\cal G}}$ has the dimension of frequency (i.e. time${}^{-1}$), as both $\nu_0$ and $n^\ast$ have the dimension of time${}^{-1}$.
Also, we find that $n_{\widehat{g}_{\cal G}}$ is inversely proportional to $n^\ast$ (the mean motion of the perturbed body) in Eq. \eqref{eqn:arika-n-omega-star}.
$n_{\widehat{g}_{\cal G}}$ actually serves as the dominant frequency of the variable $\chi$ as well as comet's argument of perihelion, $g^\ast_{\cal G}$.
This is explained as follows.
The time-dependent solution of the ordinary differential equation of $\chi (t^\ast)$ (Eq. \eqref{eqn:dchidt-3}) is expressed as Eq. \eqref{eqn:sol-A11} or Eq. \eqref{eqn:sol-A11-sn} in a closed form through Eqs. \eqref{eqn:arika-theta}--\eqref{eqn:arika-n-omega-star}.
We can convert the time-dependent solution $\chi (t^\ast)$ into the that of eccentricity $e^\ast (t^\ast)$ through the definition of $\chi$ in Eq. \eqref{eqn:def-chi}.
Then, from the conservation of the quantity $j$ defined in Eq. \eqref{eqn:constant-j}, we can obtain the time-dependent solution of orbital inclination $I^\ast_{\cal G}$ as $I^\ast_{\cal G} (t^\ast)$.
Finally, we can relate $\chi (t^\ast)$ to the time-dependent solution of argument of pericenter $g^\ast_{\cal G}$ through Eq. \eqref{eqn:constant-C-10} as
\begin{equation}
  g^\ast_{\cal G} = \frac{1}{2}\cos^{-1} \frac{Q_1(\chi)}{5\left(\chi-j^2\right)\left(1-\chi\right)} ,
  \label{eqn:cos2w-inv}
\end{equation}
where
\begin{equation}
\begin{aligned}
  Q_1(\chi) &= \left( \chi_{t^\ast=0} - \chi\right) \left( 3\chi - \frac{5j^2}{\chi_{t^\ast=0}} \right) \\
            &\quad
             + 5\left( 1 - \frac{j^2}{\chi_{t^\ast=0}} \right) \left( 1 - \chi_{t^\ast=0} \right)
                \chi \cos 2 \widehat{g}_{{\cal G},t^\ast=0} .
\end{aligned}
  \label{eqn:Q1-chi}
\end{equation}
In other words,
the periodicity of both $\chi$ and $g^\ast_{\cal G}$ are under the control of the mean motion $n_{\widehat{g}_{\cal G}}$ throughout the series of equations of Eqs.
\eqref{eqn:sol-A11}, 
\eqref{eqn:arika-theta}, 
\eqref{eqn:arika-omega_star},
\eqref{eqn:def-chi},
\eqref{eqn:constant-j}, and
\eqref{eqn:constant-C-10}.
Note that the circumstance ($n_{\widehat{g}_{\cal G}} \, \propto \, {n^\ast}^{-1}$) depicted in Eq. \eqref{eqn:arika-n-omega-star} is similar to what occurs in the classical inner \citeauthor{vonzeipel1910}--\citeauthor{lidov1961-en}--\citeauthor{kozai1962b} oscillation whose analytic solution at the quadrupole-level approximation shows that the oscillation frequency of perturbed body's argument of pericenter is inversely proportional to the mean motion of the perturbed body \citep[e.g.][]{kinoshita2007a,antognini2015}.

Among the remaining elements, $H^\ast_{\cal G} (= G^\ast \cos I^\ast_{\cal G})$ is obviously constant because the Hamiltonian $F^\ast$ in Eq. \eqref{eqn:hamiltonian-full} does not contain longitude of ascending node $h^\ast_{\cal G}$.
And, since $h^\ast_{\cal G}$ itself is governed by the equation of motion
\begin{equation}
  \frac{d h^\ast_{\cal G}}{d t^\ast} = \frac{\partial F_1^\ast}{\partial H^\ast_{\cal G}},
  \label{eqn:dhdt}
\end{equation}
the formal, time-dependent solution $h^\ast_{\cal G} (t^\ast)$ is principally obtained through quadrature such as
\begin{equation}
  h^\ast_{\cal G} (t^\ast) = \int \frac{\partial F_1^\ast}{\partial H^\ast_{\cal G}} d t^\ast .
  \label{eqn:dhdt-quadrature}
\end{equation}
See \citet{moiseev1945b-en} or \citet{lidov1961-en} for similar manifestations in the doubly-averaged circular restricted three-body problem.
From the definition of $F_1^\ast$ in Eq. \eqref{eqn:def-F0F1} and that of $C$ in Eq. \eqref{eqn:constant-C-10}, we see that $F_1^\ast$'s dependence on $H^\ast_{\cal G}$ is confined in the factor $1 - \frac{j^2}{1-{e^\ast}^2}$ in Eq. \eqref{eqn:constant-C-10} which is equivalent to $\sin^2 I_{\cal G}$ or $1 - \frac{{H^{\ast 2}_{\cal G}}}{{L^{\ast 2}}}$.
Therefore we have
\begin{equation}
  \begin{aligned}
  \frac{\partial F_1^\ast}{\partial H^\ast_{\cal G}} &=
    \frac{\nu_0 {a^\ast}^2}{4}
    \left( 1 - {e^\ast}^2 + 5 {e^\ast}^2 \sin^2 {g^\ast_{\cal G}} \right)
    \frac{\partial}{\partial H^\ast_{\cal G}} \left( 1 - \frac{{H^{\ast 2}_{\cal G}}}{{L^{\ast 2}}} \right) \\
&= -\frac{\nu_0^2 j}{2 n^\ast}
    \left( 1 - {e^\ast}^2 + 5 {e^\ast}^2 \sin^2 {g^\ast_{\cal G}} \right) .
  \end{aligned}
\label{eqn:dF1dH-with-LGH}
\end{equation}

Substituting Eq. \eqref{eqn:constant-sin2g-newarika} and the definition of $\chi$ in Eq. \eqref{eqn:def-chi} into the right-hand side of Eq. \eqref{eqn:dF1dH-with-LGH}, we obtain from Eq. \eqref{eqn:dhdt}
\begin{equation}
    \frac{d h^\ast_{\cal G}}{dt^\ast}
 =  \frac{\partial F_1^\ast}{\partial H^\ast_{\cal G}}
 = -\frac{\nu_0^2 j}{2 n^\ast} \frac{\chi^\ast_0 - j^2}{\chi - j ^2} ,
\label{eqn:dF1dH-with-chi0}
\end{equation}
where a constant quantity $\chi^\ast_0$ is defined in Eq. \eqref{eqn:def-chi0}.

We can evaluate the quadrature of Eq. \eqref{eqn:dhdt-quadrature} using Eq. \eqref{eqn:dF1dH-with-chi0} and the time-dependent solution for $\chi (t^\ast)$ expressed as Eq. \eqref{eqn:sol-A11}.
The evaluation process is detailed in \citet[][p. 1706]{higuchi2007}, and their expression of the resulting solution is as follows:
\begin{equation}
  h^\ast_{\cal G} (t^\ast)
= h^\ast_{{\cal G},t^\ast=0} - \frac{j}{2\sqrt{\alpha_2 - \alpha_0}}
                               \frac{\chi_0^\ast - j^2}{\alpha_0 - j^2}
      \Pi \left(
            \varphi, \frac{\alpha_0 - \alpha_1}{\alpha_0 - j^2}, k 
          \right),
  \label{eqn:sol-W}
\end{equation}
where $\varphi$ is defined in Eq. \eqref{eqn:arika-sinphi-general},
$\Pi$ is the incomplete elliptic integral of the third kind, and $h^\ast_{{\cal G},t^\ast=0}$ is the initial value of $h^\ast_{\cal G}$ when $t^\ast = 0$.
Consult also \citet[][her Section 2.4.2]{higuchi2020} for a rigorous way to carry out the quadrature of Eq. \eqref{eqn:dhdt-quadrature} and the mathematical properties of the solution $h^\ast_{\cal G} (t^\ast)$.

Having all the necessary time-dependent solutions of
$e^\ast(t^\ast)$,
$I^\ast_{\cal G} (t^\ast)$,
$g^\ast_{\cal G} (t^\ast)$, and
$h^\ast_{\cal G} (t^\ast)$ in our hand,
we validate the time-skip scheme as follows.
Assume the heliocentric orbital elements of a comet at the distance of $r = 800$ au at time $t = t_\mathrm{b}$ are obtained as
$a (t_\mathrm{b})$,
$e (t_\mathrm{b})$,
$I (t_\mathrm{b})$,
$g (t_\mathrm{b})$,
$h (t_\mathrm{b})$, and
$l (t_\mathrm{b})$.
The subscript b means ``border'' or ``boundary'' beyond which the planetary perturbation becomes no longer effective in our model.
Note that in what follows we ignore the difference between the canonically averaged variables such as $e^\ast$ and the osculating variables such as $e$, so we will not employ the superscript $\ast$.
Unless $e (t_\mathrm{b}) \geq 1$ or $Q (t_\mathrm{b}) = a (t_\mathrm{b}) (1 + e (t_\mathrm{b})) > Q_\mathrm{max} = 2 \times 10^5$ au at this distance (the condition under which the comet is removed from the numerical model), we calculate the length of time during which we assume this comet experiences no perturbation as follows:
\begin{equation}
  \Delta t_\mathrm{skip} = \frac{2(\pi - l (t_\mathrm{b}))}{n} ,
  \label{eqn:t-skipped}
\end{equation}
where $n$ is mean motion of the comet at time $t_\mathrm{b}$.
We also convert the orbital elements described in the heliocentric frame into those in the galactocentric frame as
$e (t_\mathrm{b}) \to e_{\cal G} (t_\mathrm{b})$,
$I (t_\mathrm{b}) \to I_{\cal G} (t_\mathrm{b})$,
$g (t_\mathrm{b}) \to g_{\cal G} (t_\mathrm{b})$, and
$h (t_\mathrm{b}) \to h_{\cal G} (t_\mathrm{b})$.
Note that since eccentricity of comets is unchanged through this coordinate conversion (i.e. $e_{\cal G} = e$), we use the symbol $e_{\cal G}$ just for formality.
Then we apply the galactic tidal force function to the converted galactocentric orbital elements over the time duration of $\Delta t_\mathrm{skip}$.
In other words, we calculate the set of orbital elements at time $t_\mathrm{b} + \Delta t_\mathrm{skip}$ as
$e_{\cal G} (t_\mathrm{b} + \Delta t_\mathrm{skip})$,
$I_{\cal G} (t_\mathrm{b} + \Delta t_\mathrm{skip})$,
$g_{\cal G} (t_\mathrm{b} + \Delta t_\mathrm{skip})$,
$h_{\cal G} (t_\mathrm{b} + \Delta t_\mathrm{skip})$
using the time-dependent solutions that we have discussed in this section.
Since the galactic tidal force does not change semimajor axis in our approximation, and since the galactic tidal force function does not deal with mean anomaly, we do not consider $a$ and $l$ here.
Finally, we convert the galactocentric orbital elements into the heliocentric ones again such as
$e_{\cal G} (t_\mathrm{b} + \Delta t_\mathrm{skip}) \to e (t_\mathrm{b} + \Delta t_\mathrm{skip})$,
$I_{\cal G} (t_\mathrm{b} + \Delta t_\mathrm{skip}) \to I (t_\mathrm{b} + \Delta t_\mathrm{skip})$,
$g_{\cal G} (t_\mathrm{b} + \Delta t_\mathrm{skip}) \to g (t_\mathrm{b} + \Delta t_\mathrm{skip})$, and
$h_{\cal G} (t_\mathrm{b} + \Delta t_\mathrm{skip}) \to h (t_\mathrm{b} + \Delta t_\mathrm{skip})$.

When we employ the time-skip scheme, we presume there is no change in orbital elements between the time $t = t_\mathrm{b}$ and $t = t_\mathrm{b} + \Delta t_\mathrm{skip}$,
i.e. we assume $x (t_\mathrm{b} + \Delta t_\mathrm{skip}) = x (t_\mathrm{b})$ where $x$ is either of the heliocentric orbital elements $e$, $I$, $g$, and $h$.
However, in general $x (t_\mathrm{b} + \Delta t_\mathrm{skip})$ is different from $x (t_\mathrm{b})$ when we apply the galactic tidal force function.
The series of panels shown in Figures
\ref{fig:oc-skipped-time-lin_s5A},
\ref{fig:oc-skipped-dist-log_s5A},
\ref{fig:oc-skipped-diff_s5A}
display the comparison between $x (t_\mathrm{b} + \Delta t_\mathrm{skip})$ and $x (t_\mathrm{b})$ which validates (or invalidates) the time-skip scheme.

\clearpage
\section{Other statistics regarding Section \ref{sec:result}\label{suppl:otherstatsinS4}}
In this section, we present some statistical results that reinforce the statements that we made in Section \ref{sec:result}.

\begin{figure}[!htbp]
  \includegraphics[width=\myfigwidth]{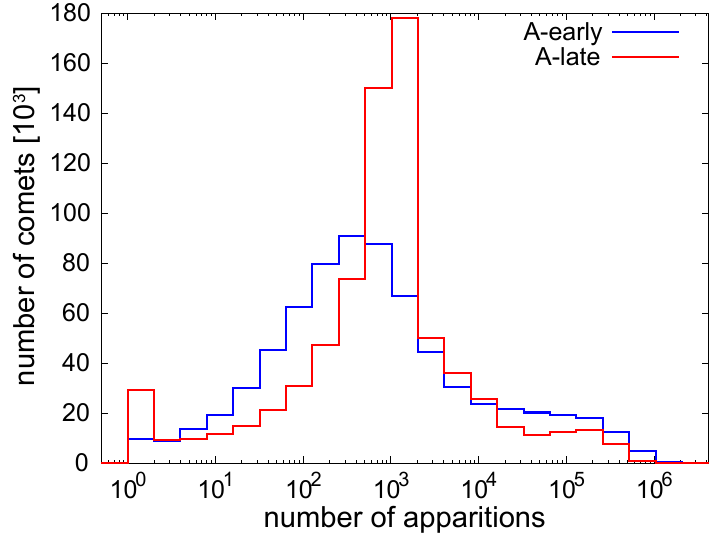}
  \caption[]{%
Histogram that shows how many apparitions each comet made during its $T_\mathrm{res}$ in A-early and A-late.
The vertical axis (the number of comets) is normalized by $10^3$.
}
\label{fig:oc-nappari_s5A}
\end{figure}

\paragraph{(1) Frequency of apparitions that each comet made}
First, as a supplement to the discussion of the comet's $T_\mathrm{res}$ in the second half of subsection \ref{ssec:lifetime}, we present in Figure \ref{fig:oc-nappari_s5A} the histogram of how many apparitions each comet made during its $T_\mathrm{res}$.
In particular, in this figure we find a concentration of comets that have just one apparition (labeled $10^0$ on the horizontal axis) during their $T_\mathrm{res}$ in A-late.
This makes the linearly-shaped concentration on the $(a_0, T_\mathrm{res})$ plane for A-late (the top right panel of Figure \ref{fig:oc-aeIqQ-Te-log_s5A}) that is slightly denser than that for A-early (the top left panel of Figure \ref{fig:oc-aeIqQ-Te-log_s5A}).
This difference comes from the difference in the averaged initial values of comets' eccentricity between A-early and A-late that we mentioned before.

\begin{figure}[!htbp]\centering
  \includegraphics[width=\myfigwidth]{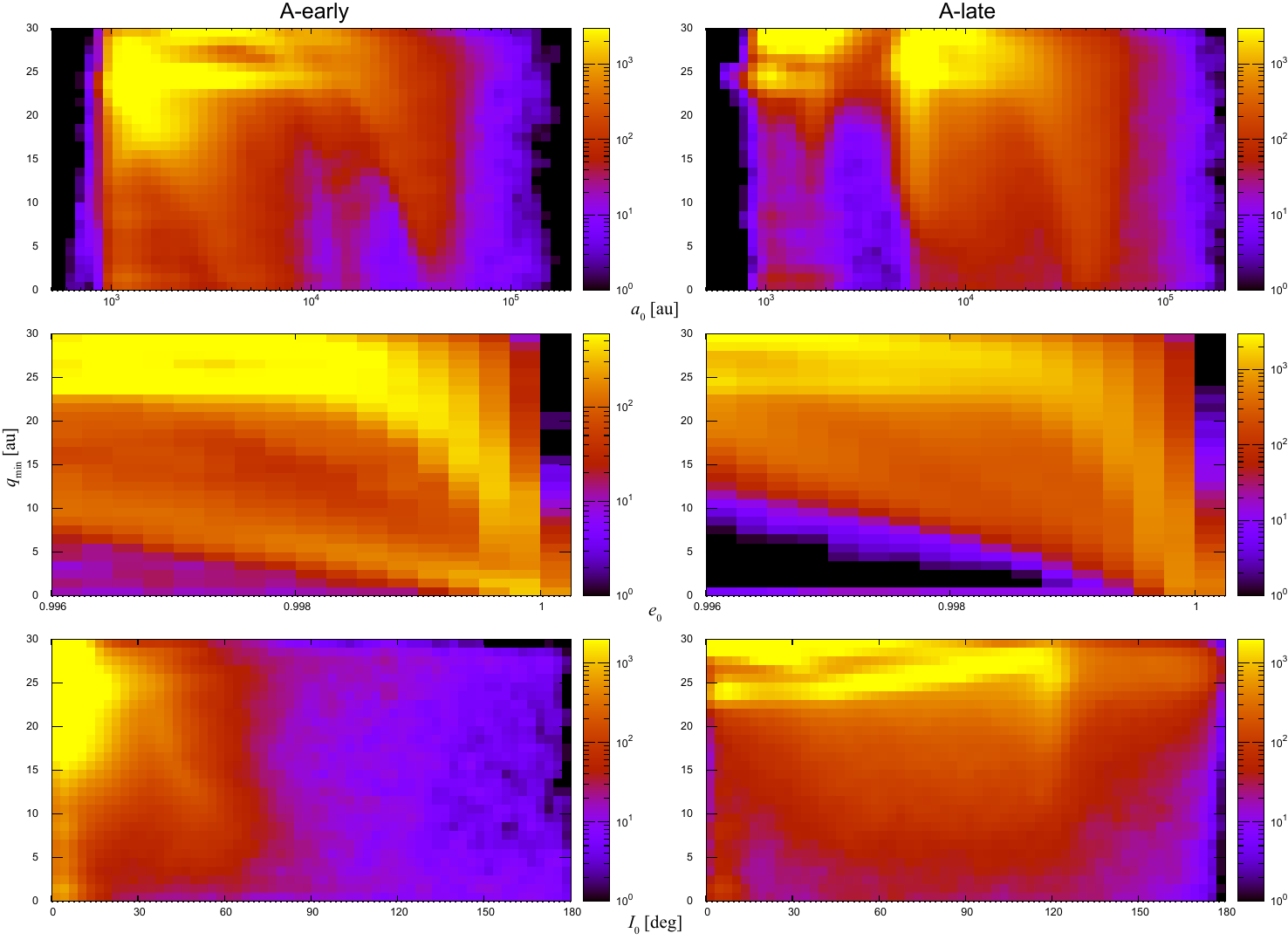}
  \caption[]{%
Dependence of the frequency distribution of $q_\mathrm{min}$ on the initial orbital elements $a_0$, $e_0$, and $I_0$ during A-early and A-late.
The color charts indicate the number of objects in the logarithmic scale.
}
\label{fig:oc-qmdep-aeI-el-log_s5A}
\end{figure}

\paragraph{(2) Dependence of the $q_\mathrm{min}$ distribution on initial orbital elements}
Next, as a supplement to the discussion of the dependence of the comet's $q_\mathrm{min}$ on each orbital element in the second half of subsection \ref{ssec:barrier}, we present Figure \ref{fig:oc-qmdep-aeI-el-log_s5A}.
This figure provides with another view of the dependence of the frequency distribution of $q_\mathrm{min}$ on the initial orbital elements $a_0$, $e_0$, $I_0$ in color diagrams.
Let us mention just a point about this figure:
$q_\mathrm{min}$ tends to be smaller when the initial eccentricity of the comets $e_0$ is higher (the panels in the middle row).
The comets with low $e_0$ can have lower relative velocity to (thus longer encounter duration with) the major planets.
Therefore, these comets are likely subject to the planet barrier, and their $q_\mathrm{min}$ tends to remain large.
On the other hand,
the comets with higher eccentricity have higher relative velocity to (thus shorter encounter duration with) the major planets.
Therefore, thy are less likely subject to the planet barrier, and their $q_\mathrm{min}$ can be small.

\begin{figure}[!htbp]\centering
  \includegraphics[width=\myfigwidth]{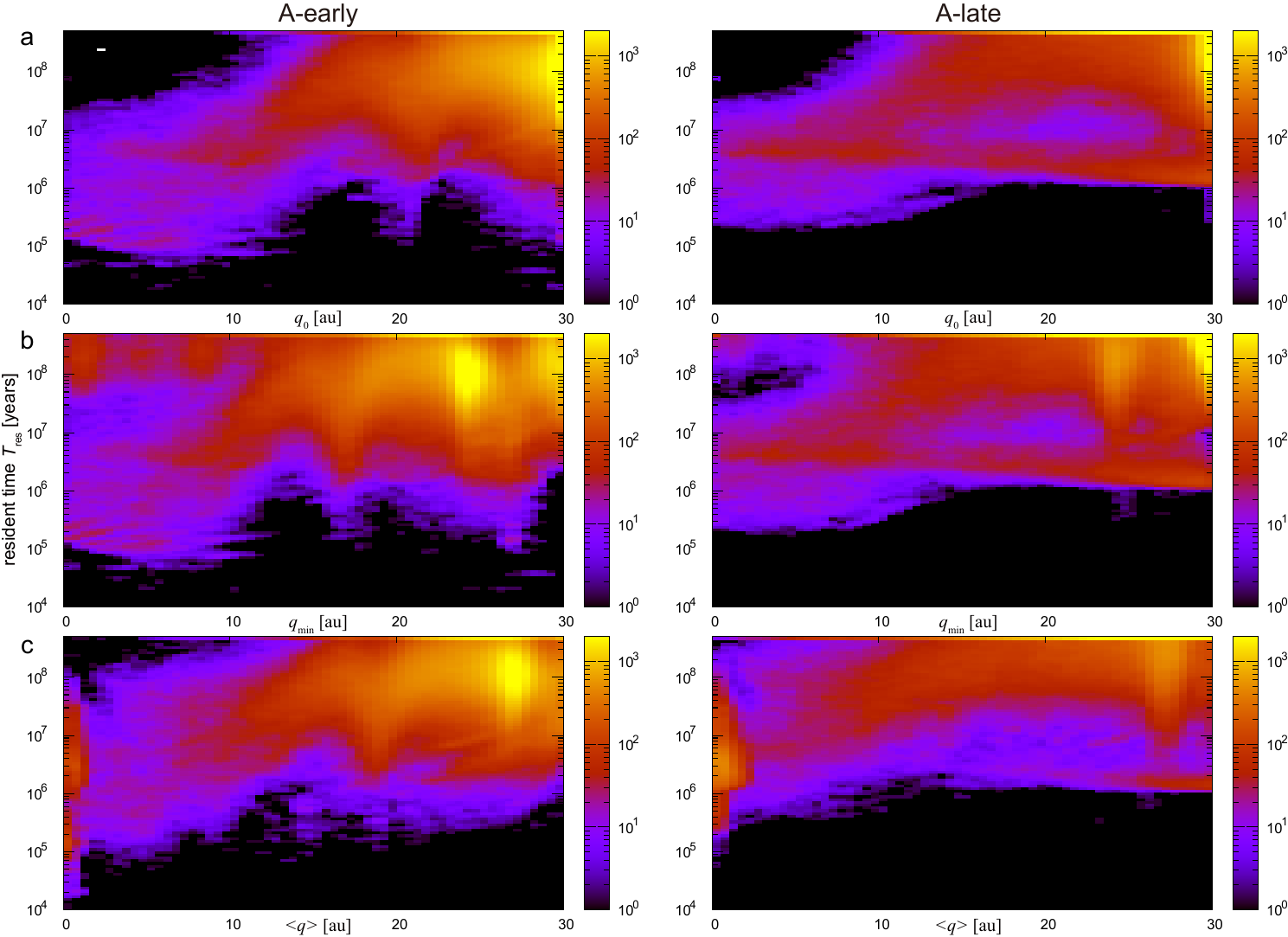}
  \caption[]{%
Relationship between comets' perihelion distance and $T_\mathrm{res}$.
The color shows the number of comets belonging to each region of
(\textsf{a}) the $(q_0, T_\mathrm{res})$ space,
(\textsf{b}) the $(q_\mathrm{min}, T_\mathrm{res})$ space, and
(\textsf{c}) the $(\left< q \right>, T_\mathrm{res})$ space.
Note that the panels in the top row (\textsf{a}) are identical to the panels at the bottom row in Figure \ref{fig:oc-aeIqQ-Te-log_s5A}.
}
\label{fig:oc-q-Te-log_s5A}
\end{figure}

\paragraph{(3) More on the dependence of $q_\mathrm{min}$ on comets' perihelion distance}
At the end of this section, let us give a set of plots about the relationship between $T_\mathrm{res}$ and perihelion distance $q$ of the comets.
Here we think of three kinds of $q$:
$q_0$ (initial value of $q$ of each comet),
$q_\mathrm{min}$ (minimum value of $q$ of each comet), and
$\left< q \right>$ (average of $q$ over the entire $T_\mathrm{res}$ of each comet).
We used the three types of $q$, and created Figure \ref{fig:oc-q-Te-log_s5A}
which shows the distribution of comets on the $(q, T_\mathrm{res})$ plane.
Considering that the color scales in Figure \ref{fig:oc-q-Te-log_s5A} are in the logarithmic scale, we find the peaks of the $T_\mathrm{res}$ distribution at
$q_0              \sim 30$     au in the top    row panels,
$q_\mathrm{min}   \sim 23$--24 au in the middle row panels, and
$\left< q \right> \sim 26$--27 au in the bottom row panels for both A-early and A-late.
The fact that the distribution peak of $q_\mathrm{min} \sim 23$--24 au in the middle row panels is much more prominent in A-early (left column) than in A-late (right column) indicates that the Neptune--Uranus barrier works more efficiently in A-early than in A-late.
We also find that the number density of comets inside $q \lesssim 10$ au is small in all the panels, implying that the Saturn--Jupiter barrier is working to some extent, in particular in A-early.
We would like to add that a similar trend can be seen in the results of our numerical simulations using the star set B (Figure \ref{fig:oc-q-Te-log_s3B}).

\clearpage
\section{Orbital distribution of the transient objects\label{suppl:transient}}
\begin{figure}[!htbp]\centering
  \includegraphics[width=\myfigwidth]{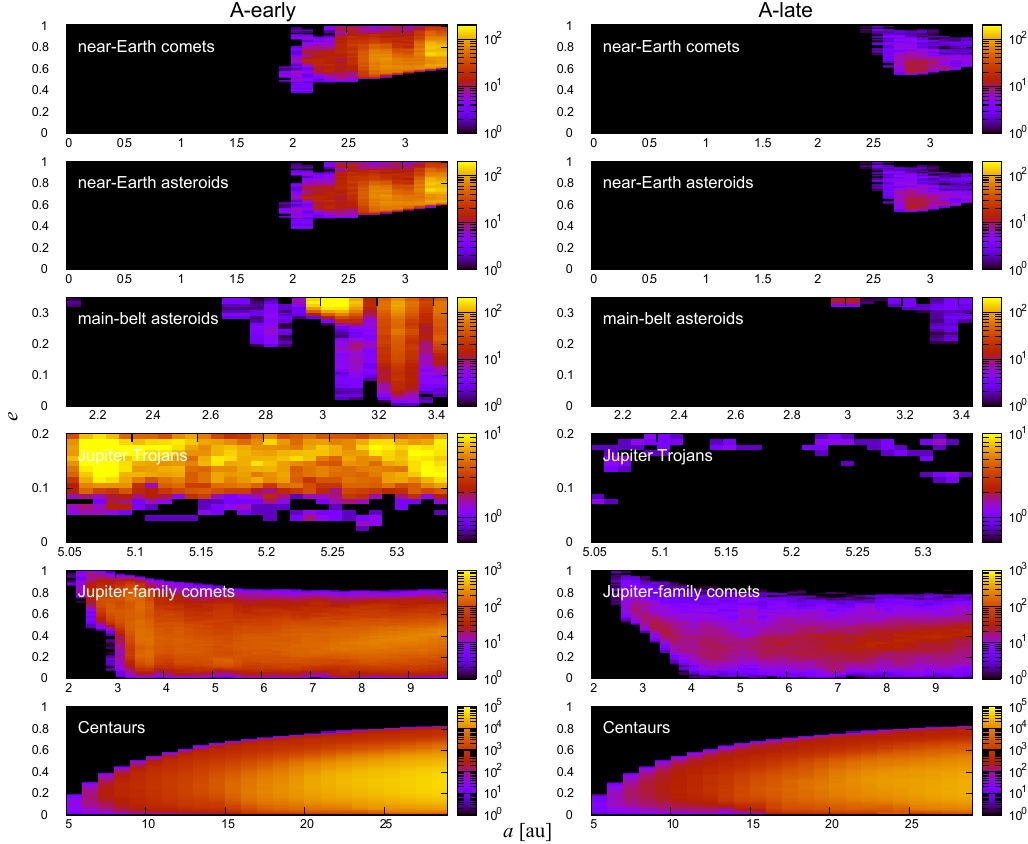}
  \caption[]{Frequency distribution of the comets that transition into the small solar system body populations plotted on the $(a,e)$ plane.
The color charts indicate the number of objects in the logarithmic scale.
}
  \label{fig:oc-group-elem-ae_s5A}
\end{figure}

\begin{figure}[!htbp]\centering
  \includegraphics[width=\myfigwidth]{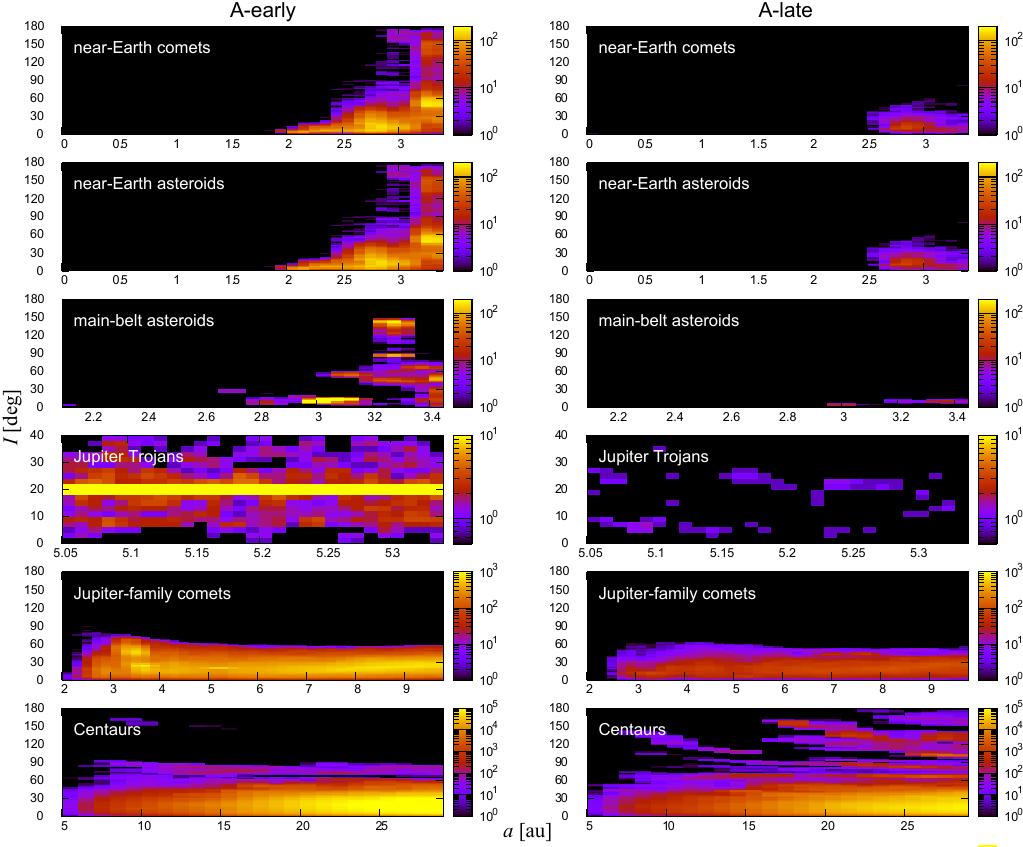}
  \caption[]{The same as Figure \ref{fig:oc-group-elem-ae_s5A}, but on the $(a,I)$ plane.}
  \label{fig:oc-group-elem-ai_s5A}
\end{figure}

\begin{figure}[!htbp]\centering
  \includegraphics[width=\myfigwidth]{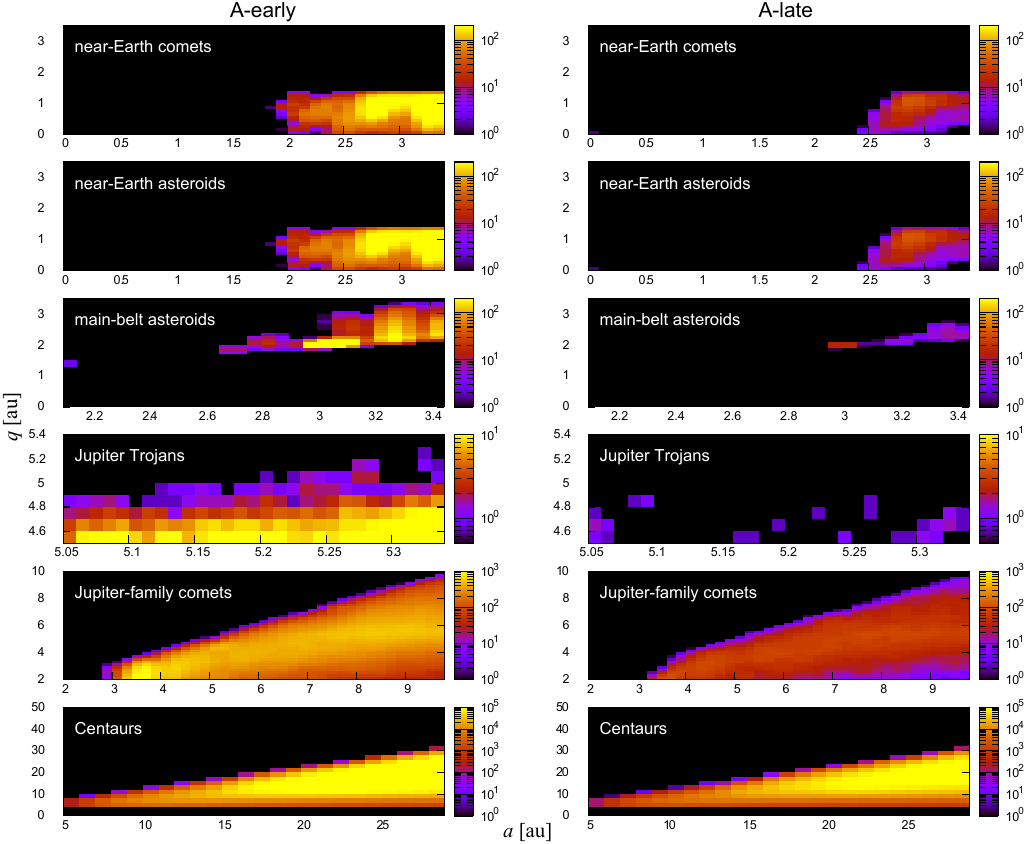}
  \caption[]{The same as Figures \ref{fig:oc-group-elem-ae_s5A} and \ref{fig:oc-group-elem-ai_s5A}, but on the $(a,q)$ plane.}
  \label{fig:oc-group-elem-aq_s5A}
\end{figure}

As supporting material to Section \ref{ssec:otherpops}, in this Appendix we show frequency distribution of the resident time of Oort Cloud comets that transition into other small solar system populations on three types of phase space: $(a,e)$, $(a,I)$, and $(a,q)$.
First we consider the following populations, and plotted the frequency distribution of the comets in these populations on Figures \ref{fig:oc-group-elem-ae_s5A}, \ref{fig:oc-group-elem-ai_s5A}, \ref{fig:oc-group-elem-aq_s5A}:
near-Earth comets,
near-Earth asteroids,
main belt asteroids,
Jupiter-family comets,
Jupiter Trojans, and
Centaurs.
See Section \ref{ssec:otherpops} for the orbital definitions of the populations in our study.
The properties described in Section \ref{ssec:otherpops} and seen in Figure \ref{fig:oc-group-rt-1_s5A} are recognized here, such as the fact that there are more Jupiter-family comets occurring in A-early than in A-late.

\begin{figure}[!htbp]\centering
  \includegraphics[width=\myfigwidth]{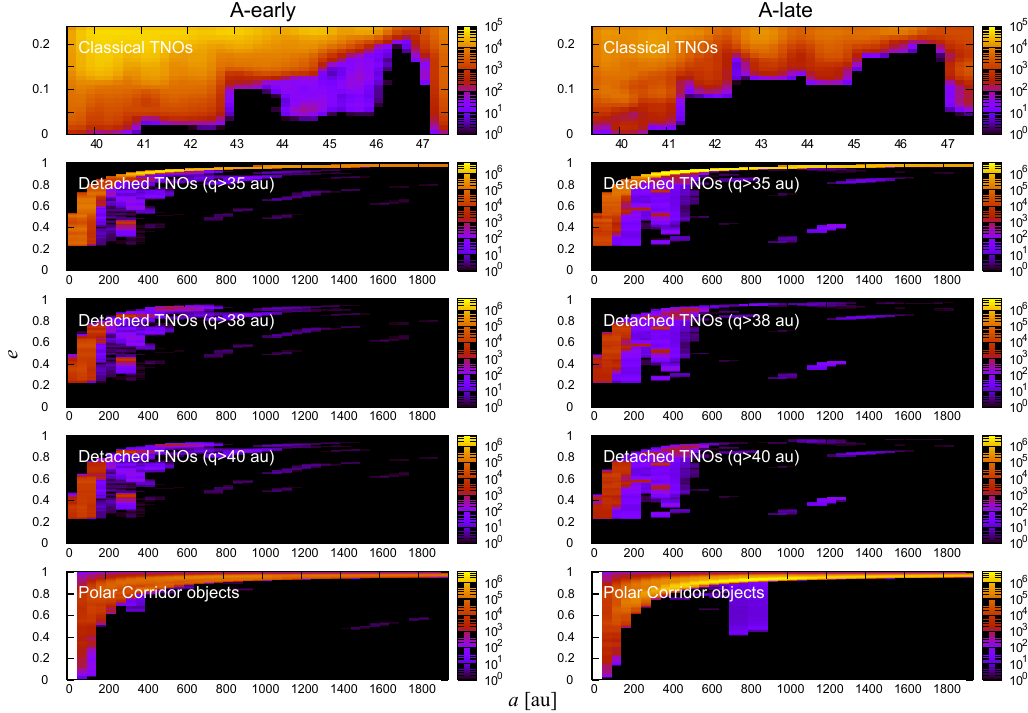}
  \caption[]{Frequency distribution of the comets that transition into another set of the small solar system body populations plotted on the $(a,e)$ plane.
The color charts indicate the number of objects in the logarithmic scale.
}
  \label{fig:oc-group-elem-ae-gladman_s5A}
\end{figure}

\begin{figure}[!htbp]\centering
  \includegraphics[width=\myfigwidth]{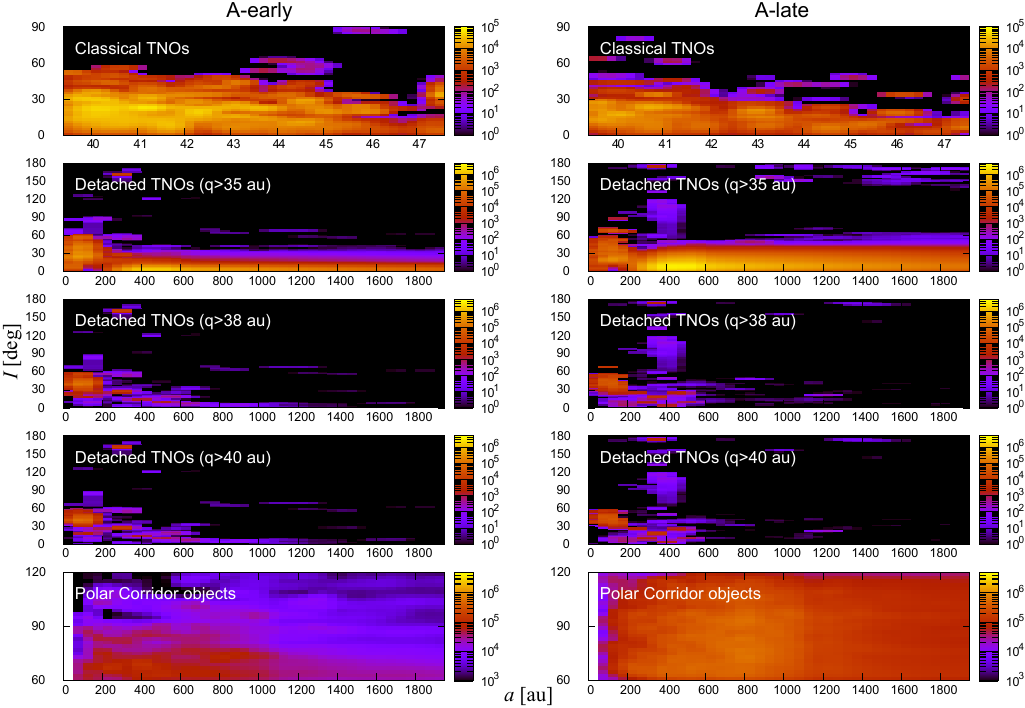}
  \caption[]{The same as Figure \ref{fig:oc-group-elem-ae-gladman_s5A}, but on the $(a,I)$ plane.}
  \label{fig:oc-group-elem-ai-gladman_s5A}
\end{figure}

\begin{figure}[!htbp]\centering
  \includegraphics[width=\myfigwidth]{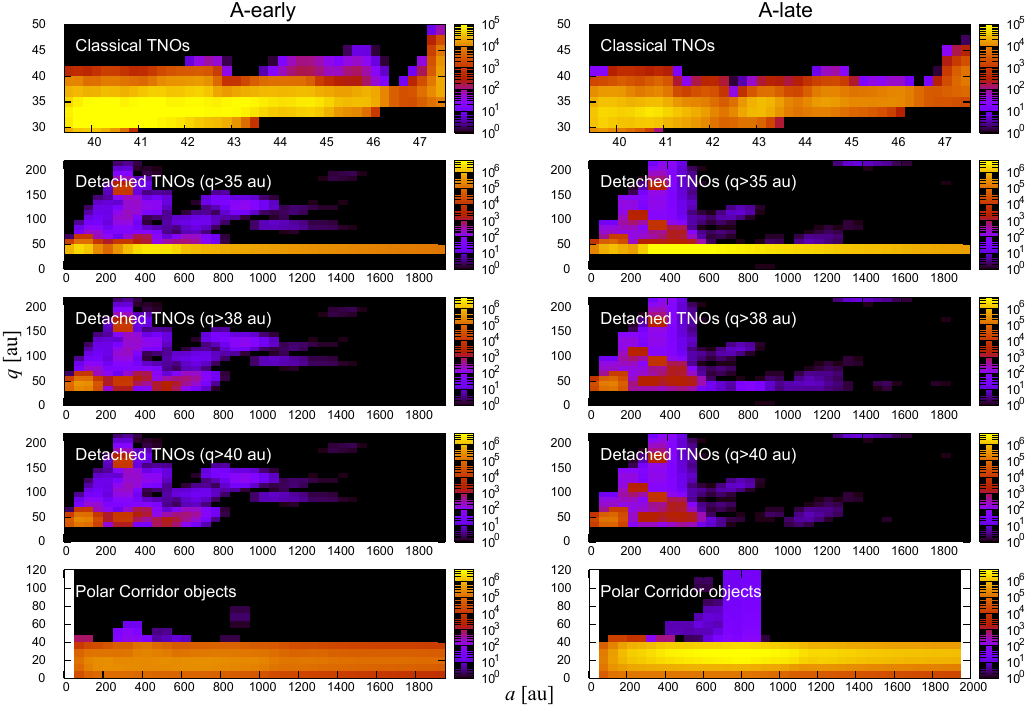}
  \caption[]{The same as Figures \ref{fig:oc-group-elem-ae-gladman_s5A} and \ref{fig:oc-group-elem-ai-gladman_s5A}, but on the $(a,q)$ plane.}
  \label{fig:oc-group-elem-aq-gladman_s5A}
\end{figure}

\begin{figure}[!htbp]
\includegraphics[width=\myfigwidth]{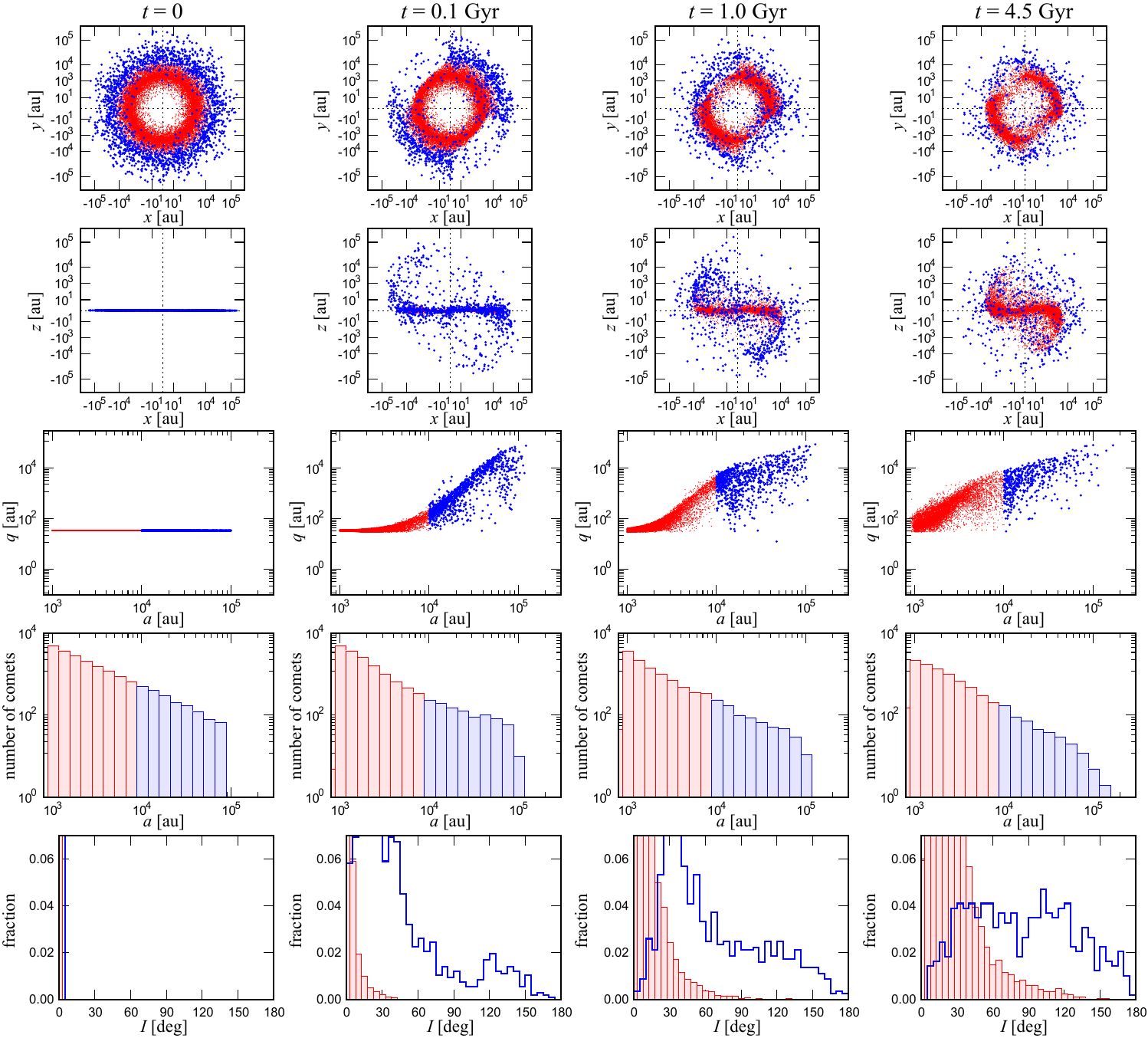}
\caption[]{%
This figure corresponds to Figure \ref{fig:oc-evol-ah_s5A}.
}
\label{fig:oc-evol-ah_s3B}
\end{figure}

\clearpage
Next we consider other populations, and plotted the frequency distribution of the comets in these populations on Figures \ref{fig:oc-group-elem-ae-gladman_s5A}, \ref{fig:oc-group-elem-ai-gladman_s5A}, \ref{fig:oc-group-elem-aq-gladman_s5A}:
the classical TNOs,
the detached TNOs ($q_\mathrm{DT} = 35$, 38, 40 au), and
the polar corridor objects.
See Section \ref{ssec:otherpops} for the orbital definitions of the populations in our study.
The properties described in Section \ref{ssec:otherpops} and seen in Figure \ref{fig:oc-group-rt-2_s5A} are recognized here, such as the fact that there are more polar corridor objects occurring in A-late than in A-early.

\clearpage
\section{Numerical results obtained through the star set B\label{suppl:star3B}}
This Appendix contains figures and tables obtained from our numerical simulation under the star set B.
Table \ref{tbl:fignumcomparison} shows the comparison of figure and table numbers between the two sets.
One of the notable differences between the two sets is the strong comet shower that occurred at time $t \sim 4.45$ Gyr in B-late (see Figures \ref{fig:oc-genoc-starmrv_s3B} and \ref{fig:oc-genocnc-combined-t_s3B}).
This shower produced a large amount of cometary flux, and caused non-negligible difference in some statistics obtained in the periods of A-late and B-late.
Compare Figure \ref{fig:oc-qmin-n1-bulk_s5A} and Figure \ref{fig:oc-qmin-n1-bulk_s3B} as an example.

\clearpage

\renewcommand{\baselinestretch}{1.2}
\begin{table}[!htbp]\centering
\caption[]{%
List of corresponding figures and tables for the star set A and B.}
\begin{tabular}{crr}
\hline
float  & \multicolumn{1}{c}{set A} & \multicolumn{1}{c}{set B} \\
\hline
Table  & \ref{tbl:starmrv-top20_s5A}                    & \ref{tbl:starmrv-top20_s3B}                            \\
Table  & \ref{tbl:ocncfate_s5A}				& \ref{tbl:ocncfate_s3B}			 \\
\hline
Figure & \ref{fig:oc-evol-ah_s5A}                       & \ref{fig:oc-evol-ah_s3B}				 \\
Figure & \ref{fig:oc-remoc_s5A}				& \ref{fig:oc-remoc_s3B}		         \\
Figure & \ref{fig:oc-genoc-starmrv_s5A}			& \ref{fig:oc-genoc-starmrv_s3B}			 \\
Figure & \ref{fig:oc-genocnc-combined-t_s5A}		& \ref{fig:oc-genocnc-combined-t_s3B}		 \\
Figure & \ref{fig:oc-aeIq0_s5A}				& \ref{fig:oc-aeIq0_s3B}			 \\
Figure & \ref{fig:oc-skipped-time-lin_s5A}		& \ref{fig:oc-skipped-time-lin_s3B}		 \\
Figure & \ref{fig:oc-skipped-elem_s5A}		        & \ref{fig:oc-skipped-elem_s3B}		 \\
Figure & \ref{fig:oc-skipped-dist-log_s5A}              & \ref{fig:oc-skipped-dist-log_s3B}              \\
Figure & \ref{fig:oc-skipped-diff_s5A}                  & \ref{fig:oc-skipped-diff_s3B}                  \\
Figure & \ref{fig:oc-t-aeIqQ_s5A}			& \ref{fig:oc-t-aeIqQ_s3B}			 \\
Figure & \ref{fig:oc-lifetime-n1-bulk_s5A_multi}	& \ref{fig:oc-lifetime-n1-bulk_s3B_multi}		 \\
Figure & \ref{fig:oc-lifetime-depend-n0-log-multi_s5A}	& \ref{fig:oc-lifetime-depend-n0-log-multi_s3B}	 \\
Figure & \ref{fig:oc-aeIqQ-Te-log_s5A}			& \ref{fig:oc-aeIqQ-Te-log_s3B}			 \\
Figure & \ref{fig:oc-qmin-n1-bulk_s5A}			& \ref{fig:oc-qmin-n1-bulk_s3B}			 \\
Figure & \ref{fig:oc-qmin-depend-n0-log-init-multi_s5A}	& \ref{fig:oc-qmin-depend-n0-log-init-multi_s3B} \\
Figure & \ref{fig:oc-eIqminTe_s5A}			& \ref{fig:oc-eIqminTe_s3B}			 \\
Figure & \ref{fig:oc-group-rt-1_s5A}			& \ref{fig:oc-group-rt-1_s3B}			 \\
Figure & \ref{fig:oc-group-rt-2_s5A}			& \ref{fig:oc-group-rt-2_s3B}			 \\
Figure & \ref{fig:oc-survivors_s5A}			& \ref{fig:oc-survivors_s3B}			 \\
Figure & \ref{fig:oc-group-elem-ae_s5A}			& \ref{fig:oc-group-elem-ae_s3B}		 \\
Figure & \ref{fig:oc-group-elem-ai_s5A}			& \ref{fig:oc-group-elem-ai_s3B}		 \\
Figure & \ref{fig:oc-group-elem-aq_s5A}			& \ref{fig:oc-group-elem-aq_s3B}		 \\
Figure & \ref{fig:oc-group-elem-ae-gladman_s5A}		& \ref{fig:oc-group-elem-ae-gladman_s3B}	 \\
Figure & \ref{fig:oc-group-elem-ai-gladman_s5A}		& \ref{fig:oc-group-elem-ai-gladman_s3B}	 \\
Figure & \ref{fig:oc-group-elem-aq-gladman_s5A}		& \ref{fig:oc-group-elem-aq-gladman_s3B}         \\
Figure & \ref{fig:oc-nappari_s5A}                       & \ref{fig:oc-nappari_s3B}                       \\
Figure & \ref{fig:oc-qmdep-aeI-el-log_s5A}		& \ref{fig:oc-qmdep-aeI-el-log_s3B}		 \\
Figure & \ref{fig:oc-q-Te-log_s5A}		        & \ref{fig:oc-q-Te-log_s3B}		         \\
\hline
\end{tabular}
\label{tbl:fignumcomparison}
\end{table}

\renewcommand{\baselinestretch}{1.5}

\clearpage
\begin{table}[!htbp]
\caption{%
This table corresponds to Table \ref{tbl:starmrv-top20_s5A}.
}
\begin{center}
\begin{tabular}{ccrccrl}
\hline
\multicolumn{1}{c}{$t$ (Gyr)} &
\multicolumn{1}{c}{$\frac{m}{rv}$} &
\multicolumn{1}{c}{$r$ (au)} &
\multicolumn{1}{c}{$v$ (au/d)} &
\multicolumn{1}{c}{$m$ ($M_\odot$)} &
\multicolumn{1}{c}{$I_\mathrm{s}$ (deg)} &
\multicolumn{1}{c}{period} \\
\hline
4.451366 & 0.03781917 &   1066.36 &  0.099184 &  4.00 &  21.148 & B-late  \\
0.659631 & 0.02630935 &  10200.09 &  0.011924 &  3.20 &  61.400 & B-early \\
2.755519 & 0.02579030 &   4144.66 &  0.037421 &  4.00 & 120.394 &         \\
4.399989 & 0.02282865 &   2129.51 &  0.019130 &  0.93 &  38.965 & B-late  \\
0.110300 & 0.02183802 & 133823.71 &  0.003080 &  9.00 & 103.820 & B-early \\
2.622523 & 0.01717702 &   1248.25 &  0.043374 &  0.93 & 157.639 &         \\
3.252644 & 0.01484295 &  12432.19 &  0.021677 &  4.00 &  30.194 &         \\
0.680333 & 0.01477814 &  23644.61 &  0.002662 &  0.93 &  68.252 & B-early \\
2.574794 & 0.01411797 &   3194.65 &  0.015299 &  0.69 & 112.917 &         \\
0.626882 & 0.01307658 &   6118.86 &  0.013748 &  1.10 & 121.599 & B-early \\
4.505654 & 0.01208604 &  41493.49 &  0.007976 &  4.00 & 123.705 & B-late  \\
1.607030 & 0.01204936 &  11609.21 &  0.022876 &  3.20 & 122.888 &         \\
2.456730 & 0.01190126 &   3686.10 &  0.004787 &  0.21 &  96.764 &         \\
0.611511 & 0.01182614 &   3841.60 &  0.020470 &  0.93 &  99.775 & B-early \\
1.757843 & 0.01012269 &   3865.44 &  0.017634 &  0.69 & 151.154 &         \\
3.113973 & 0.00973648 &  11182.65 &  0.019287 &  2.10 &  35.512 &         \\
0.710195 & 0.00955522 &  26226.28 &  0.012769 &  3.20 &  98.307 & B-early \\
0.206400 & 0.00929319 &  27140.87 &  0.012687 &  3.20 &  68.765 & B-early \\
4.984237 & 0.00915874 &  57096.88 &  0.002104 &  1.10 &  71.754 & B-late  \\
3.175928 & 0.00860164 &   5084.75 &  0.015776 &  0.69 & 153.769 &         \\
\hline
\end{tabular}
\end{center}
\label{tbl:starmrv-top20_s3B}
\end{table}

\clearpage

\begin{table}[!htbp]
\caption{%
This table corresponds to Table \ref{tbl:ocncfate_s5A}.
}
\begin{center}
\begin{tabular}{lrrrr} \hline
\multicolumn{1}{c}{fate} &
\multicolumn{1}{c}{B-early} &
\multicolumn{1}{c}{B-late}  \\
\hline               
$e>1$                & 525124 & 314427 \\
$Q > Q_\mathrm{max}$ & 177484 & 140897 \\
Sun                  &  22139 &     42 \\
Jupiter              &    175 &     26 \\
Saturn               &     59 &      9 \\
Neptune              &     26 &      5 \\
Uranus               &    103 &     10 \\
survived             &  28856 & 304655 \\
\hline                                 
 Total               & 753966 & 760071 \\
\hline
\end{tabular}
\end{center}
\label{tbl:ocncfate_s3B}
\end{table}

\clearpage

\begin{figure}[!htbp]
\includegraphics[width=\myfigwidth]{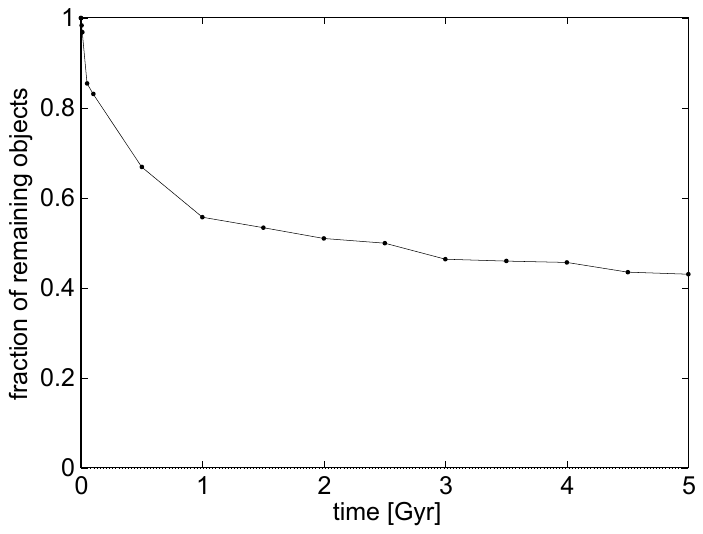}
\caption[]{%
This figure corresponds to Figure \ref{fig:oc-remoc_s5A}.
}
\label{fig:oc-remoc_s3B}
\end{figure}

\begin{figure}[!htbp]
 \includegraphics[width=\myfigwidth]{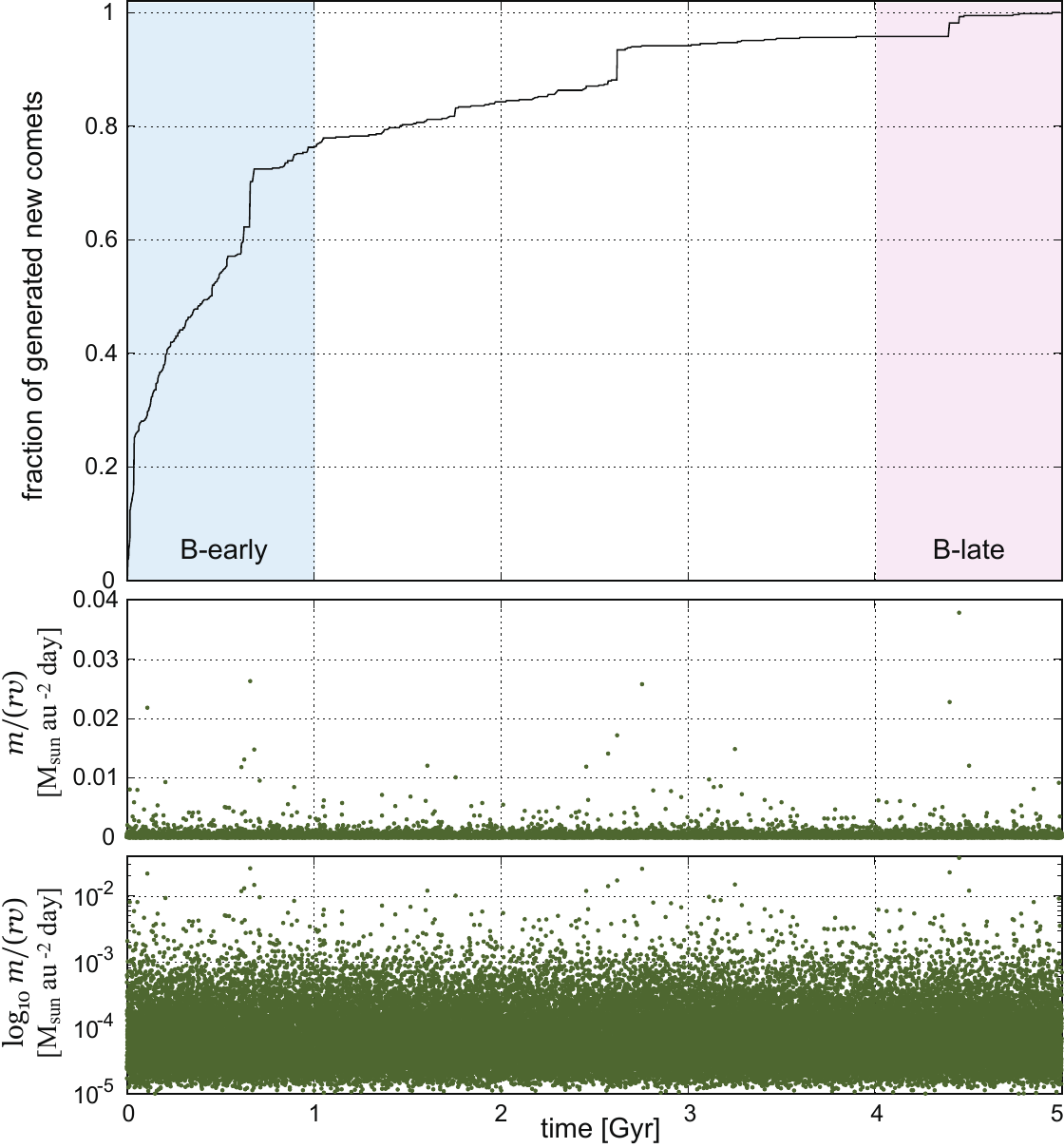}
\caption[]{%
This figure corresponds to Figure \ref{fig:oc-genoc-starmrv_s5A}.
}
 \label{fig:oc-genoc-starmrv_s3B}
\end{figure}

\begin{figure}[!htbp]
\includegraphics[width=\myfigwidth]{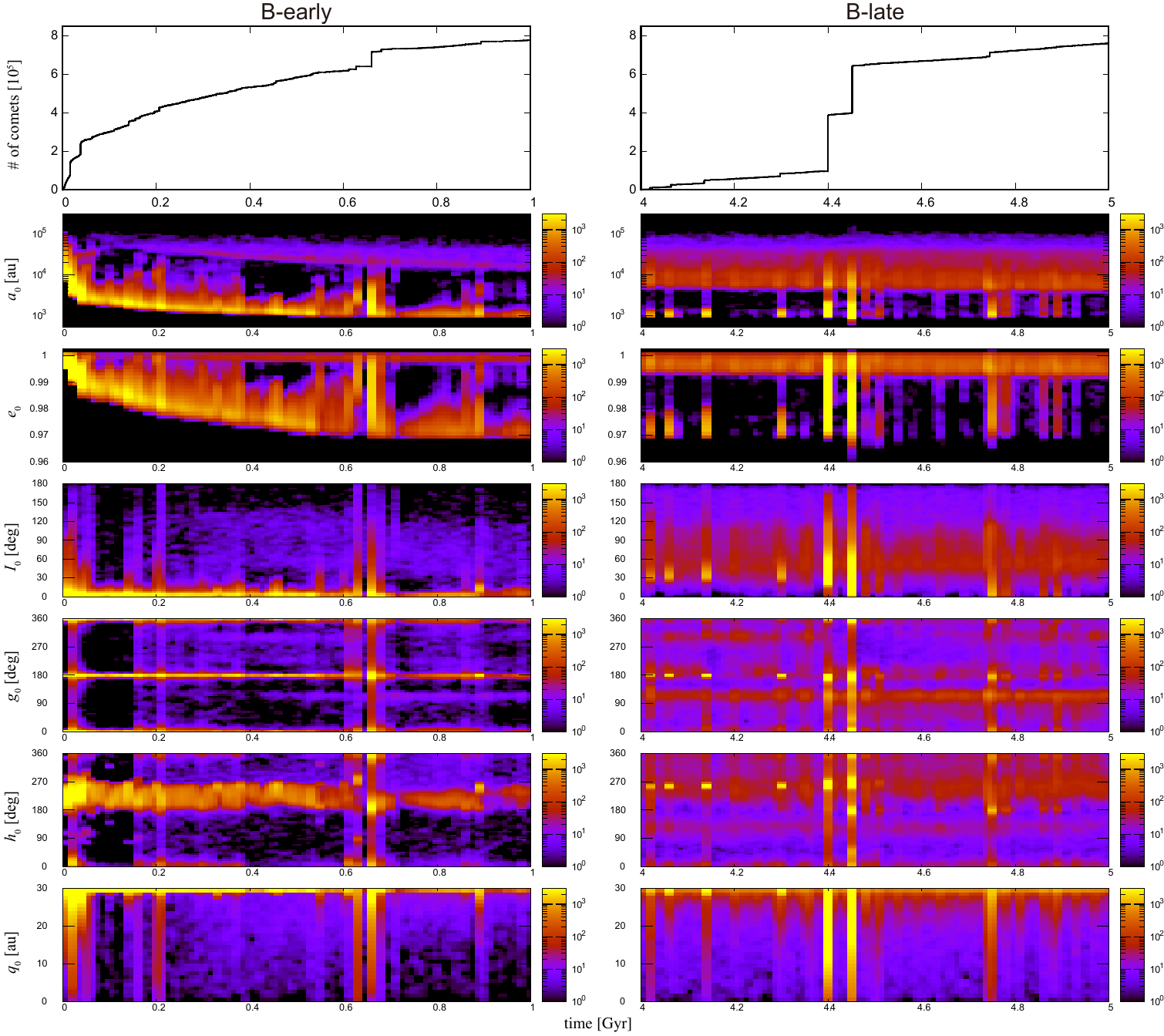}
\caption[]{%
This figure corresponds to Figure \ref{fig:oc-genocnc-combined-t_s5A}.
}
\label{fig:oc-genocnc-combined-t_s3B}
\end{figure}

\begin{figure}[!htbp]
\includegraphics[width=\myfigwidth]{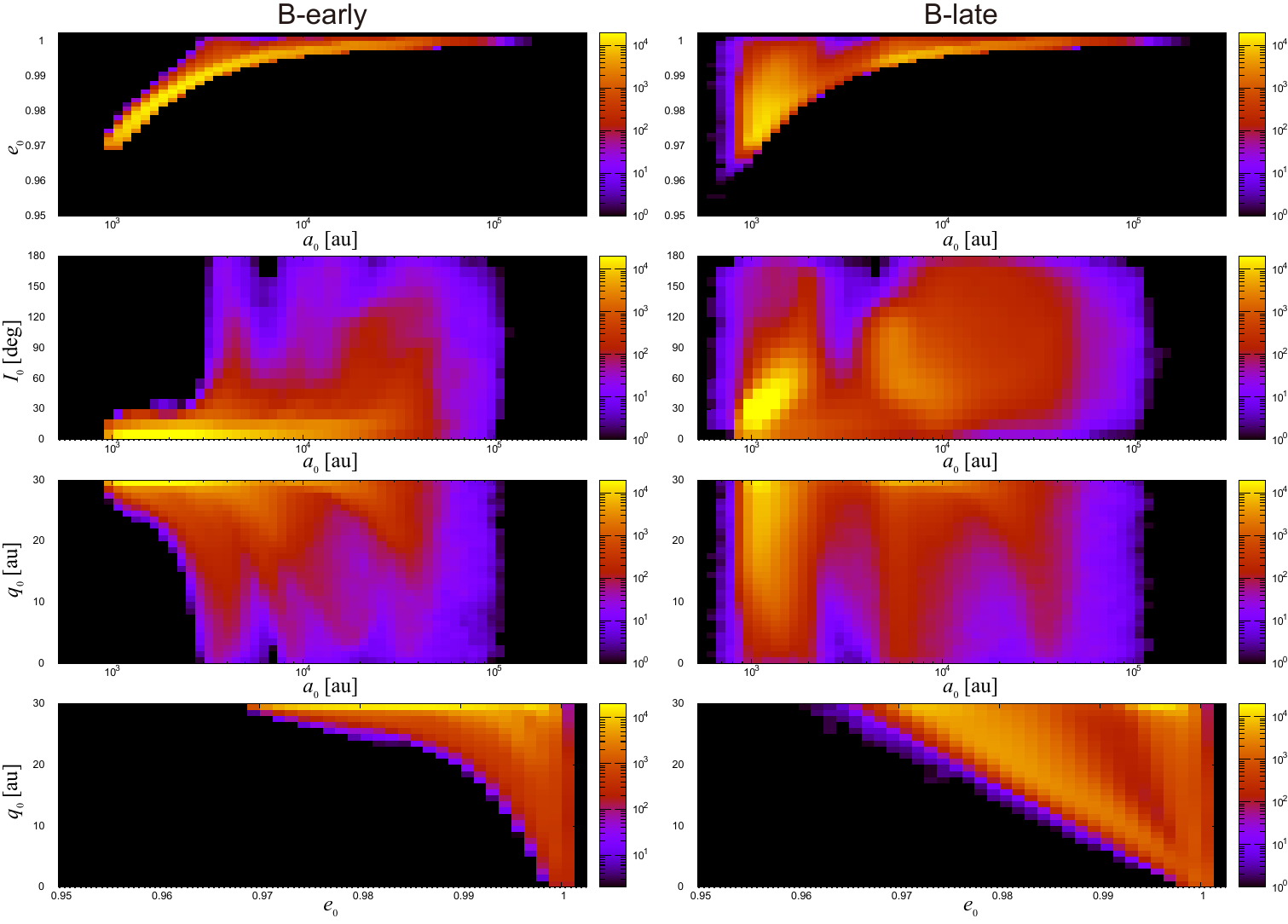}
\caption[]{%
This figure corresponds to Figure \ref{fig:oc-aeIq0_s5A}.
}
\label{fig:oc-aeIq0_s3B}
\end{figure}

\clearpage

\begin{figure}[!htbp]
\includegraphics[width=\myfigwidth]{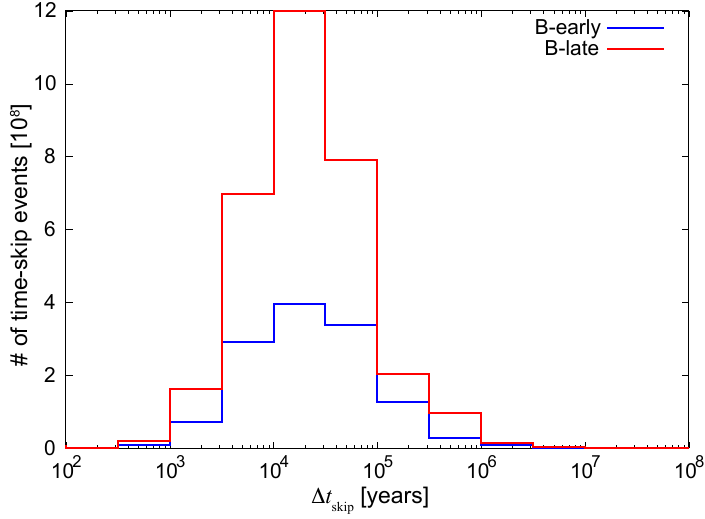}
\caption[]{%
This figure corresponds to Figure \ref{fig:oc-skipped-time-lin_s5A}.
Total number of the time-skip events in B-early is
     1,286,422,208,
and that in B-late  is
     3,201,526,024.
}
\label{fig:oc-skipped-time-lin_s3B}
\end{figure}

\begin{figure}[!htbp]
\includegraphics[width=\myfigwidth]{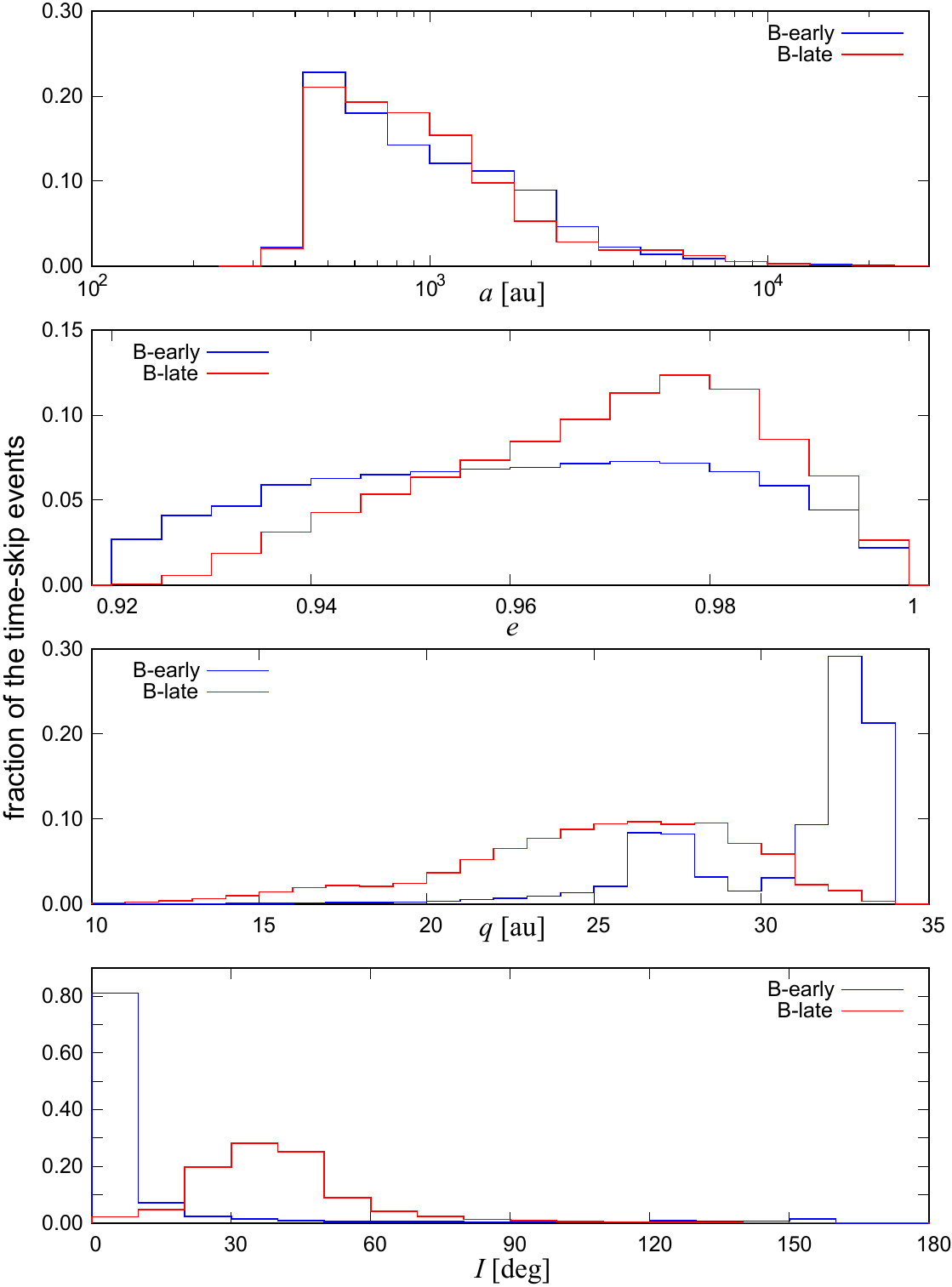}
\caption[]{%
This figure corresponds to Figure \ref{fig:oc-skipped-elem_s5A}.
}
\label{fig:oc-skipped-elem_s3B}
\end{figure}

\clearpage

\begin{figure}[!htbp]\centering
\includegraphics[width=\myfigwidth]{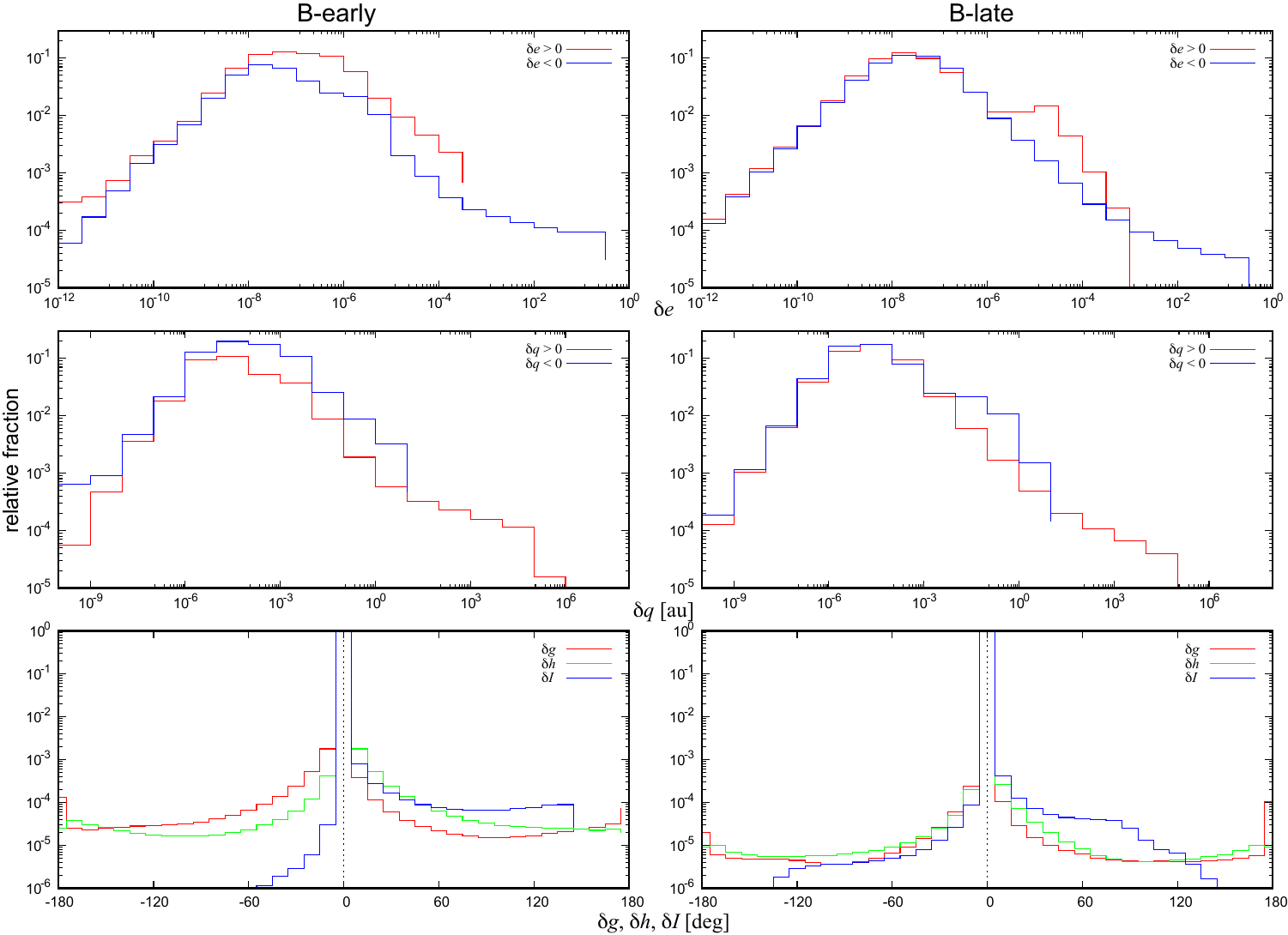}
\caption[]{%
This figure corresponds to Figure \ref{fig:oc-skipped-dist-log_s5A}.
}
\label{fig:oc-skipped-dist-log_s3B}
\end{figure}

\clearpage

\begin{figure}[!htbp]\centering
\includegraphics[width=\myfigwidth]{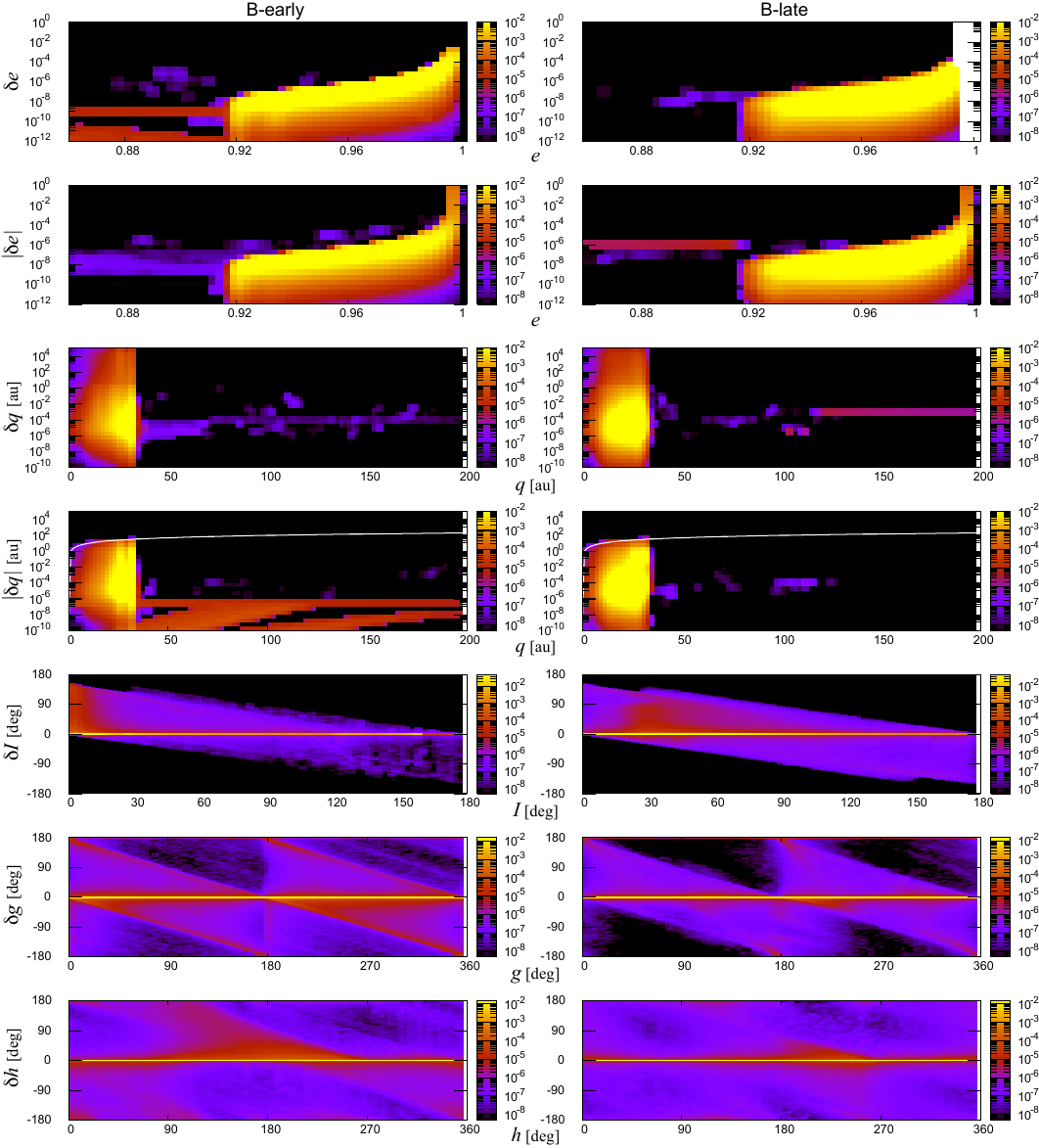}
\caption[]{%
This figure corresponds to Figure \ref{fig:oc-skipped-diff_s5A}.
}
\label{fig:oc-skipped-diff_s3B}
\end{figure}

\clearpage

\begin{figure}[!htbp]
\includegraphics[width=\myfigwidth]{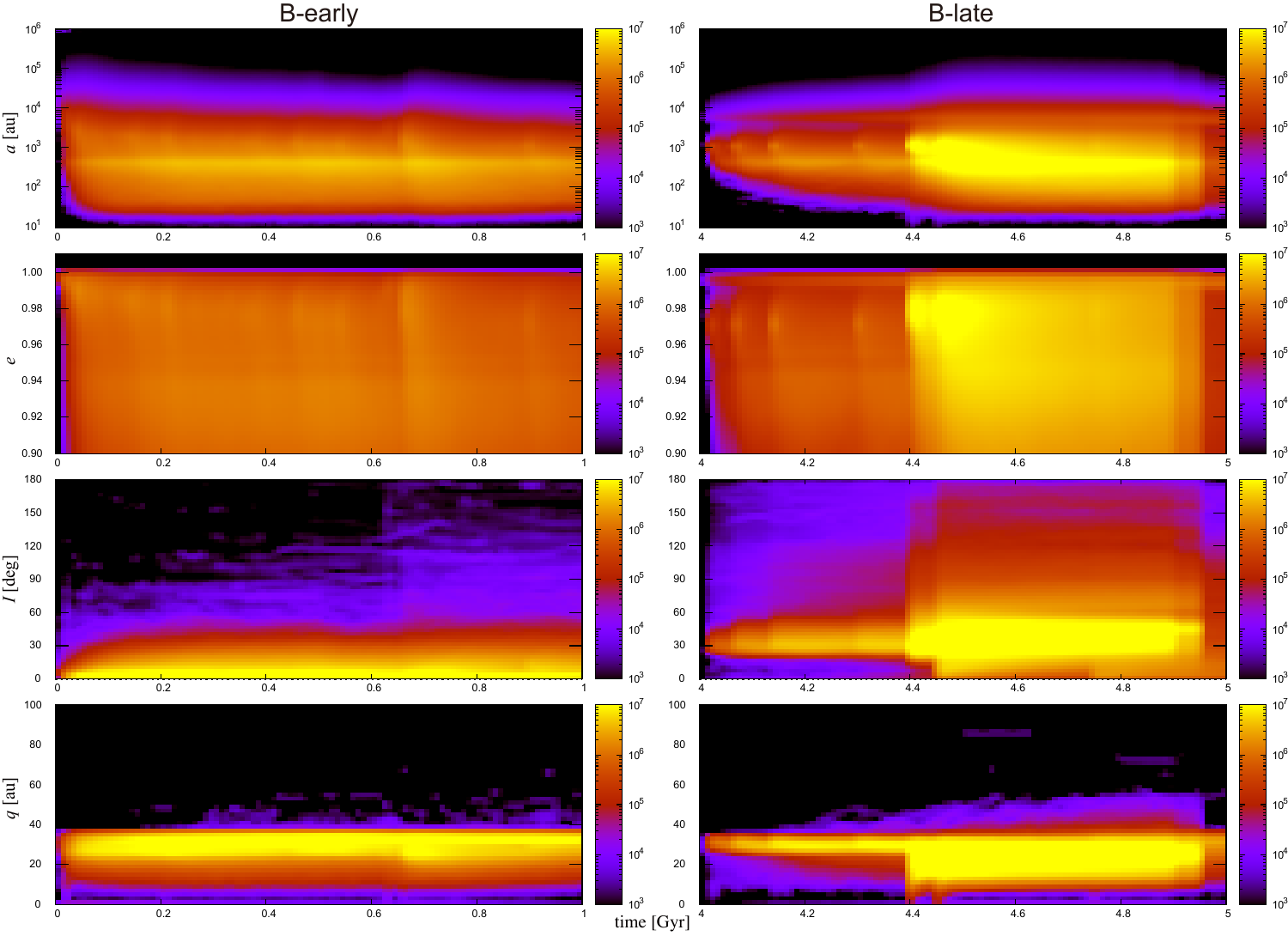}
\caption[]{%
This figure corresponds to Figure \ref{fig:oc-t-aeIqQ_s5A}.
}
\label{fig:oc-t-aeIqQ_s3B}
\end{figure}

\begin{figure}[!htbp]
 \includegraphics[width=\myfigwidth]{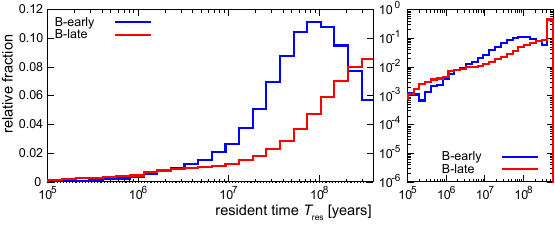}
\caption[]{%
This figure corresponds to Figure \ref{fig:oc-lifetime-n1-bulk_s5A_multi}.
}
 \label{fig:oc-lifetime-n1-bulk_s3B_multi}
\end{figure}

\begin{figure}[!htbp]
\includegraphics[width=\myfigwidth]{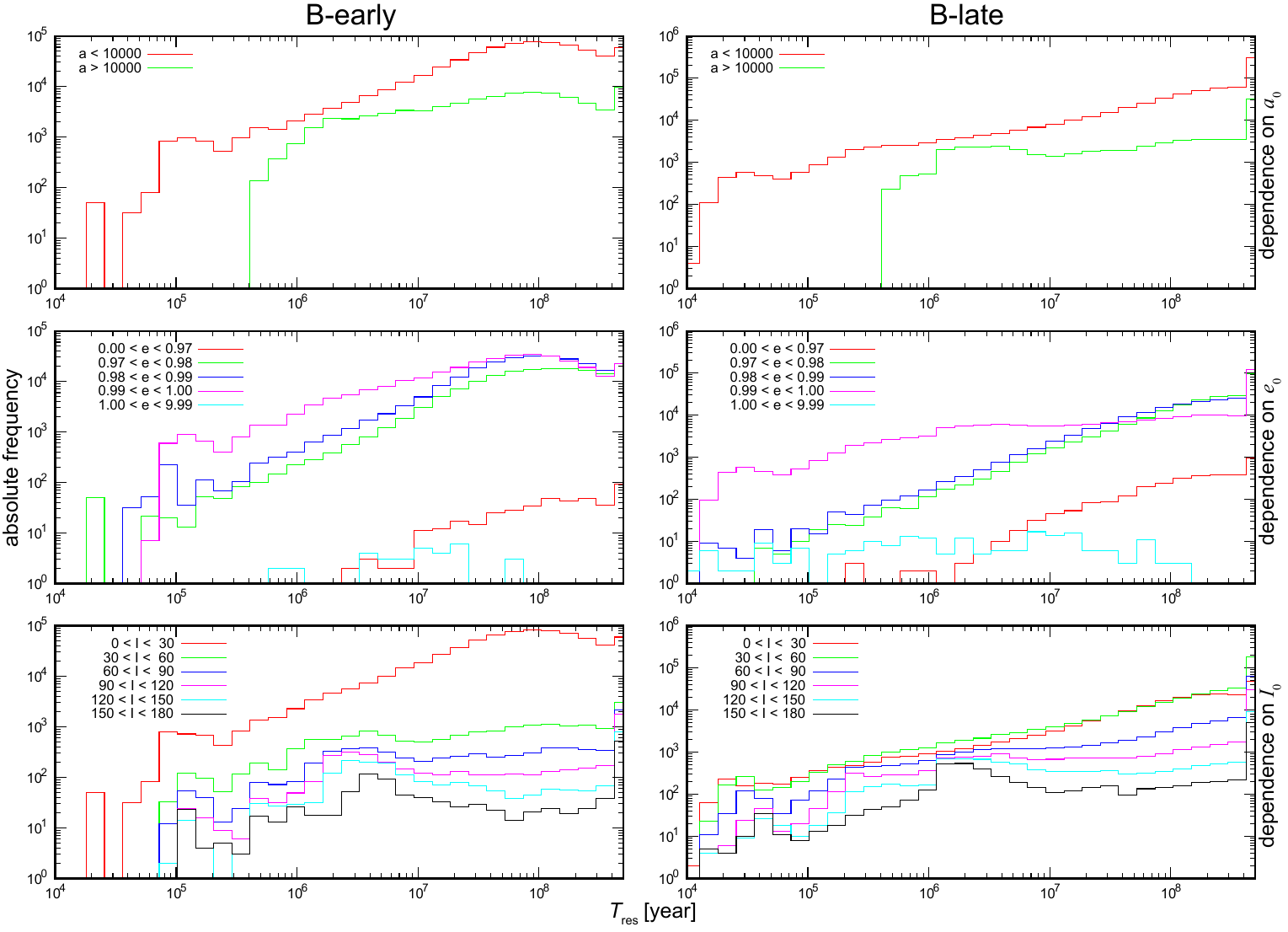}
\caption[]{%
This figure corresponds to Figure \ref{fig:oc-lifetime-depend-n0-log-multi_s5A}.
}
\label{fig:oc-lifetime-depend-n0-log-multi_s3B}
\end{figure}

\begin{figure}[!htbp]
\includegraphics[width=\myfigwidth]{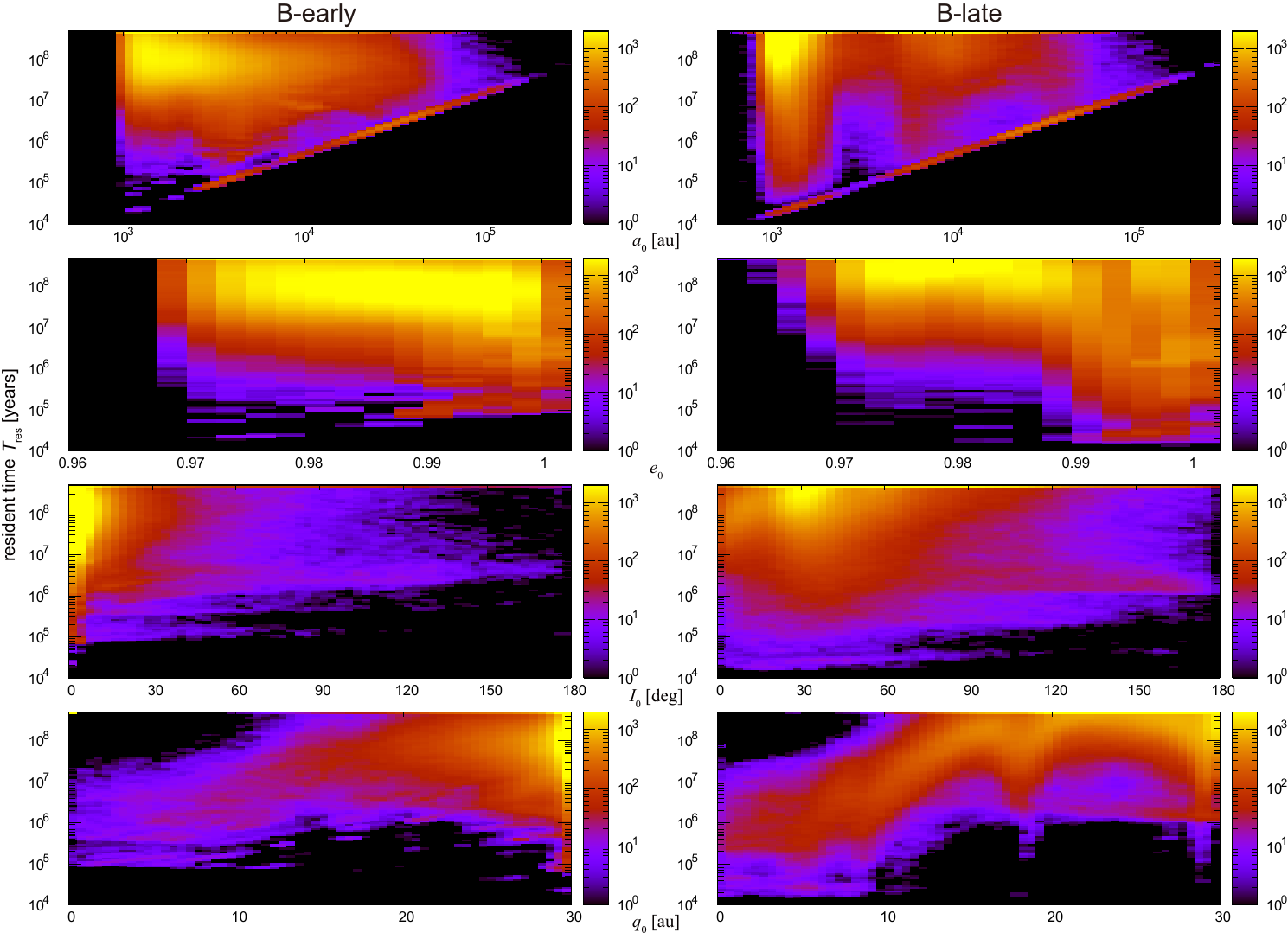}
\caption[]{%
This figure corresponds to Figure \ref{fig:oc-aeIqQ-Te-log_s5A}.
}
\label{fig:oc-aeIqQ-Te-log_s3B}
\end{figure}

\begin{figure}[!htbp]\centering
\includegraphics[width=\myfigwidth]{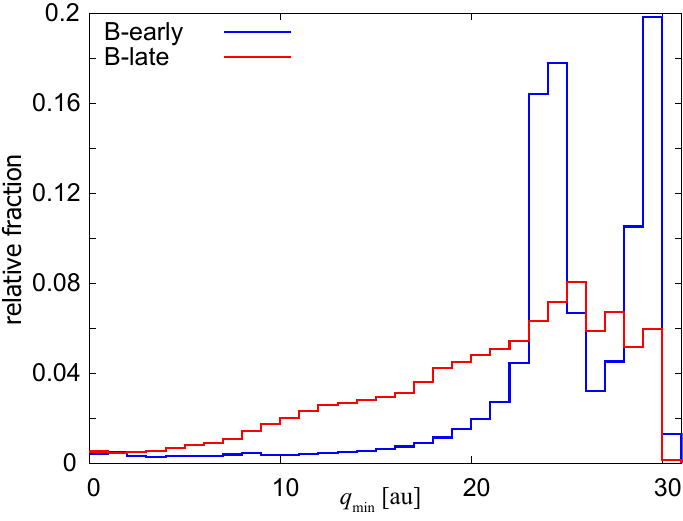}
\caption[]{%
This figure corresponds to Figure \ref{fig:oc-qmin-n1-bulk_s5A}.
}
\label{fig:oc-qmin-n1-bulk_s3B}
\end{figure}

\begin{figure}[!htbp]\centering
\includegraphics[width=\myfigwidth]{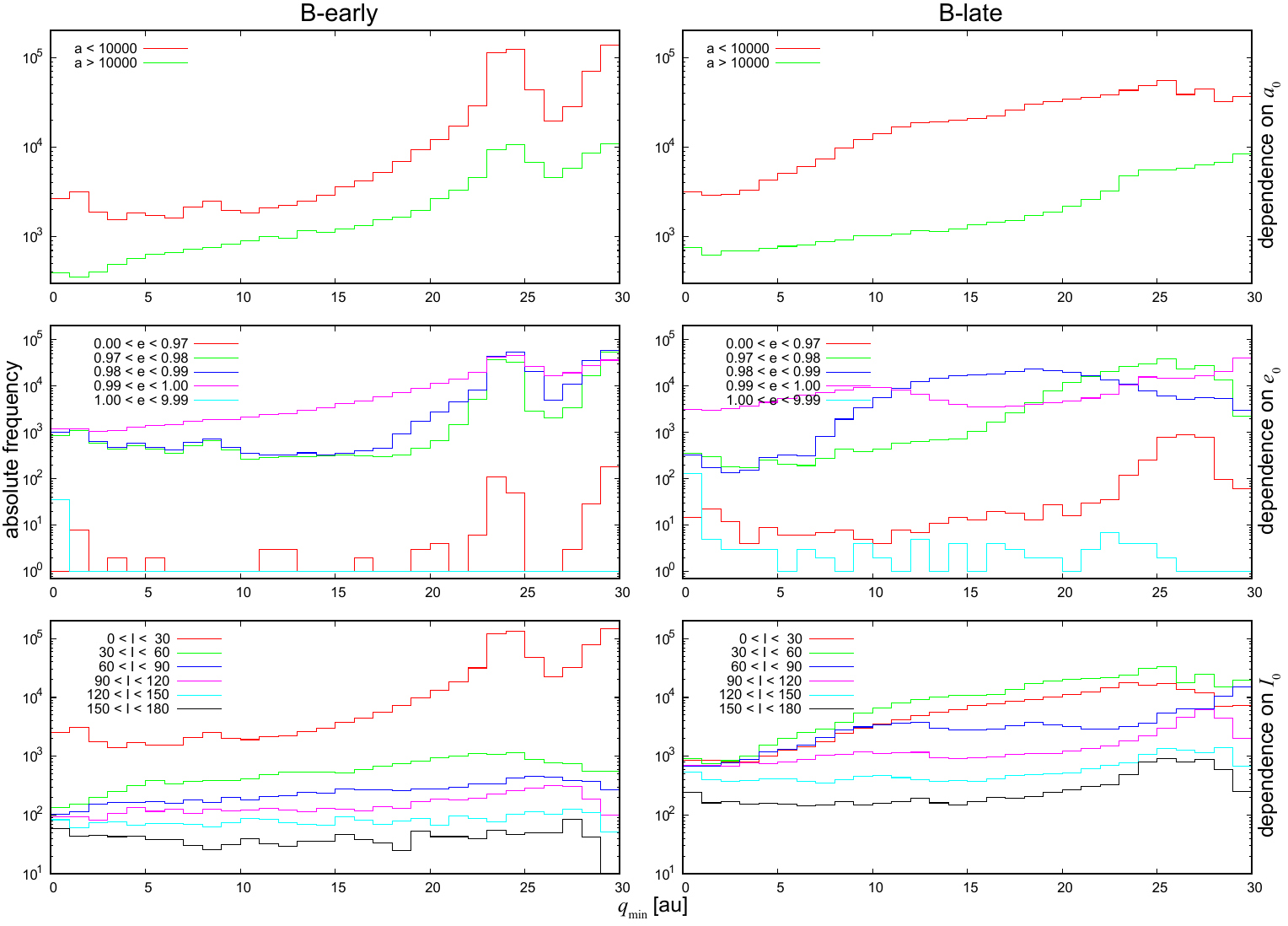}
\caption[]{%
This figure corresponds to Figure \ref{fig:oc-qmin-depend-n0-log-init-multi_s5A}.
}
\label{fig:oc-qmin-depend-n0-log-init-multi_s3B}
\end{figure}

\begin{figure}[!htbp]\centering
\includegraphics[width=\myfigwidth]{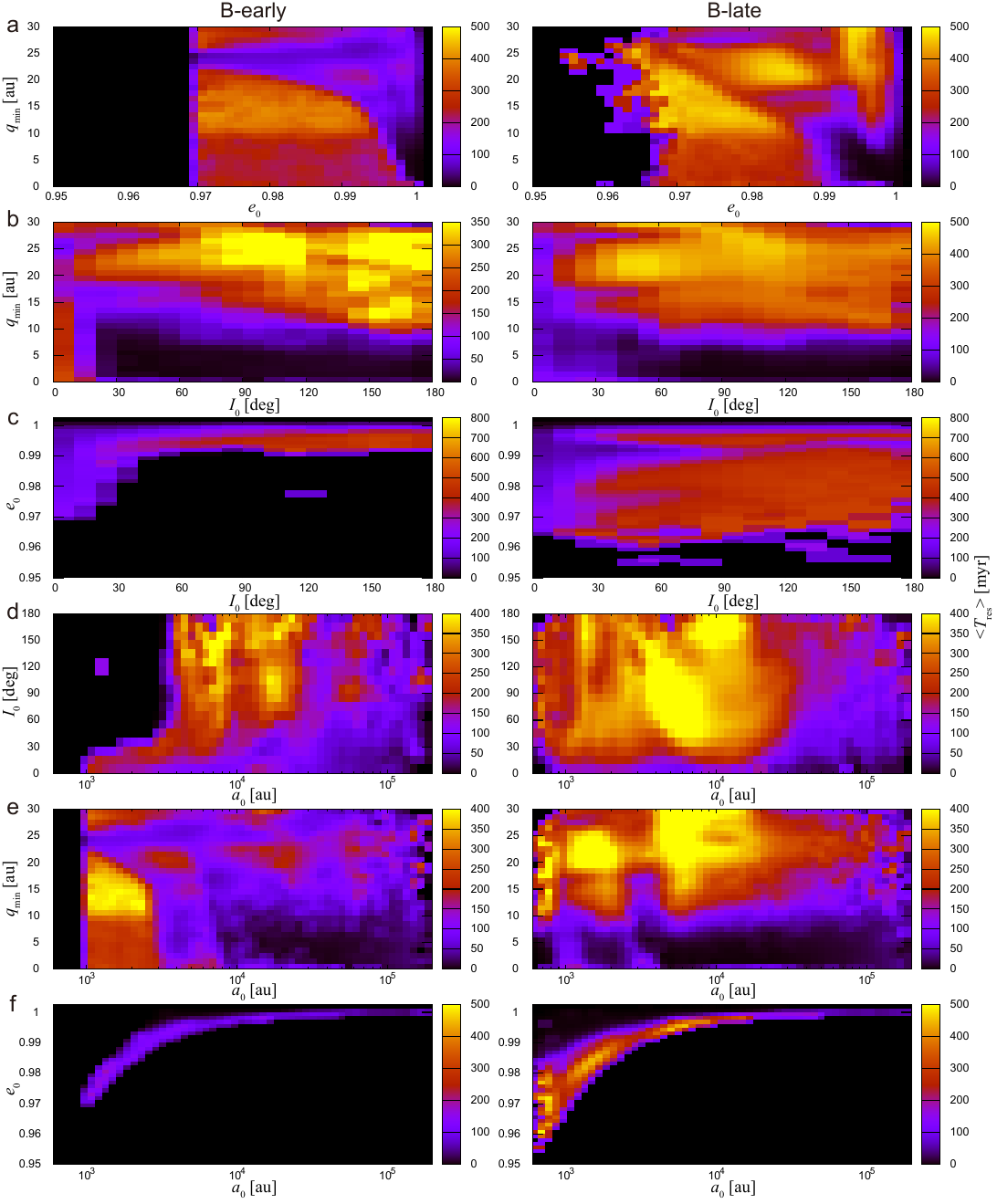}
\caption[]{%
This figure corresponds to Figure \ref{fig:oc-eIqminTe_s5A}.
}
\label{fig:oc-eIqminTe_s3B}
\end{figure}

\begin{figure}[!htbp]\centering
\includegraphics[width=\myfigwidth]{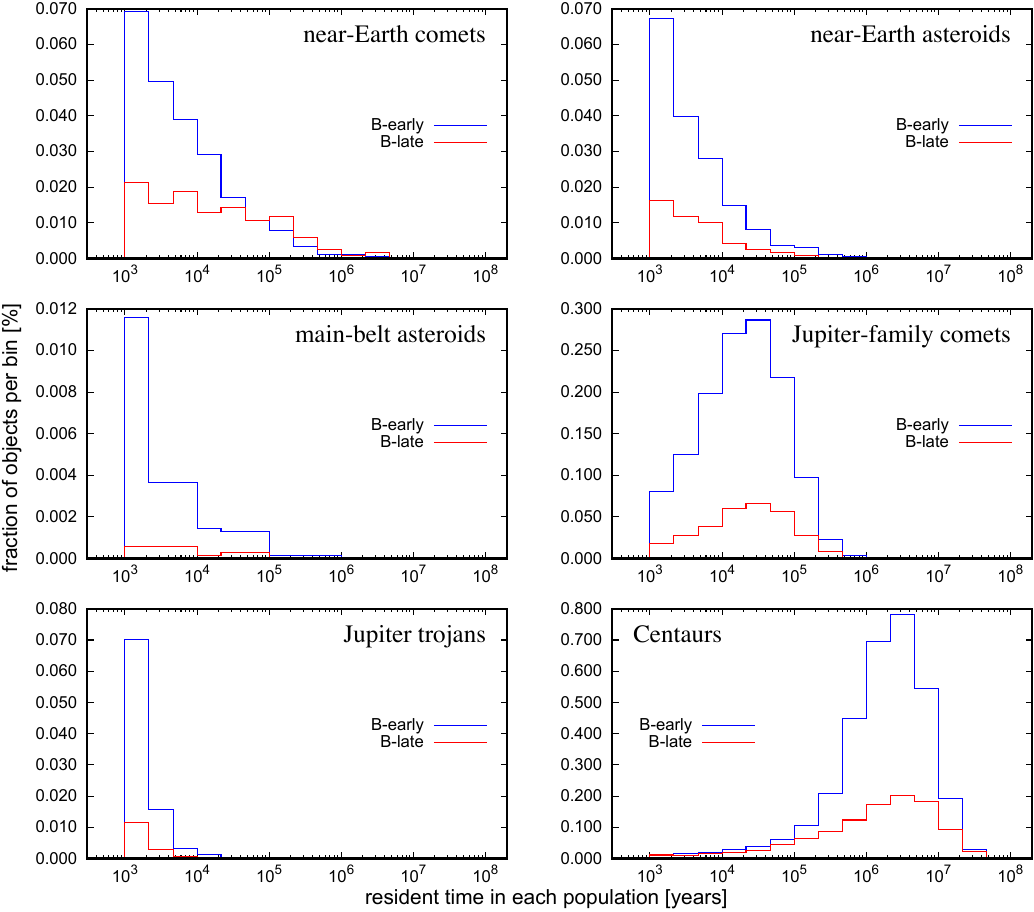}
\caption[]{%
This figure corresponds to Figure \ref{fig:oc-group-rt-1_s5A}.
The time-average value of orbital inclination of the objects that temporarily became the Jupiter-family comets  $\left< I_\mathrm{JFC} \right>$ is,
in B-early,       $\left< I_\mathrm{JFC} \right> = 23.1^\circ \pm 11.6^\circ$.
In B-late,  it is $\left< I_\mathrm{JFC} \right> = 28.9^\circ \pm 13.5^\circ$.
}
\label{fig:oc-group-rt-1_s3B}
\end{figure}

\begin{figure}[!htbp]\centering
\includegraphics[width=\myfigwidth]{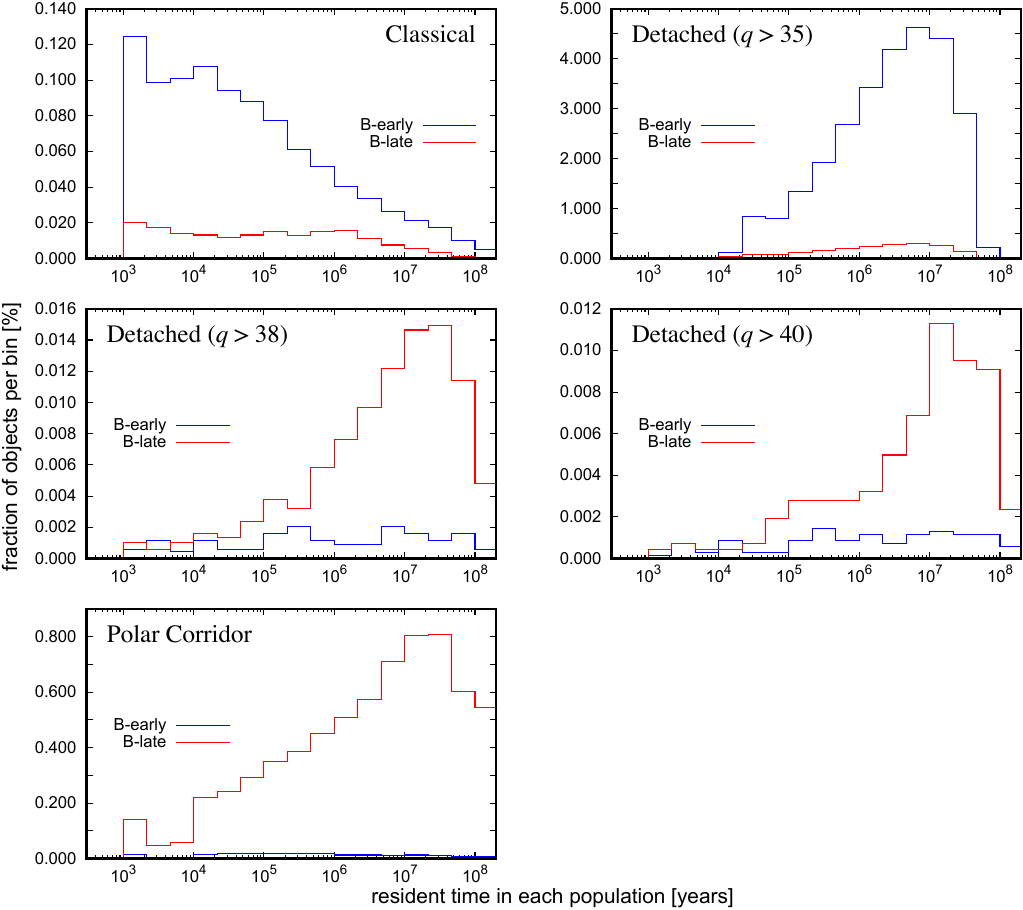}
\caption[]{%
This figure corresponds to Figure \ref{fig:oc-group-rt-2_s5A}.
}
\label{fig:oc-group-rt-2_s3B}
\end{figure}

\begin{figure}[!htbp]\centering
\includegraphics[width=\myfigwidth]{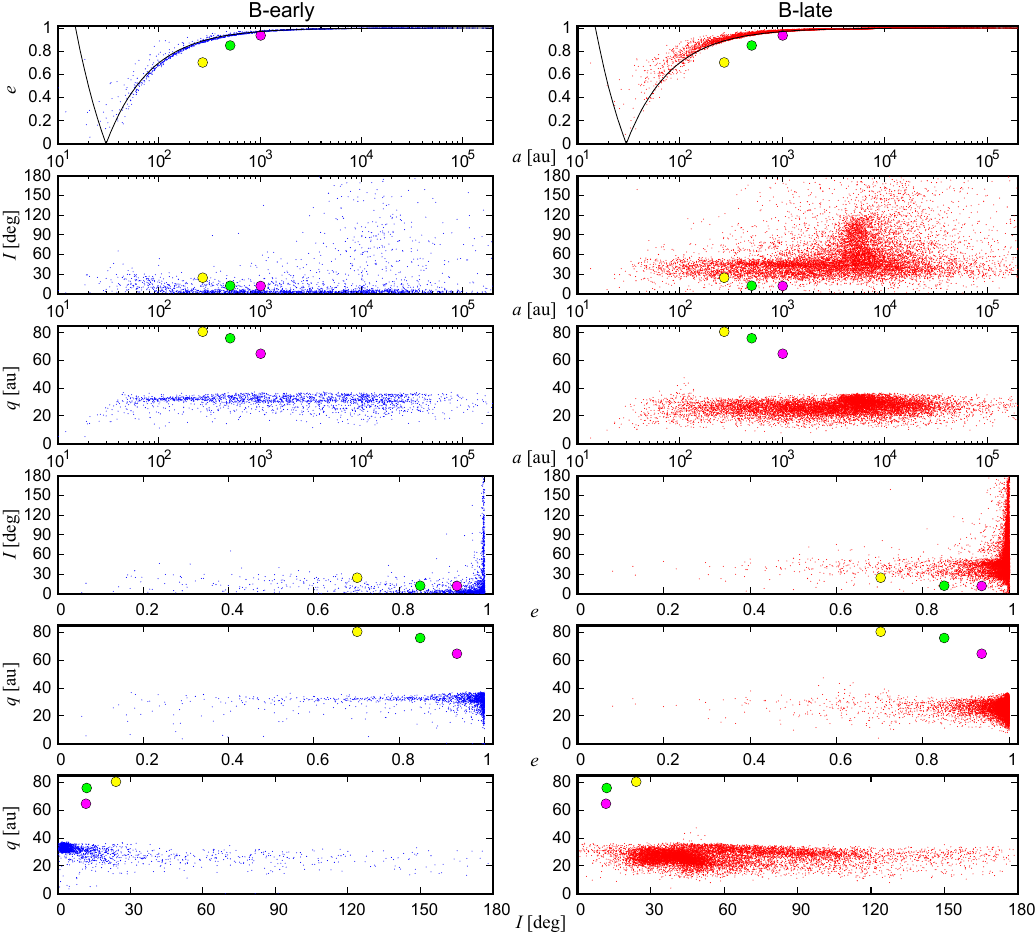}
\caption[]{%
This figure corresponds to Figure \ref{fig:oc-survivors_s5A}.
}
\label{fig:oc-survivors_s3B}
\end{figure}

\clearpage

\begin{figure}[!htbp]\centering
\includegraphics[width=\myfigwidth]{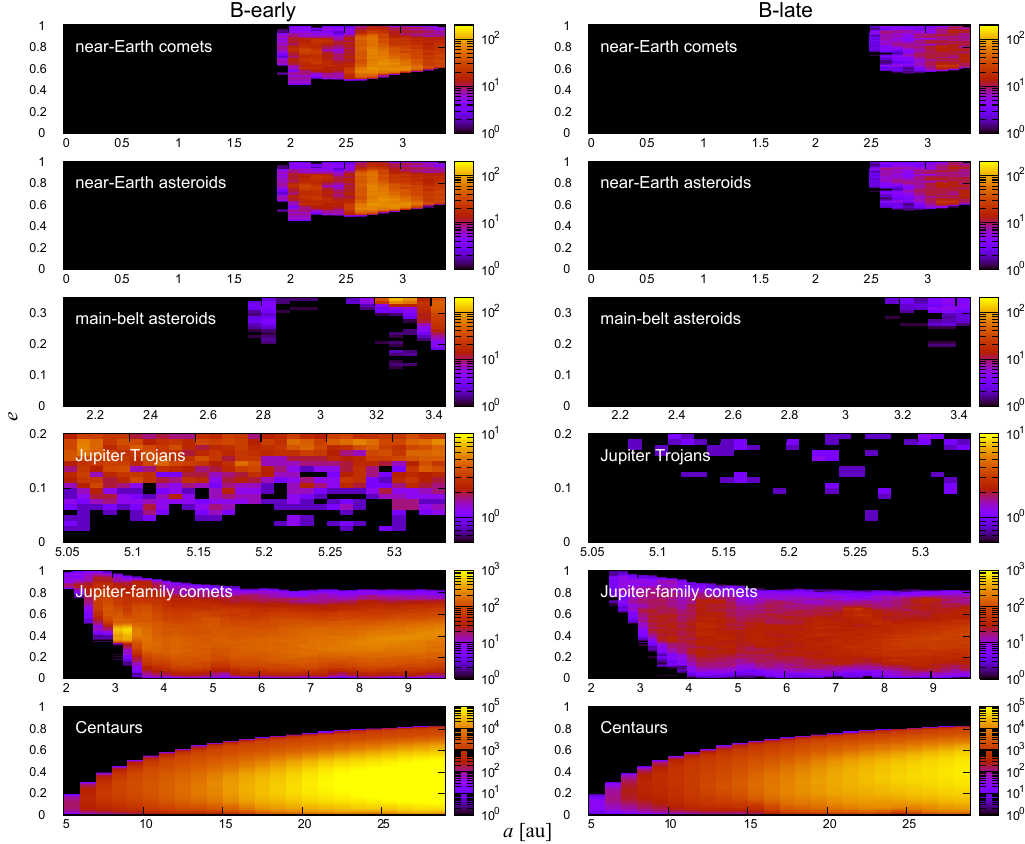}
\caption[]{%
This figure corresponds to Figure \ref{fig:oc-group-elem-ae_s5A}.
}
\label{fig:oc-group-elem-ae_s3B}
\end{figure}

\begin{figure}[!htbp]\centering
\includegraphics[width=\myfigwidth]{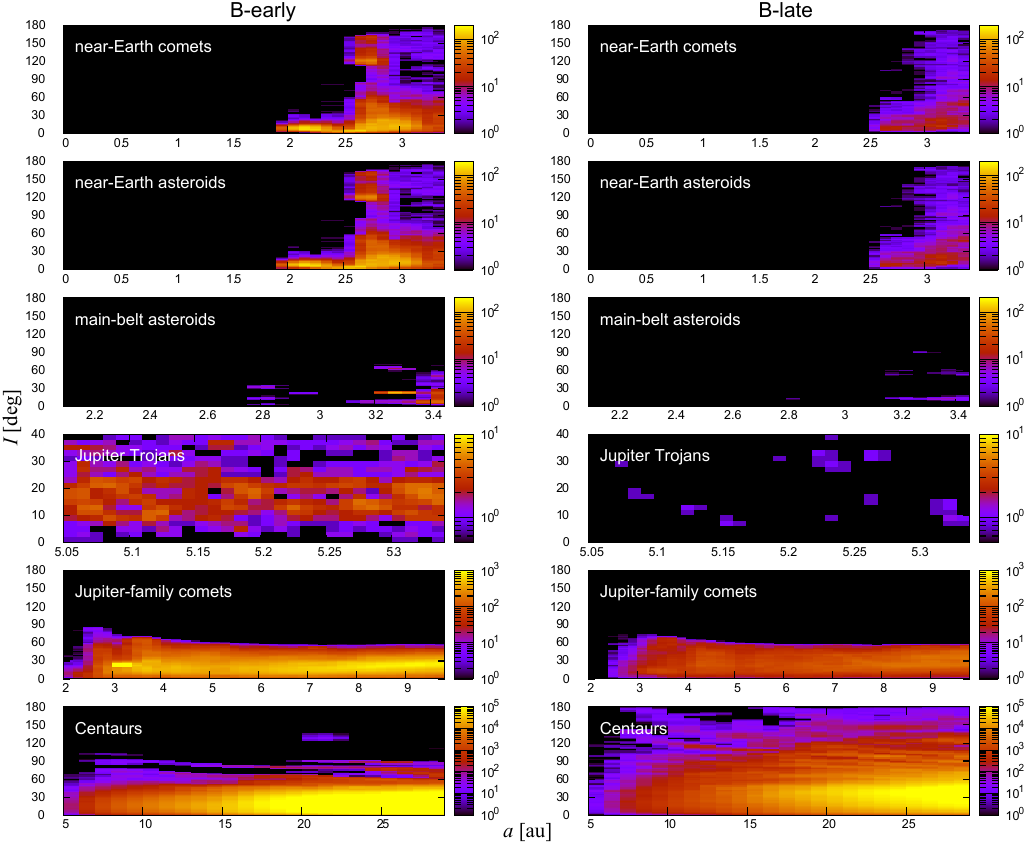}
\caption[]{%
This figure corresponds to Figure \ref{fig:oc-group-elem-ai_s5A}.
}
\label{fig:oc-group-elem-ai_s3B}
\end{figure}

\begin{figure}[!htbp]\centering
\includegraphics[width=\myfigwidth]{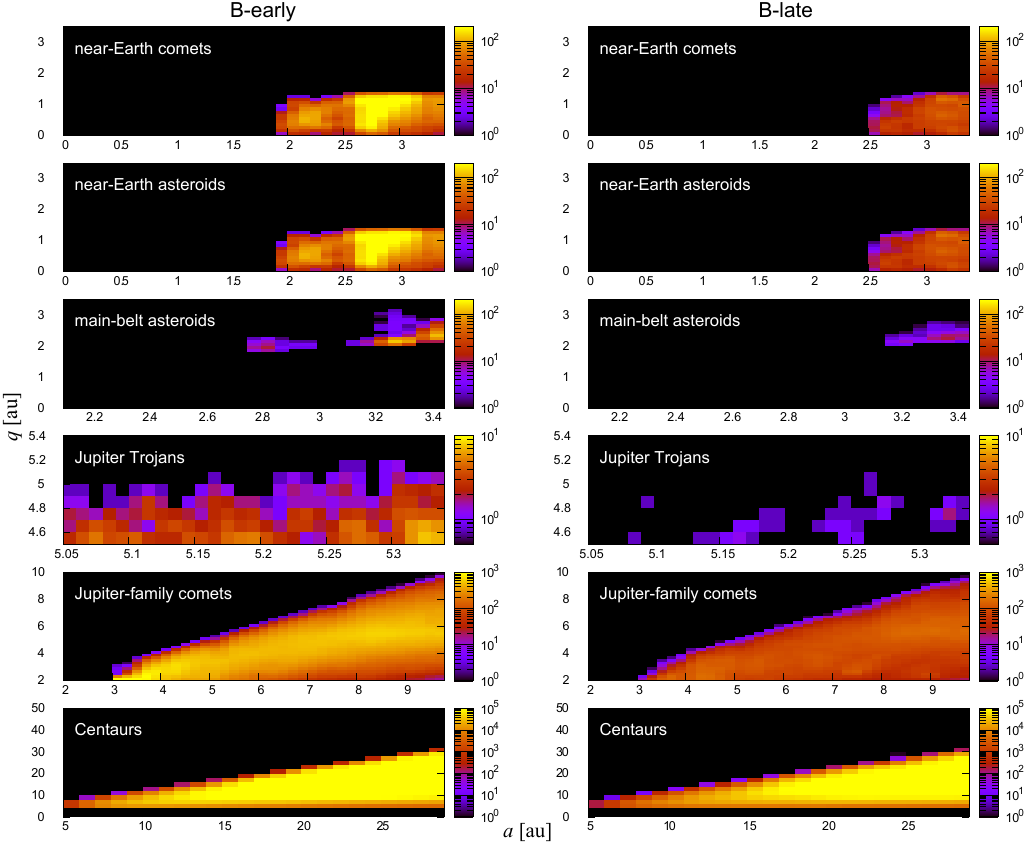}
\caption[]{%
This figure corresponds to Figure \ref{fig:oc-group-elem-aq_s5A}.
}
\label{fig:oc-group-elem-aq_s3B}
\end{figure}

\begin{figure}[!htbp]\centering
\includegraphics[width=\myfigwidth]{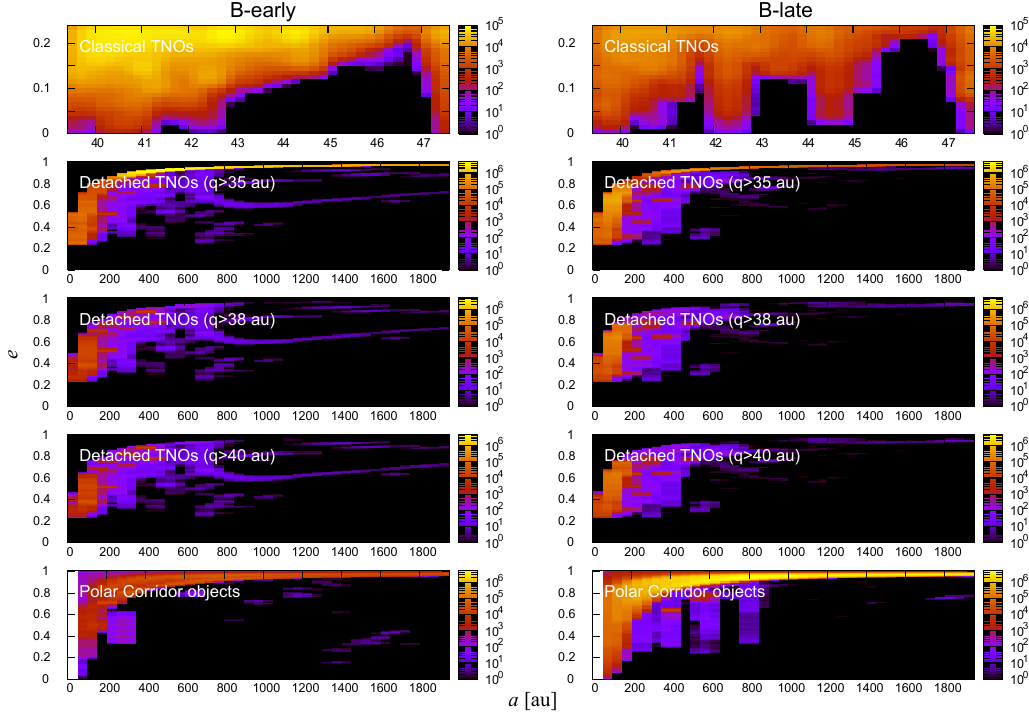}
\caption[]{%
This figure corresponds to Figure \ref{fig:oc-group-elem-ae-gladman_s5A}.
}
\label{fig:oc-group-elem-ae-gladman_s3B}
\end{figure}

\begin{figure}[!htbp]\centering
\includegraphics[width=\myfigwidth]{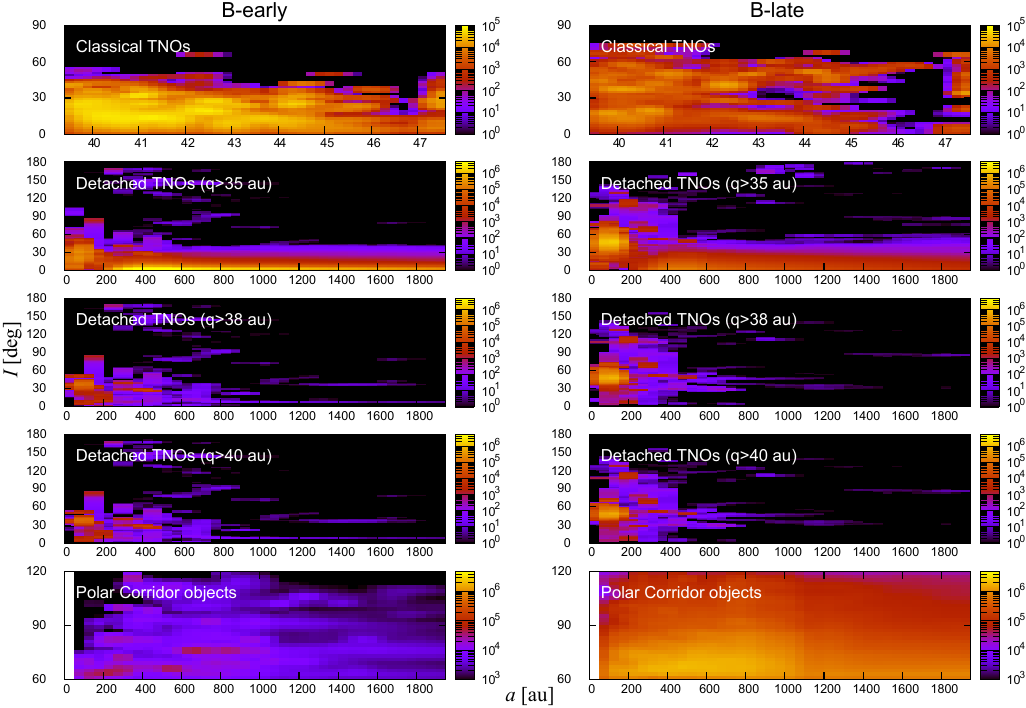}
\caption[]{%
This figure corresponds to Figure \ref{fig:oc-group-elem-ai-gladman_s5A}.
}
\label{fig:oc-group-elem-ai-gladman_s3B}
\end{figure}

\begin{figure}[!htbp]\centering
\includegraphics[width=\myfigwidth]{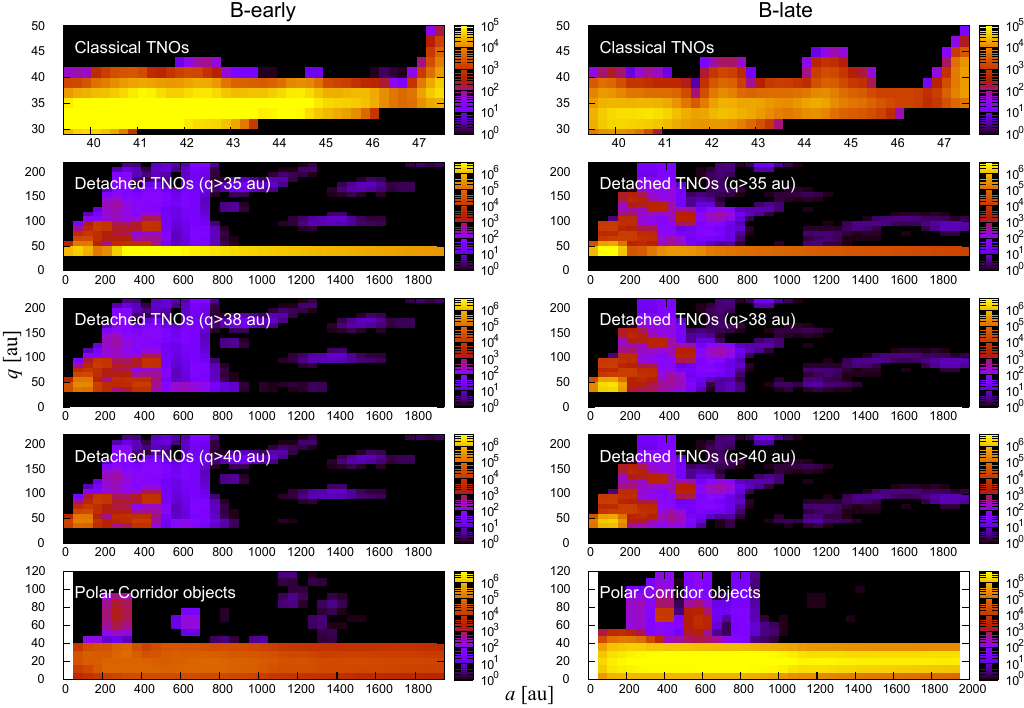}
\caption[]{%
This figure corresponds to Figure \ref{fig:oc-group-elem-aq-gladman_s5A}.
}
\label{fig:oc-group-elem-aq-gladman_s3B}
\end{figure}

\begin{figure}[!htbp]
  \includegraphics[width=\myfigwidth]{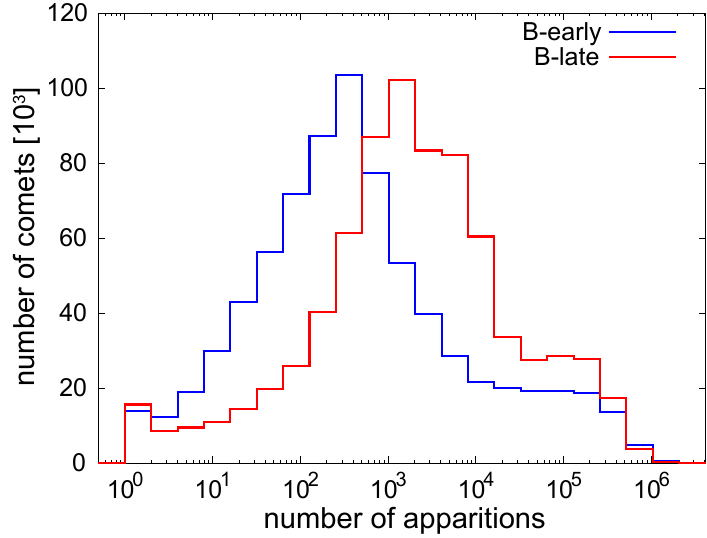}
  \caption[]{%
This figure corresponds to Figure \ref{fig:oc-nappari_s5A}.
}
\label{fig:oc-nappari_s3B}
\end{figure}

\begin{figure}[!htbp]\centering
\includegraphics[width=\myfigwidth]{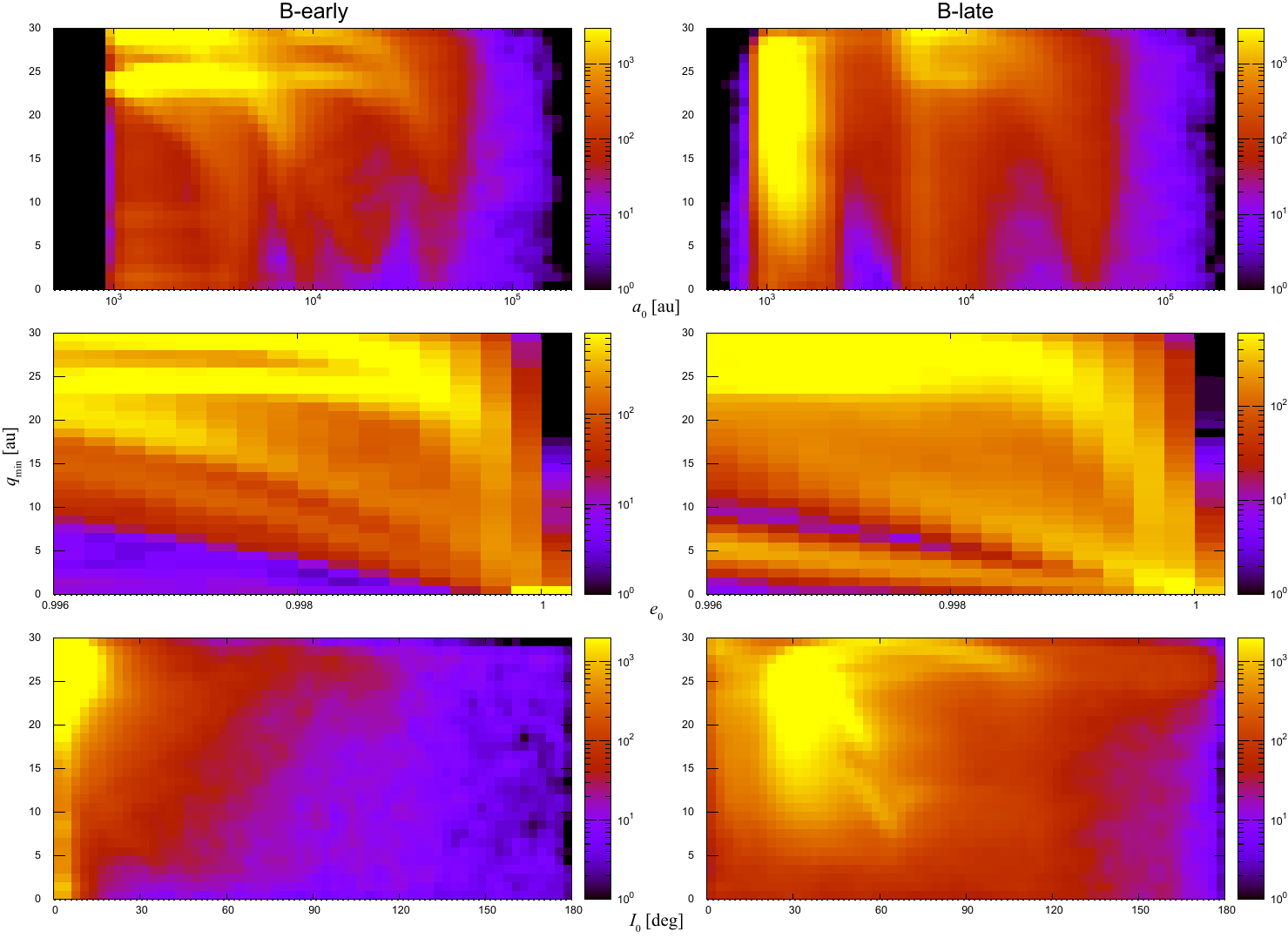}
\caption[]{%
This figure corresponds to Figure \ref{fig:oc-qmdep-aeI-el-log_s5A}.
}
\label{fig:oc-qmdep-aeI-el-log_s3B}
\end{figure}

\begin{figure}[!htbp]\centering
\includegraphics[width=\myfigwidth]{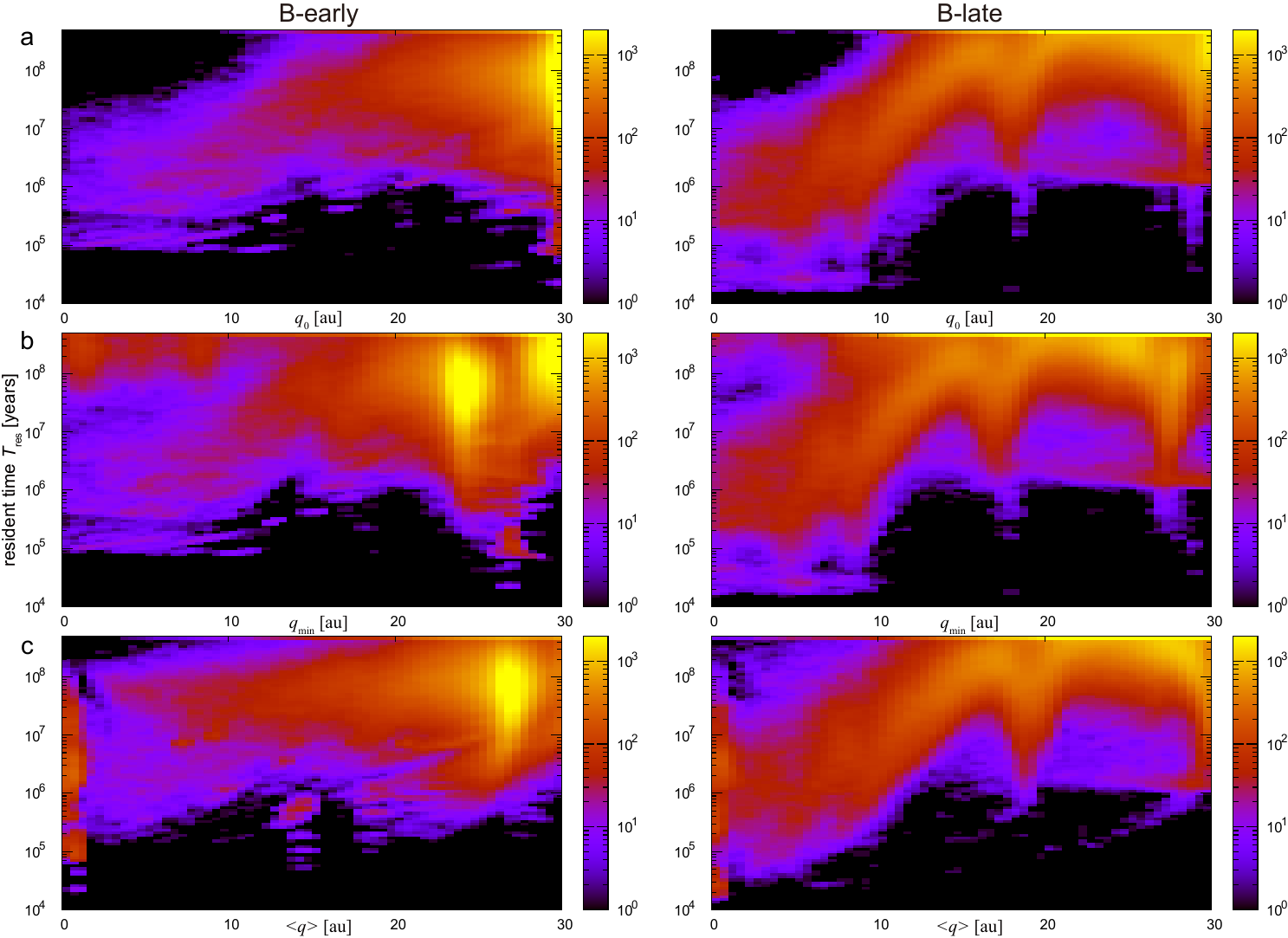}
\caption[]{%
This figure corresponds to Figure \ref{fig:oc-q-Te-log_s5A}.
}
\label{fig:oc-q-Te-log_s3B}
\end{figure}

\clearpage
\addcontentsline{toc}{section}{References}

\input{4th_arxiv_bbl}
\end{document}

%% file: pream4.tex
\usepackage{amssymb}

\long\def\comment#1{}
\comment{
   Usage:
     \comment{ comments....}
}

\makeatletter
\def\DD{\@ifnextchar [{\D@D}{\D@D[]}}
\def\D@D[#1]#2#3{\frac{d^{#1} #2}{d #3{}^{#1}}}
\def\DP{\@ifnextchar [{\D@P}{\D@P[]}}
\def\D@P[#1]{\@ifnextchar [{\D@Pi[#1]}{\D@PD[#1]}}
\def\D@PD[#1]#2#3{\frac{\partial^{#1} #2}{\partial #3{}^{#1}}}
\def\D@Pi[#1][#2]#3#4{\left( \frac{\partial^{#1} #3}{\partial #4{}^{#1}}\right)_{#2}}
\makeatother

\newcommand*\patchAmsMathEnvironmentForLineno[1]{%
  \expandafter\let\csname old#1\expandafter\endcsname\csname #1\endcsname
  \expandafter\let\csname oldend#1\expandafter\endcsname\csname end#1\endcsname
  \renewenvironment{#1}%
     {\linenomath\csname old#1\endcsname}%
     {\csname oldend#1\endcsname\endlinenomath}}%
\newcommand*\patchBothAmsMathEnvironmentsForLineno[1]{%
  \patchAmsMathEnvironmentForLineno{#1}%
  \patchAmsMathEnvironmentForLineno{#1*}}%
\AtBeginDocument{%
\patchBothAmsMathEnvironmentsForLineno{equation}%
\patchBothAmsMathEnvironmentsForLineno{align}%
\patchBothAmsMathEnvironmentsForLineno{flalign}%
\patchBothAmsMathEnvironmentsForLineno{alignat}%
\patchBothAmsMathEnvironmentsForLineno{gather}%
\patchBothAmsMathEnvironmentsForLineno{multline}%
}

\usepackage{lineno}

\usepackage{xcolor}

\ifdvipdfmgraphicx
  \usepackage[dvipdfmx]{graphicx}
\else
  \usepackage{graphicx}
\fi

\usepackage{url}
\ifUseOverleaf
  \usepackage{hyperref}
\else
  \ifdvipdfmgraphicx
    \usepackage[dvipdfmx]{hyperref}
  \else
    \usepackage{hyperref}
  \fi
\fi

\hypersetup{
    colorlinks,
    linkcolor={red!50!black},
    citecolor={blue!50!black},
    urlcolor={blue!80!black}
}
\ifUseOverleaf
  \usepackage{color}
\else
  \ifdvipdfmgraphicx
    \usepackage[dvipdfmx]{color}
  \else
    \usepackage{color}
  \fi
\fi

\usepackage{amsmath,amsfonts,amssymb,mathrsfs,bm,bbm}

\usepackage{mathabx}

\usepackage{soul}

 \definecolor{applegreen}{rgb}{0.55,0.71,0.0}
 \definecolor{forestgreen}{rgb}{0.13,0.55,0.13}
 \definecolor{pinegreen}{rgb}{0.0,0.47,0.44}
 \definecolor{upforestgreen}{rgb}{0.0,0.27,0.13}
 \definecolor{vividviolet}{rgb}{0.62,0.0,1.0}
 \definecolor{lightcyan}{rgb}{0.717647, 0.933071, 0.996078} 
 \definecolor{scarlet}{rgb}{0.7421875, 0.00390625, 0.09765625}

\soulregister\Hl{7}
\soulregister\ref7
\soulregister\cite7
\soulregister\citet7
\soulregister\pageref7

\usepackage{here}

\usepackage{bigfoot}

\newcounter{revcomcounter}
\usepackage{natbib}

\usepackage{ascmac}

\usepackage[yyyymmdd,hhmmss]{datetime}

\usepackage[OT2,T1]{fontenc}
\usepackage[english,russian]{babel}

\usepackage{textcomp}

\usepackage{ulem}
 \DeclareRobustCommand{\erase}{\bgroup\markoverwith{\textcolor{red}{\rule[.5ex]{2pt}{1.6pt}}}\ULon}

\usepackage{marginnote}
 at 7truept

\usepackage{orcidlink}  

\usepackage{relsize}

\usepackage{multicol}

\usepackage{wrapfig}

\usepackage{framed}

\usepackage{array}

\makeatletter
\def\mathcolor#1#{\@mathcolor{#1}}
\def\@mathcolor#1#2#3{%
  \protect\leavevmode
  \begingroup
    \color#1{#2}#3%
  \endgroup
}
\makeatother

 \usepackage{wasysym}

\makeatletter
\DeclareFontFamily{U}{tipa}{}
\DeclareFontShape{U}{tipa}{m}{n}{<->tipa10}{}
\newcommand{\arc@char}{{\usefont{U}{tipa}{m}{n}\symbol{62}}}%
\newcommand{\arc}[1]{\mathpalette\arc@arc{#1}}
\newcommand{\arc@arc}[2]{%
  \sbox0{$\m@th#1#2$}%
  \vbox{
    \hbox{\resizebox{\wd0}{\height}{\arc@char}}
    \nointerlineskip
    \box0
  }%
}
\makeatother

\usepackage{cancel}

\usepackage{tikz}

\def\beq{\begin{equation}}
\def\eeq{\end{equation}}
\def\bea{\begin{aligned}}
\def\eea{\end{aligned}}

\newcommand{\lw}[1]{\smash{\lower2.ex\hbox{#1}}}
\newcommand{\llw}[1]{\smash{\lower4.ex\hbox{#1}}}

\newcommand{\barray}[1]{\begin{array}{#1}}
\newcommand{\earray}{\end{array}}

\makeatletter
\newsavebox{\@brx}
\newcommand{\llangle}[1][]{\savebox{\@brx}{\(\m@th{#1\langle}\)}%
  \mathopen{\copy\@brx\kern-0.5\wd\@brx\usebox{\@brx}}}
\newcommand{\rrangle}[1][]{\savebox{\@brx}{\(\m@th{#1\rangle}\)}%
  \mathclose{\copy\@brx\kern-0.5\wd\@brx\usebox{\@brx}}}
\makeatother

{\end{list}}

\usepackage{lscape}

\makeatletter
\def\Hline{%
\noalign{\ifnum0=`}\fi\hrule \@height 0.75pt \futurelet
\reserved@a\@xhline}
\makeatother

